\documentclass[12pt]{article}

\usepackage{geometry, fullpage, authblk}
\usepackage{xr-hyper}
\usepackage{times}
\usepackage{hyperref}[]
\hypersetup{
	colorlinks=true,
	linkcolor=darkblue,
	filecolor=magenta,      
	urlcolor=darkblue,
	citecolor=blue,
}
\usepackage[normalem]{ulem}
\usepackage{multicol}

\usepackage{url}

\usepackage{bigints}
\usepackage{amsfonts,amsmath,amssymb,amsthm, bbm}
\usepackage{wrapfig}
\usepackage{verbatim,float,url}
\usepackage{graphicx}
\usepackage{sidecap}
\usepackage{subfigure}
\usepackage{subcaption}
\usepackage[skip=0pt]{caption}
\usepackage{natbib}
\usepackage[english]{babel}
\usepackage{cancel}

\usepackage{color}
\definecolor{darkblue}{RGB}{10, 10, 200}

\usepackage[dvipsnames]{xcolor}
\usepackage{cleveref}
\usepackage{enumerate}
\usepackage{multirow}
\usepackage{amsmath}

\usepackage{collcell}
\makeatletter
\newcolumntype{G}{>{\collectcell\@gobble}c<{\endcollectcell}@{}}
\makeatother

\usepackage{bm}

\usepackage{mathtools}

\usepackage[dvipsnames]{xcolor}
\definecolor{reddark}{rgb}{0.68, 0.0, 0.0}
\definecolor{darkblue}{rgb}{0.0, 0.0, 0.64}
\definecolor{orange2}{rgb}{0.98, 0.6, 0.01}

\usepackage{pifont}

\newtheorem{proposition}{Proposition}

\usepackage{mdwlist}

\theoremstyle{definition}

\usepackage{tikz}  
\usetikzlibrary{trees}
\usetikzlibrary{arrows.meta}
\usetikzlibrary{decorations.pathreplacing}
\usetikzlibrary{bayesnet}

 \usepackage{titlesec}
\titlespacing*\section{0pt}{0pt plus 2pt minus 2pt}{0pt plus 2pt minus 2pt}
\titlespacing*\subsection{0pt}{0pt plus 2pt minus 2pt}{0pt plus 2pt minus 2pt}
\titlespacing*\subsubsection{0pt}{0pt plus 2pt minus 2pt}{0pt plus 2pt minus 2pt}

\usepackage{ctable}
\usepackage{colortbl}
\usepackage{dsfont}
\usepackage{mathtools}

\renewcommand{\&}{and}
 
\usepackage{ulem}

\usepackage[usenames,dvipsnames]{xcolor}

\usepackage[ruled,vlined]{algorithm2e}
\SetKwInput{kwInit}{Initialize}

\theoremstyle{plain}
\allowdisplaybreaks
\usepackage{bm}

\title{Flexible variational approximations for stochastic volatility-managed portfolios}

\newcommand{\abs}{

This paper investigates volatility-managed portfolios through a variational Bayes approach that smooths stochastic volatility forecasts. Our algorithm exhibits competitive performance relative to established methods both in terms of inferential accuracy and computational efficiency. Analyzing equity factors and characteristic-based portfolios, we show that smoothing reduces excess leverage and turnover, significantly improving risk-adjusted returns after transaction costs. Importantly, our smoothed stochastic volatility method achieves positive net performance while avoiding extreme negative outcomes, a combination no competing approach attains. We demonstrate that superior forecasting accuracy does not guarantee superior portfolio performance, revealing a fundamental trade-off between model precision and economic utility.

\vspace{0.1in}

\noindent\textbf{Keywords:} Bayesian inference; Portfolio optimization; Smoothing splines; Wavelets.

}

\author[1]{Nicolas Bianco
}
\author[2]{Mauro Bernardi}
\affil[1]{Scientific Computing Center, Karlsruhe Institute of Technology, Germany}
\affil[2]{Department of Statistical Sciences, University of Padova, Italy}

\begin{document}
	
\maketitle
\thispagestyle{empty}

\centerline{\bf Abstract}
\medskip
\abs
\normalsize
\setlength{\parskip}{.2cm }
\setlength{\parindent}{0.5cm}

\clearpage
\pagenumbering{arabic}


\section{Introduction}\label{sec:intro}
The well-documented empirical evidence that variance is highly predictable at short horizons, yet variance forecasts are only weakly related to future returns, has led to the adoption of volatility-managed portfolios as a practical risk-management tool \citep{moreira2017volatility}. The most common implementation scales investment exposure by the inverse of the previous month's realized variance, effectively increasing risk exposure when volatility is low and decreasing it when volatility is high. 

While this approach exploits time-variation in volatility without requiring return predictability, it comes with practical drawbacks. Realized variance estimates, while statistically unbiased, contain substantial high-frequency noise that generates economically costly but informationally irrelevant portfolio adjustments, leading to high liquidity needs and trading costs. Figure \ref{fig:Leverage heatmaps} highlights that realized variance often require over tenfold capital investment in the original portfolios, a phenomenon pervasive across equity strategies.

These costs are particularly burdensome for investors, creating a wedge between the maximum (unconstrained) Sharpe ratio and the risk-adjusted returns obtained under limits-to-arbitrage, which prevents highly leveraged allocations in practice. One common approach to mitigating these costs is the introduction of leverage constraints \citep{cederburg2020performance}. While effective in reducing liquidity demands, such constraints are often set arbitrarily. Their effectiveness depends heavily on the properties of the underlying volatility estimates: more erratic estimates make constraints more binding, whereas smoother estimates provide greater capital efficiency.

This paper presents a novel approach to improve the risk-return profile of volatility-managed portfolios. Rather than relying on ad-hoc leverage constraints, we propose a Variational Bayes (VB) framework that smooths the predictive density of a stochastic volatility model. Our smoothing approach mitigates the effect of transient volatility fluctuations while preserving responses to persistent volatility changes. This reduces turnover and leverage spikes while maintaining the economic benefits of volatility timing.

It is important to emphasize that our approach does not seek to redesign the original equity factors underlying volatility scaling. For the purposes of this paper, selecting a factor or long-short portfolio to invest in is treated as an upstream task within the investment decision-making process. Instead, we focus on stabilizing capital expenditures under portfolio-specific time-varying volatility.

Our contributions are threefold. First, we develop a flexible VB algorithm that enables smoothing the predictive density of an otherwise standard stochastic volatility model. The algorithm balances computational efficiency with accuracy, demonstrating competitive performance relative to established Markov Chain Monte Carlo (MCMC) \citep{stochvol_package} and variational inference \citep{chan_yu2022}, and optimization-based inference \citep{stochvolTMB} approaches. 

Second, across a wide set of portfolios spanning almost a century of monthly returns, our smoothed stochastic volatility (SSV) approach delivers systematic improvements in risk-adjusted returns after transaction costs. Joint significance tests and complete performance distribution analysis demonstrate that SSV achieves a 45.8\% success rate—close to the theoretical 50\% breakeven point—while all competing methods exhibit significantly lower success rates (ranging from 4.2\% to 33.3\%) and occasional extreme negative performances.

Third, we provide evidence on the trade-off between forecasting accuracy and economic utility in volatility-managed portfolios. We show that superior forecasting accuracy -- as proxied by the out-of-sample $R^2$ -- does not guarantee superior profitability. This finding supports our central thesis that optimizing for economic utility rather than statistical accuracy often proves beneficial in practical portfolio management \citep{bianchi2024rethinking}.

In addition to \citet{moreira2017volatility}, our work contributes to the growing literature exploring the economic value of volatility-managed portfolios \citep[e.g.,][]{harvey2018impact,bongaerts2020conditional,cederburg2020performance,liu2019volatility,barroso2021limits,wang2021downside,bovzovic2024vix,demiguel2024multifactor,schwarz2025performance}. We emphasize how regularizing volatility forecasts improve the transaction cost-adjusted performance. 

A second strand of literature we contribute to focuses on the Bayesian estimation of stochastic volatility models. Bayesian methods were first introduced by \citet{kim_etal.1998,jacquier2002bayesian,shephard_pitt.2004}.\footnote{Established frequentist approaches include the generalized method of moments (GMM) \citep{melino_turnbull.1990}, quasi-maximum likelihood (QML) \citep{harvey_etal.1994,ruiz.1994}, and the efficient method of moments (EMM) \citep{gallant_etal.1997}.} Our contribution extends this literature by employing variational Bayes inference to flexibly smooth the predictive density of the latent volatility state, regardless of the underlying persistence. Our approach is general and adaptable, allowing for various smoothness assumptions without altering the structure of the underlying stochastic volatility model.

Finally, we contribute to the growing adoption of approximate inference methods in economics and finance. While variational inference has become a popular approach in applied macroeconomics \citep{gefang_koop_poon2019,gunawan2021variational,chan_yu2022,loaiza2022fast,koop2023bayesian}, its applications in finance is limited \citep{bernardi2024variational,loaiza2024efficient}. Our study bridges this gap and highlights the potential of approximate methods for addressing practical challenges in investment decision modelling.

\section{Model specification and inference}
\label{sec:univ_model}

Consider a standard univariate stochastic volatility (SV) model, as described in \citet{taylor1994modeling}. Without loss of generality, the model can be expressed in a state-space form as follows:
\begin{align}
    y_t &= \mathbf{x}_t^\intercal\beta + \exp(h_t/2)\varepsilon_t, \quad \varepsilon_t \sim \mathsf{N}(0,1), \label{eq:svm1a} \\
    h_t &= c + \rho (h_{t-1} - c) + u_t, \quad u_t \sim \mathsf{N}(0,\eta^2), \label{eq:svm1b}
\end{align}
where \( y_t \in \mathbb{R} \), \( \mathbf{x}_t \in \mathbb{R}^p \), and \( h_t = \log\sigma_t^2 \) represent, respectively, the log-return, a set of covariates, and the log-volatility of returns at time \( t \), for \( t = 1, 2, \dots, n \). The error terms \( \varepsilon_t \) and \( u_t \) are mutually independent Gaussian white noise processes. 

For simplicity, we consider the latent state in Eq.~\eqref{eq:svm1b} as a conventional AR(1) process, with unconditional mean \( c \), persistence \( \rho \), and conditional variance \( \eta^2 \). With \( |\rho| < 1 \), the initial state \( h_0 \) can be sampled from the marginal distribution \( h_0 \sim \mathsf{N}\left(c, \eta^2/(1-\rho^2)\right) \). It is worth highlighting that our framework is general and can accommodate higher-order autoregressive processes for the latent stochastic volatility \( h_t \), as long as they remain stationary. The AR(1) assumption is motivated by the desire for direct comparability with the existing literature \citep[e.g.,][]{jacquier2002bayesian,chan_yu2022}.

Finally, although we focus on the univariate case, the methodology developed here extends naturally to multivariate SV settings: via a Cholesky decomposition of the time-varying covariance matrix, a multivariate SV model reduces to a sequence of univariate SV models, to which our estimation approach applies directly. We leave this extension for future work, as the case study in Section~\ref{sec:appl} only requires the univariate setting.

\subsection{Variational Bayes}
\label{subsec:variational inference}
A variational Bayes (VB) approach minimizes the Kullback-Leibler (KL) divergence between an approximating density \(q(\boldsymbol{\vartheta})\) and the true posterior density \(p(\boldsymbol{\vartheta}|\mathbf{y})\) \citep{Blei.2017}. As shown by \citet{ormerod2010explaining}, minimizing this KL divergence is equivalent to maximizing the Evidence Lower Bound (ELBO), denoted as \(\underline{p}(\mathbf{y};q)\):
\begin{equation}
\label{eq:vb_elbo_optim}
q^*(\boldsymbol{\vartheta}) = \arg\max_{q(\boldsymbol{\vartheta}) \in \mathcal{Q}} \log \underline{p}(\mathbf{y};q) \quad \text{where} \quad
\underline{p}(\mathbf{y};q) = \int q(\boldsymbol{\vartheta}) \log \left\{ \frac{p(\mathbf{y},\boldsymbol{\vartheta})}{q(\boldsymbol{\vartheta})} \right\} \, d\boldsymbol{\vartheta}.
\end{equation}
Here, \(q^*(\boldsymbol{\vartheta}) \in \mathcal{Q}\) represents the optimal variational density, and \(\mathcal{Q}\) is a family of candidate distributions. The choice of \(\mathcal{Q}\) is critical, as it determines the computational and practical feasibility of the approach. 

We adopt the Mean-Field Variational Bayes (MFVB) method, which assumes a factorization of the variational density \(q(\boldsymbol{\vartheta}) = \prod_{j=1}^d q(\boldsymbol{\vartheta}_j),\) where \(\{\boldsymbol{\vartheta}_1, \dots, \boldsymbol{\vartheta}_d\}\) is a partition of the parameter vector \(\boldsymbol{\vartheta}\). This factorization simplifies the optimization problem by enabling the derivation of the optimal variational density for each component:
\begin{equation}
q^*(\boldsymbol{\vartheta}_j) \propto \exp\left\{ \mathbb{E}_{-\boldsymbol{\vartheta}_j} \left[ \log p(\mathbf{y}, \boldsymbol{\vartheta}) \right] \right\}, \label{eq:mfvb}
\end{equation}
where \(\mathbb{E}_{-\boldsymbol{\vartheta}_j}\) denotes the expectation with respect to the variational density over all parameters except \(\boldsymbol{\vartheta}_j\). \citet{ormerod2010explaining} show that this can be approximated as:
\begin{equation}
\log q^*(\boldsymbol{\vartheta}_j) \propto \mathbb{E}_{-\boldsymbol{\vartheta}_j} \left[ \log p(\boldsymbol{\vartheta}_j \mid \text{rest}) \right],
\end{equation}
where \(p(\boldsymbol{\vartheta}_j \mid \text{rest})\) is the full conditional distribution of \(\boldsymbol{\vartheta}_j\). Consequently, the optimal densities \(q^*(\boldsymbol{\vartheta}_j), \ j=1, \ldots, d\), can be estimated using a coordinate ascent algorithm. Additional restrictions may be imposed to the extent that \(q^*(\boldsymbol{\vartheta}_j)\) is too complex to handle. For instance, assuming that \(q^*(\boldsymbol{\vartheta}_j)\) belongs to a specific parametric family results in a semi-parametric approach \citep{rohde_wand2016}.

For the AR(1) stochastic volatility model in Eq.\eqref{eq:svm1a}-Eq.\eqref{eq:svm1b}, the parameter vector is \(\boldsymbol{\vartheta} = (\mathbf{h}^\intercal, \mathbf{\beta}^\intercal, c, \rho, \eta^2)^\intercal\), representing the latent states and model parameters. The joint density \(p(\mathbf{y}, \boldsymbol{\vartheta})\) forms the basis of the mean-field factorization, and the variational density can be written as \(q(\boldsymbol{\vartheta}) = q(\mathbf{h}) q(\mathbf{\beta}) q(c) q(\rho) q(\eta^2)\). In the following sections, we describe each component of \(q(\boldsymbol{\vartheta})\) in detail. The full set of proofs is provided in Supplementary Material \ref{app:params}.

\subsection{Optimal variational densities} 

Assume that \(\mathbf{y}\) is an \(n\)-dimensional vector, \(\mathbf{X}\) a \(p \times n\) matrix, and \(\mathbf{\beta}\) a \(p\)-dimensional vector of parameters whose prior distribution is \(\mathbf{\beta} \sim \mathsf{N}_p\left(\mathbf{\mu}_\beta, \mathbf{\Sigma}_\beta\right)\). Proposition~\ref{eq:prop1} provides the corresponding optimal variational density.

\begin{proposition}\label{eq:prop1}
The optimal variational density for \(\mathbf{\beta}\) is \(q^\ast(\mathbf{\beta}) \equiv \mathsf{N}_p(\mathbf{\mu}_{q(\beta)}, \mathbf{\Sigma}_{q(\beta)})\), with
\begin{align}\label{eq:up_beta}
\mathbf{\Sigma}_{q(\beta)} &= \left(\mathbf{X}^\intercal\mathbf{H}^{-1}\mathbf{X} + \mathbf{\Sigma}_\beta^{-1}\right)^{-1}, \\
\mathbf{\mu}_{q(\beta)} &= \mathbf{\Sigma}_{q(\beta)}\left(\mathbf{X}^\intercal\mathbf{H}^{-1}\mathbf{y} + \mathbf{\Sigma}_\beta^{-1}\mathbf{\mu}_\beta\right),
\end{align}
where \(\mathbf{H} = \mathsf{Diag}\left(\mathbb{E}_q\left[\mathrm{e}^{\mathbf{h}}\right]\right)\) is a diagonal matrix with elements that depend on the optimal density for the latent log-volatilities \(q\left(\mathbf{h}\right)\) (see Proposition~\ref{eq:prop5}).
\end{proposition}

Supplementary Material \ref{eq:prop_beta} provides the proof and the analytical derivation of \(q^\ast(\mathbf{\beta})\). The derivation leverages the specification of the full conditional distribution \(p\left(\mathbf{\beta} \mid \text{rest}\right)\), similar to a conventional Gibbs sampler.

We note that the marginal distribution \(p\left(\mathbf{h}\right)\) of the joint vector \(\mathbf{h}^\intercal = (h_0, h_1, \ldots, h_n)\) admits a Gaussian Markov Random Field (GMRF) representation \(\mathbf{h} \sim \mathsf{N}_{n+1}(c\boldsymbol{\iota}_{n+1}, \eta^2\mathbf{Q}^{-1})\), which preserves the time dependence structure of the autoregressive process. Specifically, the matrix \(\mathbf{Q}\) is a tridiagonal precision matrix with diagonal elements \(q_{1,1} = q_{n+1,n+1} = 1\) and \(q_{i,i} = 1 + \rho^2\) for \(i = 2, \ldots, n\), and off-diagonal elements \(q_{i,j} = -\rho\) if \(|i-j| = 1\), and \(0\) elsewhere \citep[see][]{rue_held.2005}. 

We exploit this representation to obtain the approximating density for the latent state and the corresponding parameters \(c, \rho, \eta^2\). Specifically, we assume a standard Gaussian prior \(c \sim \mathsf{N}\left(\mu_c, \sigma^2\right)\) for the unconditional mean of the latent state. Proposition~\ref{eq:prop2} provides the corresponding optimal variational density.

\begin{proposition}\label{eq:prop2}
The optimal variational density for \(c\) is \(q^\ast(c) \equiv \mathsf{N}(\mu_{q(c)}, \sigma^2_{q(c)})\), where:
\begin{equation}\begin{aligned}\label{eq:up_c}
    \sigma^2_{q(c)} &= \left(\mu_{q(1/\eta^2)} \boldsymbol{\iota}_{n+1}^\intercal \mathbf{\mu}_{q(\mathbf{Q})} \boldsymbol{\iota}_{n+1} + 1/\sigma^2_c\right)^{-1}, \\
    \mu_{q(c)} &= \sigma^2_{q(c)} \left(\mu_{q(1/\eta^2)} \boldsymbol{\iota}_{n+1}^\intercal \mathbf{\mu}_{q(\mathbf{Q})} \mathbf{\mu}_{q(\mathbf{h})} + \mu_c / \sigma^2_c\right),
\end{aligned}\end{equation}
with \(\mu_{q(\mathbf{Q})}\) the element-wise expectation of the tridiagonal matrix \(\mathbf{Q}\) specified above.
\end{proposition}

Supplementary Material \ref{eq:prop_c} provides the proof and the analytical derivation of \(q^\ast(c)\). Proposition~\ref{eq:prop3} provides the optimal variational density of the autoregressive parameter \(\rho\).

\begin{proposition}\label{eq:prop3}
The optimal variational density for \(\rho\) can be written as:
\begin{equation}\label{eq:up_rho}
    \log q^\ast(\rho) \propto \frac{1}{2} \log(1-\rho^2) - \frac{1}{2} \mu_{q(1/\eta^2)} \left(\sum_{t=1}^{n-1} a_t\right) \left(\rho - \frac{\sum_{t=0}^{n-1} b_t}{\sum_{t=1}^{n-1} a_t}\right)^2, \quad \rho \in (-1,1),
\end{equation}
with
\begin{align}\label{eq:ab_eq}
    a_t &= \mathbb{E}_q\left[(h_t - c)^2\right] = (\mu_{q(h_t)} - \mu_{q(c)})^2 + \sigma^2_{q(h_t)} + \sigma^2_{q(c)}, \\
    b_t &= \mathbb{E}_q\left[(h_t - c)(h_{t+1} - c)\right] = (\mu_{q(h_t)} - \mu_{q(c)})(\mu_{q(h_{t+1})} - \mu_{q(c)}) + \sigma_{q(h_t, h_{t+1})} + \sigma^2_{q(c)},
\end{align}
where \(\sigma_{q(h_t, h_{t+1})}\) denotes the covariance between \(h_t\) and \(h_{t+1}\) under the optimal variational density for the latent state \(\mathbf{h}\) (see Proposition~\ref{eq:prop5}). Thus, the normalizing constant and the first two moments can be found by Monte Carlo methods, sampling from a Gaussian distribution with mean \(\frac{\sum_{t=0}^{n-1} b_t}{\sum_{t=1}^{n-1} a_t}\) and precision \(\mu_{q(1/\eta^2)} \left(\sum_{t=1}^{n-1} a_t\right)\).
\end{proposition}
Supplementary Material \ref{eq:prop_rho} provides the complete proof and analytical derivation for \(q^\ast(\rho)\). We assume a standard Inverse-Gamma uninformative prior \(\eta^2 \sim \mathsf{IG}(A,B)\), with \(A = B = 0.001\), for the latent state variance \(\eta^2\). Proposition~\ref{eq:prop4} provides the optimal variational density. 

\begin{proposition}\label{eq:prop4}
The optimal variational density for \(\eta^2\) is \(q^\ast(\eta^2) \equiv \mathsf{IG}(A_{q(\eta^2)}, B_{q(\eta^2)})\), where:
\begin{equation}\begin{aligned}\label{eq:up_eta2}
    A_{q(\eta^2)} &= A + \frac{n+1}{2}, \\
    B_{q(\eta^2)} &= B + \frac{1}{2} (\mathbf{\mu}_{q(\mathbf{h})} - \mu_{q(c)} \boldsymbol{\iota}_{n+1})^\intercal \mathbf{\mu}_{q(\mathbf{Q})} (\mathbf{\mu}_{q(\mathbf{h})} - \mu_{q(c)} \boldsymbol{\iota}_{n+1}) \\
    &\qquad\quad + \frac{1}{2} \left(\mathsf{tr}\left\{\mathbf{\Sigma}_{q(\mathbf{h})} \mathbf{\mu}_{q(\mathbf{Q})}\right\} + \sigma^2_{q(c)} \boldsymbol{\iota}_{n+1}^\intercal \mathbf{\mu}_{q(\mathbf{Q})} \boldsymbol{\iota}_{n+1}\right),
\end{aligned}\end{equation}
where \(\mathbf{\mu}_{q(\mathbf{h})}, \mathbf{\Sigma}_{q(\mathbf{h})}\) are the mean vector and covariance matrix of the optimal variational density for the latent state \(\mathbf{h}\) (see Proposition~\ref{eq:prop5}). Note that \(\mu_{q(1/\eta^2)} = A_{q(\eta^2)} / B_{q(\eta^2)}\).
\end{proposition}
Supplementary Material \ref{eq:prop_eta2} provides the complete proof and analytical derivation.

An important point concerns our approach to achieving smoothness in volatility estimates. Rather than relying on shrinkage priors \citep{bitto2019achieving,knaus2023dynamic} -- which often introduce additional hierarchical layers and computational complexity -- our method achieves smoothness by deliberately distorting the true posterior, thus defining its closest smooth approximation in the Kullback-Leibler sense, while maintaining a weakly-informative prior specification $\eta^2 \sim \mathsf{IG}(0.001,0.001)$.

This approach provides controlled smoothness without altering the underlying economic model or imposing restrictive prior beliefs about $\eta^2$. In addition, the degree of smoothness can be calibrated through the choice of the basis matrix \(\mathbf{W}\) rather than through prior hyperparameters, offering greater flexibility and interpretability. 

While both approaches ultimately introduce smoothness on the dynamic of the latent process, our method operates through the approximation mechanism rather than the prior specification. As demonstrated in Figure \ref{fig:sim_par_hat} in Supplementary Material \ref{app:simulation}, the smooth variational approximation leads to underestimation of \(\eta^2\), effectively mimicking the behavior of a tight shrinkage prior but with more direct control over the smoothing mechanism.

\subsection{Smoothing stochastic volatility}
\label{subsec:MatrixW}

Consistent with the representation \(\mathbf{h} \sim \mathsf{N}_{n+1}(c\boldsymbol{\iota}_{n+1}, \eta^2 \mathbf{Q}^{-1})\), the variational density \(q(\mathbf{h})\) is assumed to take the form \(\mathbf{h} \sim \mathsf{N}_{n+1}(\mathbf{\mu}_{q(\mathbf{h})}, \mathbf{\Omega}_{q(\mathbf{h})}^{-1})\) with mean vector \(\mathbf{\mu}_{q(\mathbf{h})} = \mathbf{W} \mathbf{f}_{q(\mathbf{h})}\) and variance-covariance matrix \(\mathbf{\Sigma}_{q(\mathbf{h})} = \mathbf{\Omega}_{q(\mathbf{h})}^{-1}\). The choice of \(\mathbf{\mu}_{q(h)}\) as a linear projection \(\mathbf{W} \mathbf{f}_{q(h)}\), with \(\mathbf{f}_{q(h)} \in \mathbb{R}^{k}\) the projection coefficients and \(\mathbf{W}\) an \((n+1) \times k\) deterministic matrix, has a direct effect on the estimates of the latent log-volatility state. This is a key feature of our estimation strategy since it allows us to customize the smoothness of the volatility forecasts without changing the underlying model assumptions. Before discussing different choices of \(\mathbf{W}\), we first introduce the optimal variational density \(q^\ast(\mathbf{h})\).

Given that the approximating density is multivariate normal, we can leverage the results in \citet{rohde_wand2016} to solve 
\[
\widehat{\boldsymbol{\xi}} = \arg\max_{\xi} \left\{\mathbb{E}_q(\log p(\mathbf{y}, \mathbf{h})) - \mathbb{E}_q(\log q(\mathbf{h}))\right\}
\]
to find the parameters \(\boldsymbol{\xi} = \left(\mathbf{f}_{q(h)}, \mathbf{\Sigma}_{q(h)}\right)\) based on a simple coordinate ascent algorithm. Proposition~\ref{eq:prop5} provides the corresponding iterative updating scheme. Supplementary Material \ref{eq:prop_h_het} provides the complete proof and analytical derivations.

\begin{proposition}\label{eq:prop5}
Let $\mathbf{\mu}_{q(\mathbf{s})}=(\mu_{q(s_1)},\ldots,\mu_{q(s_n)})^\intercal$ with $\mu_{q(s_t)} = (y_t-\mathbf{X}_t^\intercal\mathbf{\mu}_{q(\beta)})^2+\mathsf{tr}\left\{\mathbf{\Sigma}_{q(\beta)}\mathbf{X}_t\mathbf{X}_t^\intercal\right\}$, and $\mathbf{\mu}_{q(\beta)},\mathbf{\Sigma}_{q(\beta)}$ denote the variational mean and covariance from Proposition \ref{eq:prop1}. Assuming a representation $\mathbf{h}\sim\mathsf{N}_{n+1}(\mathbf{\mu}_{q(h)},\mathbf{\Omega}_{q(h)}^{-1})$, with mean vector $\mathbf{\mu}_{q(h)}=\mathbf{W}\mathbf{f}_{q(h)}$ and covariance matrix $\mathbf{\Sigma}_{q(h)}=\mathbf{\Omega}_{q(h)}^{-1}$, an iterative algorithm can be set as:
\begin{align}
	\mathbf{\Sigma}_{q(h)}^{new} &= \left[\nabla_{\boldsymbol{\mu}_{q(h)}\boldsymbol{\mu}_{q(h)}}^2 S(\mathbf{\mu}_{q(h)}^{old},\mathbf{\Sigma}_{q(h)}^{old})\right]^{-1}, \\
	\mathbf{f}_{q(h)}^{new} &= \mathbf{f}_{q(h)}^{old} + \mathbf{W}^+\,\mathbf{\Sigma}_{q(h)}^{new}\nabla_{\boldsymbol{\mu}_{q(h)}} S(\mathbf{\mu}_{q(h)}^{old},\mathbf{\Sigma}_{q(h)}^{old}),\\
	\mathbf{\mu}_{q(h)}^{new} &= \mathbf{W}\mathbf{f}_{q(h)}^{new},
\end{align}
with $\mathbf{W}^+=(\mathbf{W}^\intercal\mathbf{W})^{-1}\mathbf{W}^\intercal$ the left Moore–Penrose pseudo-inverse of $\mathbf{W}$, and $S(\mathbf{\mu}_{q(h)},\mathbf{\Sigma}_{q(h)})$ equal to $\mathbb{E}_q(\log p(\mathbf{h},\mathbf{y}))$ (see Supplementary Material \ref{eq:seq_appendix}), such that,
\begin{equation}\begin{aligned}
	\nabla_{\boldsymbol{\mu}_{q(h)}} S(\mathbf{\mu}_{q(h)},\mathbf{\Sigma}_{q(h)}) &= -\frac{1}{2}[0,\boldsymbol{\iota}_n^\intercal]^\intercal+\frac{1}{2}[0,\mathbf{\mu}_{q(\mathbf{s})}^\intercal]^\intercal\odot\mathrm{e}^{-\boldsymbol{\mu}_{q(h)}+\frac{1}{2}\mathsf{diag}\left(\boldsymbol{\Sigma}_{q(\mathbf{h})}\right)} \\
	&\qquad-\mu_{q(1/\eta^2)}\mathbf{\mu}_{q(\mathbf{Q})}(\mathbf{\mu}_{q(h)}-\mu_{q(c)}\boldsymbol{\iota}_{n+1}),\\
	\nabla_{\boldsymbol{\mu}_{q(h)}\boldsymbol{\mu}_{q(h)}}^2 S(\mathbf{\mu}_{q(h)},\mathbf{\Sigma}_{q(h)}) &=-\frac{1}{2}\mathsf{Diag}\Bigg[[0,\mathbf{\mu}_{q(\mathbf{s})}^\intercal]^\intercal\odot\mathrm{e}^{-\boldsymbol{\mu}_{q(h)}+\frac{1}{2}\mathsf{diag}\left(\boldsymbol{\Sigma}_{q(\mathbf{h})}\right)}\Bigg]
	-\mu_{q(1/\eta^2)}\mathbf{\mu}_{q(\mathbf{Q})},
\end{aligned}\end{equation}
where $\boldsymbol{\iota}_n$ is an n-dimensional vector of ones, $\mu_{q\left(1/\eta^2\right)}$ is the variational mean of $1/\eta^2$, $\mathbf{\mu}_{q(\mathbf{Q})}$ is the element-wise variational mean of $\mathbf{Q}$, and $\odot$ denotes the Hadamard product.
\end{proposition}
Algorithm~\ref{code:VBSV} in Supplementary Material \ref{app:algorithm} provides a pseudo-code for implementing our variational Bayes approach merging from Proposition \ref{eq:prop1} to Proposition \ref{eq:prop5}.
            
While we assume Gaussian errors for analytical tractability, our framework can accommodate heavy-tailed conditional distributions (e.g., Student-t errors) and asymmetric effects through numerical integration of the expectation terms, following extensions discussed in \citet{omori2007stochastic} and \citet{huber2023introducing}. The key insight is that our approximation relies solely on computing $\mathbb{E}_q[\log p(h,y)]$ and its derivatives. As long as these quantities can be computed, the method remains applicable regardless of the specific distribution of the $\varepsilon_t$. Heavy-tailed extensions would require numerical integration, but the core methodology remains unchanged.

It is worth highlighting that our variational Bayes approach expands the global approximation proposed by \citet{chan_yu2022} along three main dimensions. First, we relax the assumption that the initial distribution \(q(h_0)\) is independent of the trajectory of the latent state \(q(\mathbf{h}_1)\); that is, we do not assume \(q(\mathbf{h}) = q(h_0)q(\mathbf{h}_1)\). Section~\ref{sec:sim_study} shows that this relaxation has key implications for estimation accuracy.

Second, the latent covariance matrix \(\mathbf{\Sigma}_{q(h)}\) is not fixed conditional on \(\mathbf{\mu}_{q(h)}\), but is left unrestricted and estimated jointly with \(\mathbf{\mu}_{q(h)}\). By generalizing the estimate of \(\mathbf{\Sigma}_{q(h)}\), the variational density is expected to provide a better approximation of the true posterior. Supplementary Material \ref{app:h_homo} provides a full discussion of the more restrictive case in which \(\mathbf{\mu}_{q(h)} = \mathbf{W} \mathbf{f}_{q(h)}\) and \(\mathbf{\Sigma}_{q(h)} = \tau^2 \Gamma^{-1}\), where \(\boldsymbol{\xi} = (\mathbf{f}, \tau^2, \Gamma)\) is the collection of the parameters to be estimated.

Third, our latent state is a more general AR(1) dynamic instead of a random walk. While the latter reduces the parameter space, it imposes a strong form of non-stationarity in the log-volatility process which implies an unbounded volatility scaling asymptotically. 

\paragraph{Choosing $\mathbf{W}$.}

\begin{figure}[!ht]
\centering
	\subfigure[Daubechies wavelet basis matrix]{\includegraphics[width=.5\textwidth]{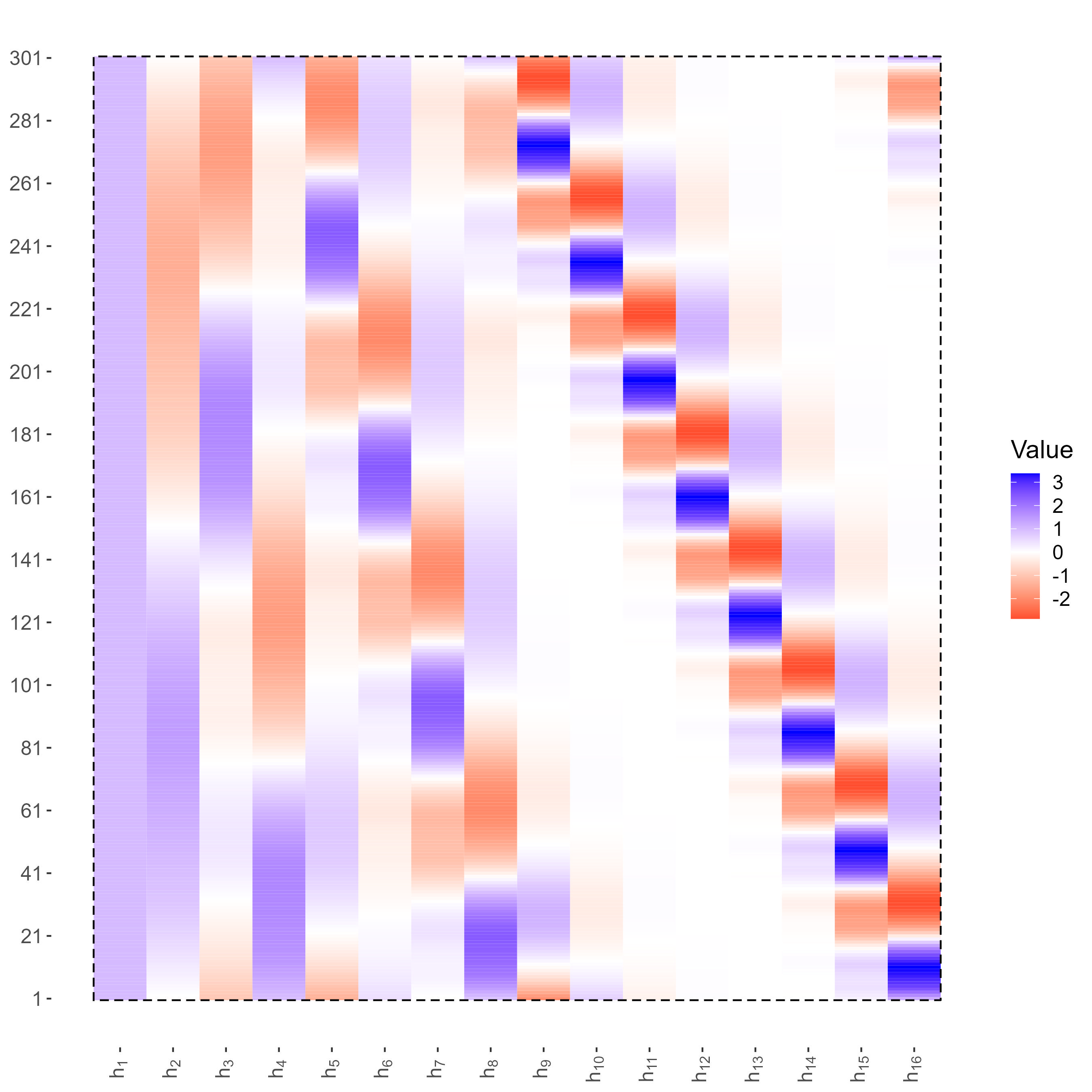}}
	\subfigure[Daubechies wavelet basis functions]{\includegraphics[width=.47\textwidth]{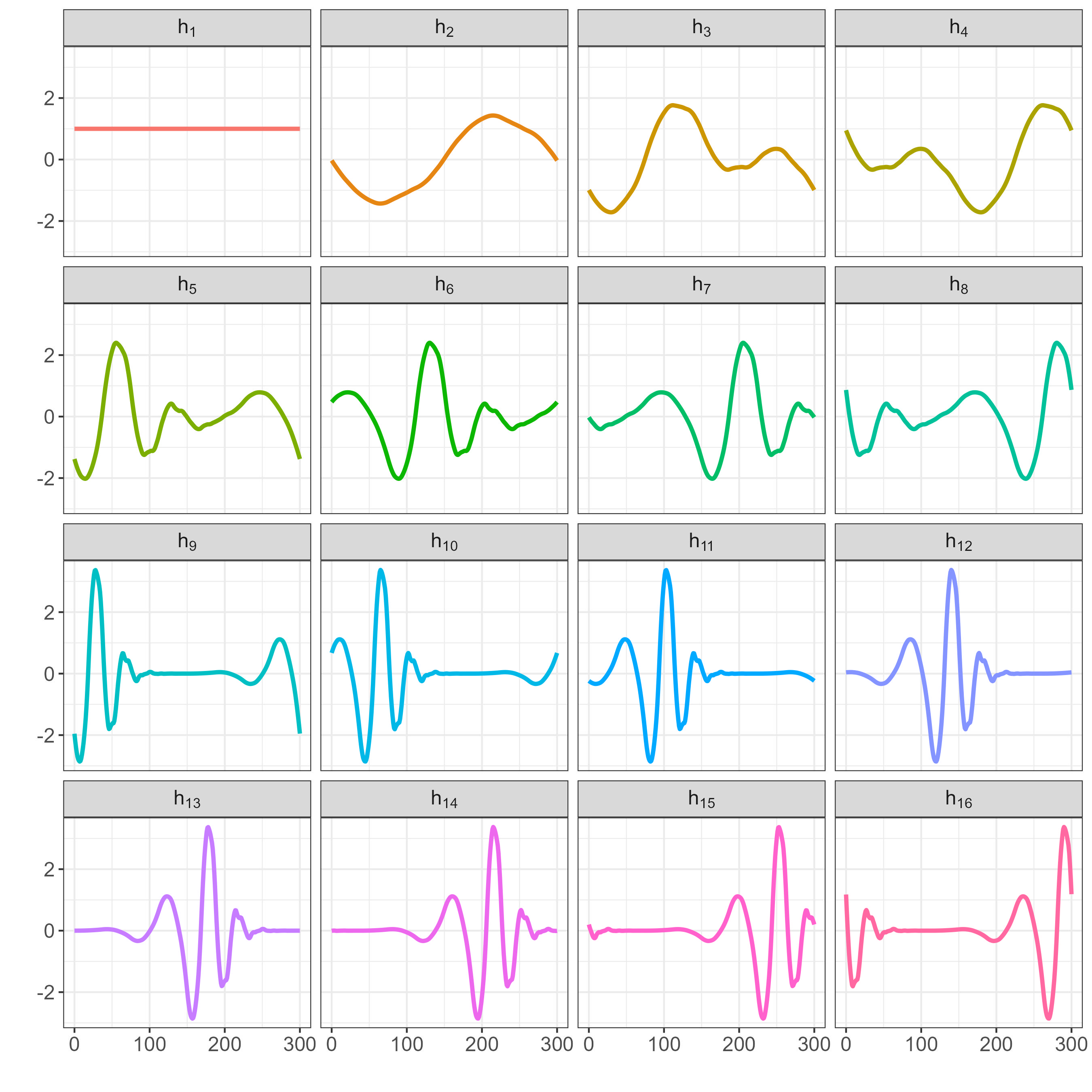}}
    \caption{\footnotesize \bf Wavelet Smoothing\normalfont. Illustration of \(\mathbf{W}\) in the case of wavelet basis functions. The right panel displays the columns of \(\mathbf{W}\). The figure corresponds to the case where \(l=5\), resulting in a resolution of \(R=16\).}
	\label{fig:form_of_W}
\end{figure}

Proposition~\ref{eq:prop5} demonstrates that the choice of the matrix \(\mathbf{W}\) and the corresponding estimates of the projection coefficients \(\mathbf{f}_{q(h)} \in \mathbb{R}^{k}\) directly influence the posterior estimates of the latent stochastic volatility. Figure~\ref{fig:form_of_W} illustrates the structure of \(\mathbf{W}\) when constructed using Daubechies wavelet basis functions, which is the main specification employed in both the simulation study and the empirical analysis. The right panel shows the columns of \(\mathbf{W}\).\footnote{Our framework is flexible and can accommodate different smoothing assumptions through the choice of \(\mathbf{W}\). For example, Supplementary Material \ref{sec:smoothing} discusses an alternative specification using B-spline basis functions.}

The wavelet basis matrix \(\mathbf{W}\) is constructed over equally spaced grids on \([0,n]\) with a length \(R\), referred to as the resolution. The resolution is defined as \(R = 2^{l-1}\), where \(l\) specifies the level and, consequently, the degree of smoothness. At level \(l\), the number of basis functions equals \(R\). These functions are created by dilations and shifts of a more general {\it mother} function. {We employ Daubechies wavelets which naturally decompose volatility into different frequency components, allowing us to preserve economically relevant low-frequency volatility trends while filtering high-frequency noise that generates excessive turnover \citep{ramsey2002wavelets}. Throughout this paper, we assume \(l=5\), which corresponds to a resolution of \(R=16\). Results with \(l=4\) and \(l=6\) are qualitatively similar.}\footnote{For the interested reader the results are available upon request.}

Figure~\ref{fig:choice_of_W} in Supplementary Material~\ref{sec:smoothing} highlights how the smoothing via \(\mathbf{W}\) influences both the conditional mean and variance of \(q^\ast(\mathbf{h})\). This enables us to regularize not only the conditional mean but also the entire predictive distribution of the latent volatility state. 

\subsection{Predictive density of the volatility state}
\label{subsec:varianceprediction}

Consider the joint posterior distribution of the latent state and parameters, \(p(\mathbf{h}, \widetilde{\boldsymbol{\vartheta}} \mid \mathbf{y})\), where \(\widetilde{\boldsymbol{\vartheta}} = (\mathbf{\beta}, c, \rho, \eta^2)\). Define the likelihood for the new latent state \(h_{t+1}\) given the information set up to time \(t\), \(\mathbf{y} = \{y_{1:t}\}\), as \(p(h_{t+1} \mid \mathbf{y}, \mathbf{h}, \widetilde{\boldsymbol{\vartheta}})\). The predictive density then takes the form:

\begin{equation}\label{eq:pred_post}
    p(h_{t+1} \mid \mathbf{y}) = \int p(h_{t+1} \mid \mathbf{y}, \mathbf{h}, \widetilde{\boldsymbol{\vartheta}}) p(\mathbf{h}, \widetilde{\boldsymbol{\vartheta}} \mid \mathbf{y}) \, d\mathbf{h} d\widetilde{\boldsymbol{\vartheta}}.
\end{equation}

Given a variational density \(q(\boldsymbol{\vartheta}) = q(\mathbf{h}, \widetilde{\boldsymbol{\vartheta}}) = q(\mathbf{h})q(\widetilde{\boldsymbol{\vartheta}})\) that approximates \(p(\mathbf{h}, \widetilde{\boldsymbol{\vartheta}} \mid \mathbf{y})\), we follow \citet{gunawan2021variational} and derive the variational predictive distribution as:
\begin{align}
    q(h_{t+1} \mid \mathbf{y}) &= \int p(h_{t+1} \mid \mathbf{y}, \mathbf{h}, \widetilde{\boldsymbol{\vartheta}}) q(\mathbf{h}) q(\widetilde{\boldsymbol{\vartheta}}) \, d\mathbf{h} d\widetilde{\boldsymbol{\vartheta}} \nonumber \\
    &= \int p(h_{t+1} \mid h_t, \widetilde{\boldsymbol{\vartheta}}) q(h_t) q(\widetilde{\boldsymbol{\vartheta}}) \, dh_t d\widetilde{\boldsymbol{\vartheta}}, \label{eq:integral}
\end{align}
where the second equality follows from the Markov property of the latent state. In the context of volatility timing, the primary interest is in forecasting the variance \(\sigma_{t+1}^2\) rather than the log-volatility \(h_{t+1}\). Since \(h_t = \log \sigma_t^2\), the density of the conditional variance is readily obtained as:
\begin{equation*}
    q(\sigma_{t+1}^2 \mid \mathbf{y}) = \frac{\partial h_{t+1}}{\partial \sigma_{t+1}^2} q(h_{t+1} \mid \mathbf{y}) = \frac{1}{\sigma_{t+1}^2} q(h_{t+1} \mid \mathbf{y}).
\end{equation*}

The integral in Eq.~\eqref{eq:integral} cannot be solved analytically but can be approximated through Monte Carlo integration. This approach takes advantage of the fact that we can efficiently sample from the variational densities \(q(h_t)\) and \(q(\widetilde{\boldsymbol{\vartheta}})\). A simulation-based approximation of the variational predictive distribution \(q(\sigma^2_{t+1} \mid \mathbf{y})\) is therefore obtained by averaging \(p(h_{t+1} \mid h_t^{(i)}, \widetilde{\boldsymbol{\vartheta}}^{(i)})\) over samples \(h_t^{(i)} \sim q^\ast(h_t)\) and \(\widetilde{\boldsymbol{\vartheta}}^{(i)} \sim q^\ast(\widetilde{\boldsymbol{\vartheta}})\), for \(i = 1, \ldots, N\), drawn from the optimal variational densities. This yields the estimator:
\[
\widehat{q}(\sigma^2_{t+1} \mid \mathbf{y}) = \frac{1}{\sigma_{t+1}^2} \frac{1}{N} \sum_{i=1}^N p(h_{t+1} \mid h_t^{(i)}, \widetilde{\boldsymbol{\vartheta}}^{(i)}).
\]

\section{Simulation study}
\label{sec:sim_study}

We evaluate the statistical underpinnings of our variational Bayes (VB) approach by comparing it to three successful methods for univariate stochastic volatility models: the MCMC method proposed by \citet{stochvol_package}, the variational approximation introduced by \citep[][henceforth {\tt CY}]{chan_yu2022}, which represents our main competitor, {and the optimization-based inference of \citep[][henceforth {\tt TMB}]{stochvolTMB} built on Template Model Builder for fast and efficient estimation.} Since none of these benchmarks incorporates smoothing of the predictive density of the latent state, the baseline comparison assumes \(\mathbf{W} = \mathbf{I}_{n+1}\), meaning that no smoothing is imposed on the variational density \(q(\mathbf{h})\).

In addition to the baseline, we assess the performance of two alternative smoothing approaches. The first uses \(\mathbf{W}\) as a Daubechies wavelet basis matrix with \(l=5\) ({\tt VBW}), and the second uses \(\mathbf{W}\) as a B-spline basis with knots equally spaced every ten time points ({\tt VBS}).\footnote{The choice of equally spaced knots for the B-spline basis and the value of \(l\) for the wavelet basis matrix ensures both approaches provide a comparable degree of smoothness.}
 
Furthermore, we consider a more tightly parameterized covariance structure for the variational density of the latent state, \(\mathbf{\Sigma}_{q(h)} = \tau^2 \mathbf{\Gamma}^{-1}\), where \(\tau^2 \in \mathbb{R}^+\) ({\tt VBH}). This mirrors the simplification proposed by \citet{chan_yu2022} and further streamlines the estimation of \(q^*(\mathbf{h})\). Supplementary Material \ref{app:h_homo} provides a detailed discussion of this parameterization.

\begin{figure}[!ht]
	\subfigure[MSE when \(\rho=0.98\)]{\includegraphics[width=.32\textwidth]{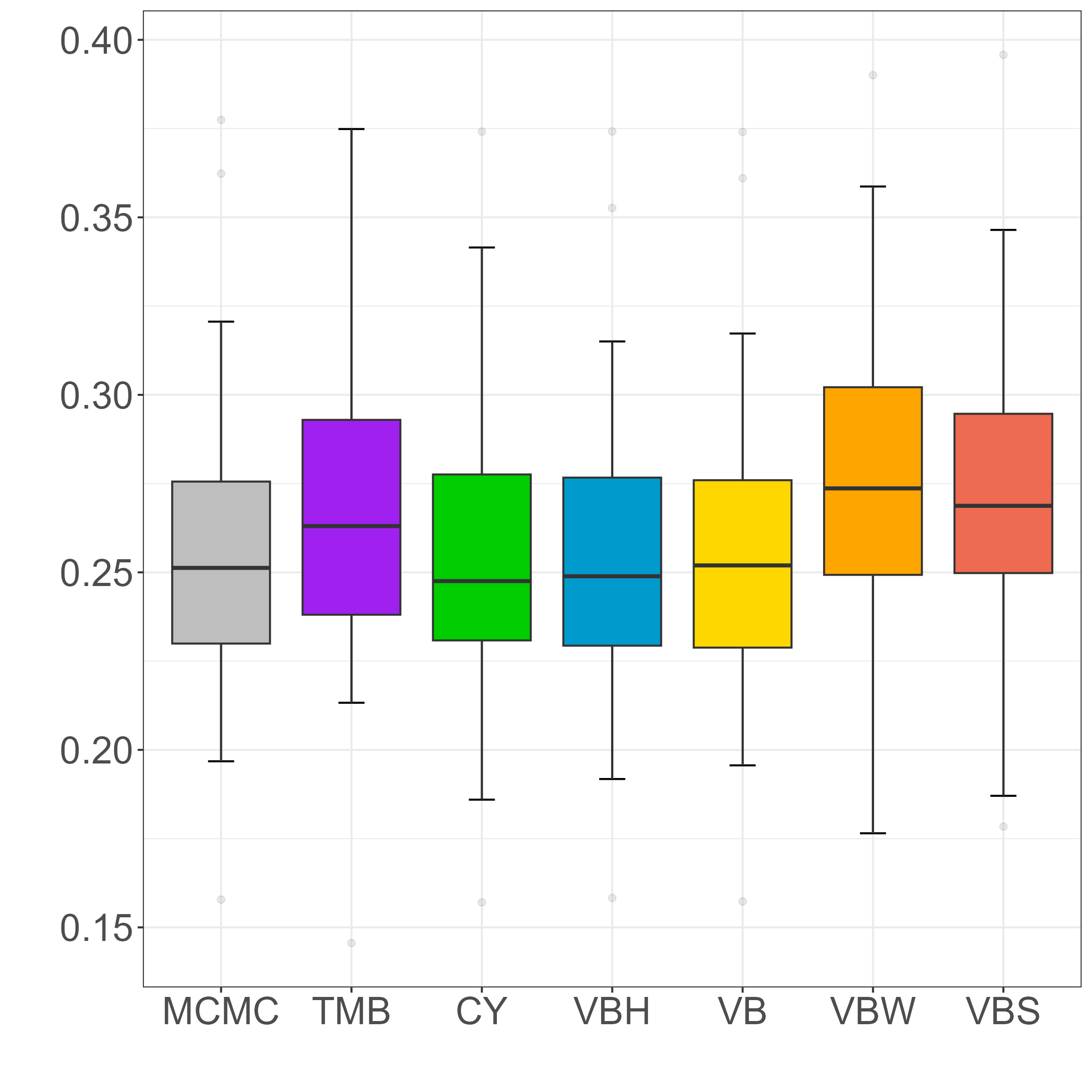}}
	\subfigure[Global acc. when \(\rho=0.98\)]{\includegraphics[width=.32\textwidth]{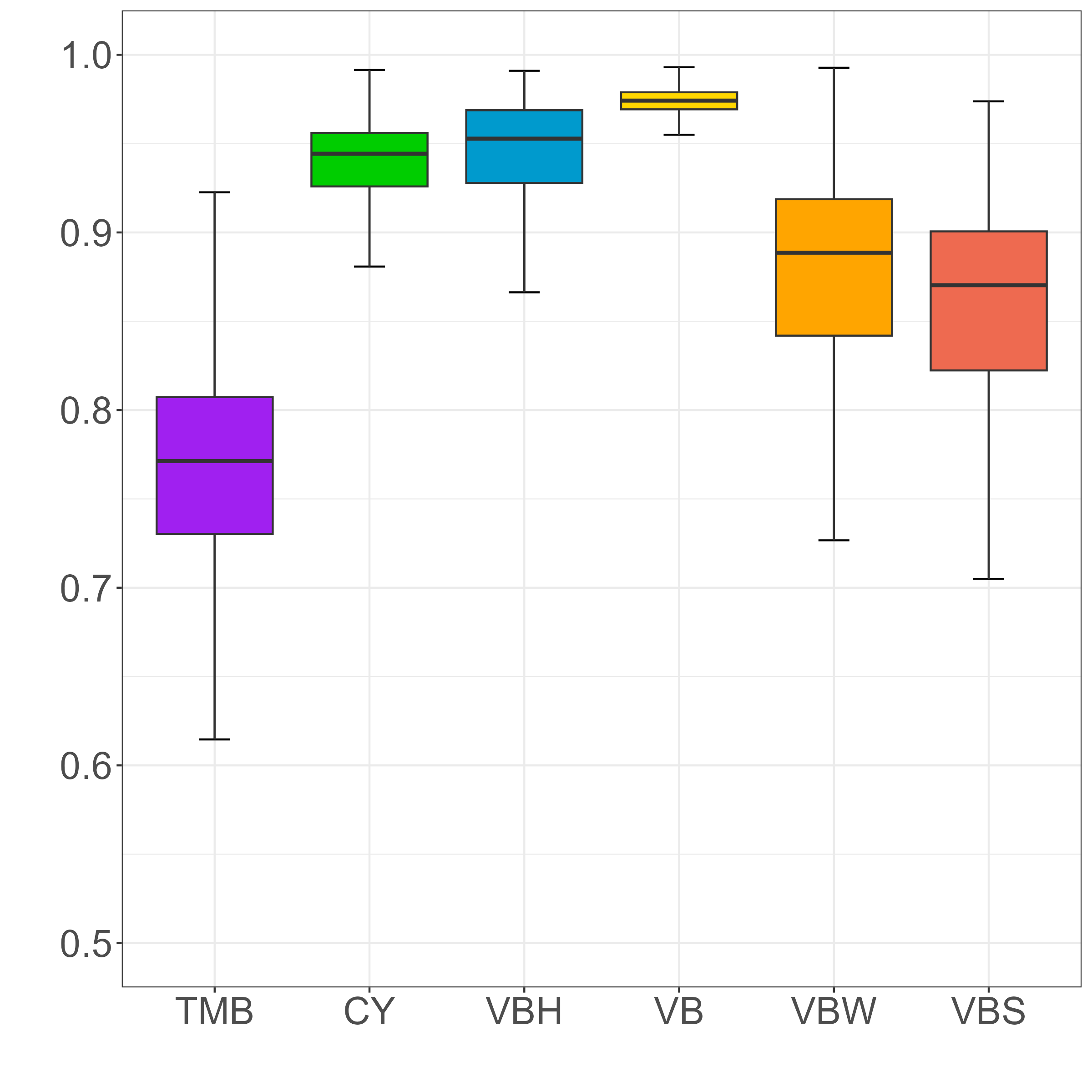}}
	\subfigure[Comp. time when \(\rho=0.98\)]{\includegraphics[width=.32\textwidth]{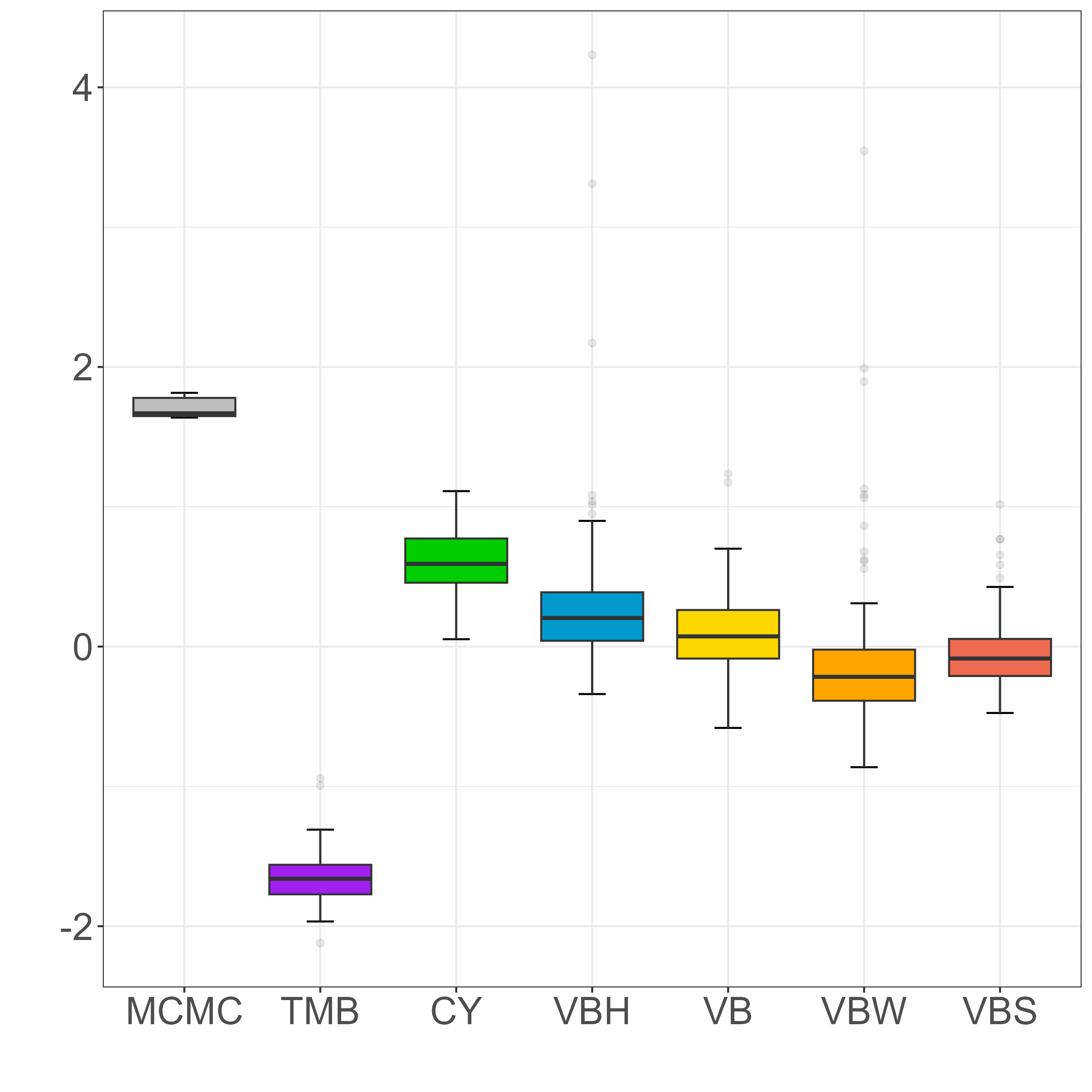}} \\
	\subfigure[MSE when \(\rho=0.70\)]{\includegraphics[width=.32\textwidth]{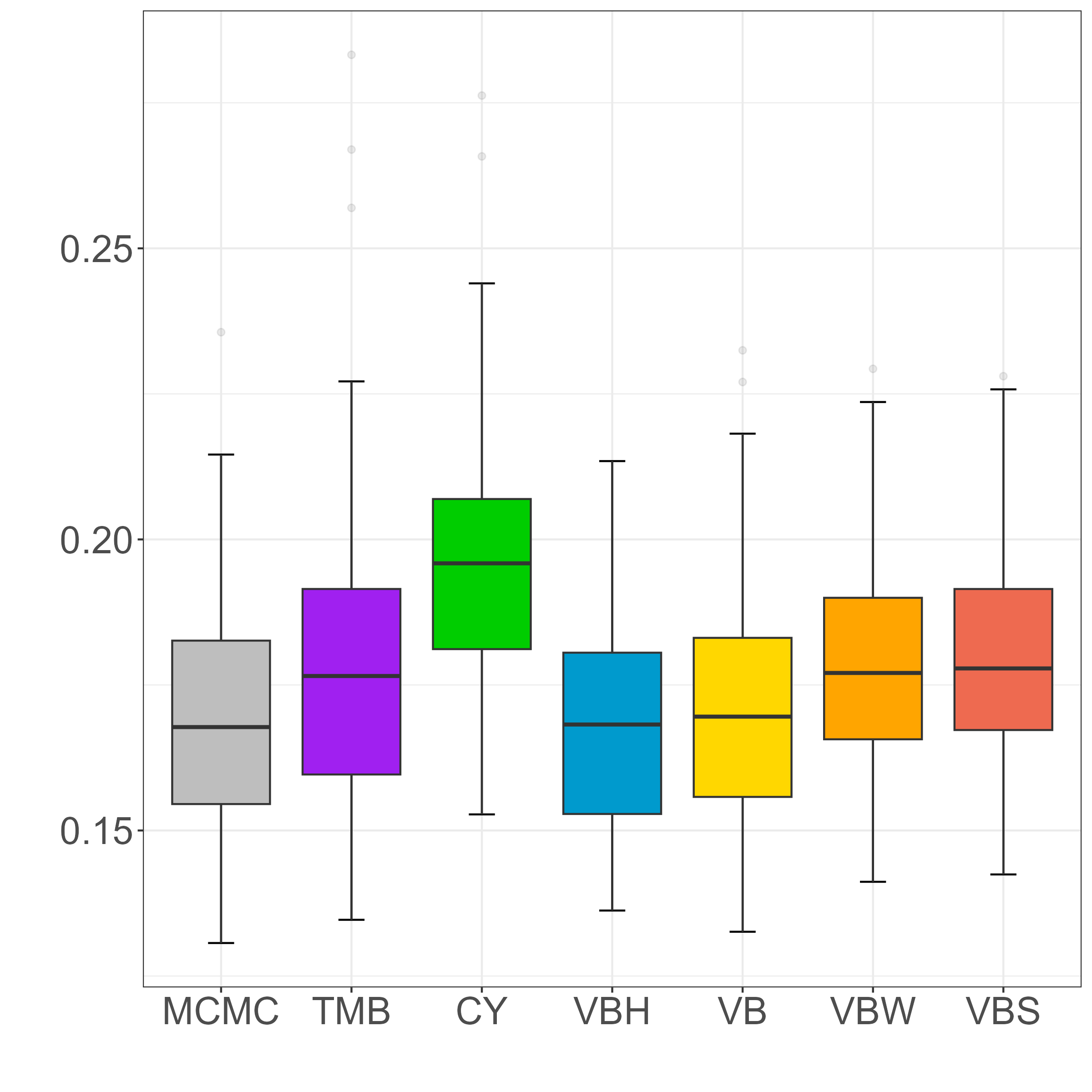}}
	\subfigure[Global acc. when \(\rho=0.70\)]{\includegraphics[width=.32\textwidth]{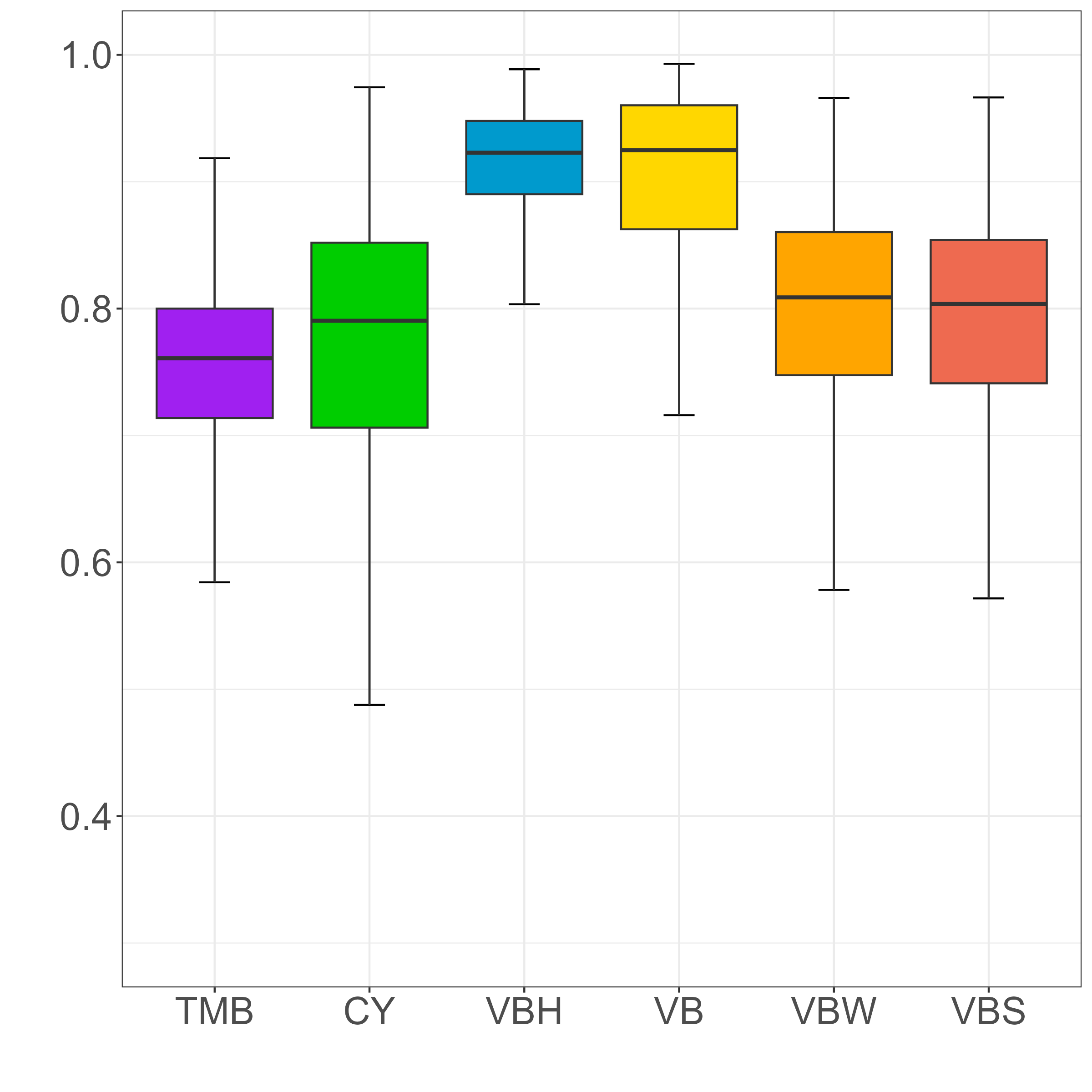}}
	\subfigure[Comp. time when \(\rho=0.70\)]{\includegraphics[width=.32\textwidth]{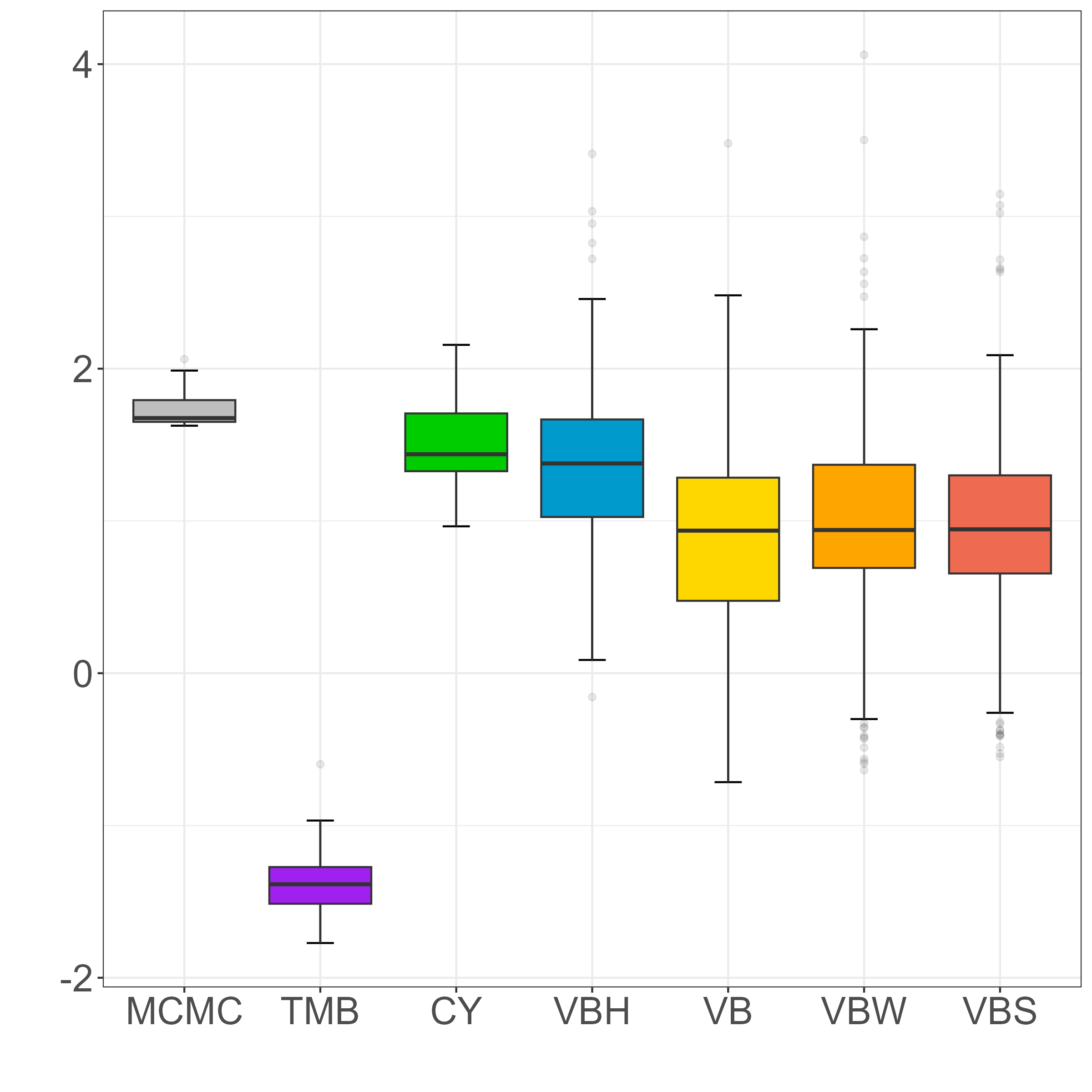}}
    \caption{\footnotesize \bf Estimation Accuracy\normalfont. This figure reports the mean squared error (left panels), global estimation accuracy compared to {\tt MCMC} as defined in Eq.~\eqref{eq:accuracy} (middle panels), and computational time (log-minute scale) across methods. Results are shown for \(\rho = 0.98\) (top panels) and \(\rho = 0.7\) (bottom panels).}
	\label{fig:simaccuracy}
\end{figure}
For each of the \(M=100\) replications, we consider \(n=600\) time observations for a zero-mean process \(y_t\) -- consistent with the shortest time series in the empirical application -- with \(c=0\), \(\eta^2=0.1\), and two different levels of persistence for the latent state \(h_t\): \(\rho \in \{0.70, 0.98\}\). Figure~\ref{fig:simaccuracy} reports the mean squared error (MSE, left columns) and a measure of global estimation accuracy (middle columns) for \(\mathbf{h}\) compared to {\tt MCMC} as defined in \citet{wand_ormerod.2011}. The MSE is calculated as \(\mbox{MSE} = n^{-1}\sum_{t=1}^n(h_t - \widehat{h}_t)^2,\) where \(h_t\) and \(\widehat{h}_t\) are the simulated log-volatility and its corresponding estimate, respectively. 

{We note that closeness to ``true'' values does not automatically validate a Bayesian method. To address this issue, we follow \citet{wand2011mean} and calculate a global inference accuracy measure. This is based on the common assumption that the {\tt MCMC} posterior obtained with a sufficiently large number of draws closely proxies for the \textit{true} posterior distribution of the latent process:}

\begin{equation}
\mathcal{ACC} = \left\{1 - 0.5 \int |q(\mathbf{h}) - p(\mathbf{h} \mid \mathbf{y})| \, d\mathbf{h} \right\}, \label{eq:accuracy}
\end{equation}
where \(p(\mathbf{h} \mid \mathbf{y})\) is the {\tt MCMC} posterior and \(q(\mathbf{h})\) is the competing approximation. This allows us to compare the posterior inference obtained via {\tt MCMC}, {\tt VB}, and the asymptotic distribution of the MLE obtained with {\tt TMB}, under the same underlying model specification.

The left panels of Figure~\ref{fig:simaccuracy} show that {\tt MCMC}, {\tt CY}, and {\tt VB} produce almost equivalent posterior estimates when \(\rho = 0.98\). {As expected, since \(\mathbf{W}\) represents a deliberate distortion on the variational density \(q(\mathbf{h})\), the smoothed estimates {\tt VBW} and {\tt VBS} exhibit higher MSE compared to the baseline {\tt VB}. This is not unexpected since our smoothing intervention deliberately sacrifices some statistical accuracy to achieve economic stability. We discuss the tradeoff between forecasting accuracy and economic utility in more detail in Section \ref{sec:additional_results}.}

For \(\rho = 0.7\), {the estimation accuracy of the latent states for {\tt CY} substantially deteriorates, with MSE significantly higher than that of {\tt MCMC} and {\tt VB}.} This occurs because the approximation proposed by \citet{chan_yu2022} assumes that the latent volatility state follows a random walk, which leads to lower accuracy when \(\rho \ll 1\). This limitation is also reflected in poor global approximation accuracy (middle panels). Overall, the closest approximation to {\tt MCMC} is achieved by the baseline {\tt VB}.

{The MSE results for {\tt TMB} indicate consistently poorer performance in latent log-volatility estimation compared to {\tt MCMC} and {\tt VB}, for both \(\rho = 0.98\) and \(\rho = 0.7\), translating into lower global accuracy for the asymptotic distribution of the MLE estimator.} Examining the smoothed approaches, when \(\rho = 0.7\), both {\tt VBW} and {\tt VBS} perform similarly to {\tt CY}. This occurs because these approaches impose high persistence in the estimates of the latent state -- either through \(\mathbf{W}\) (for {\tt VBW} and {\tt VBS}) or via the random walk assumption for \(\mathbf{h}\) (in the case of {\tt CY}).\footnote{Figure~\ref{fig:sim_par_hat} in Supplementary Material~\ref{app:simulation} provides further insights, showing that imposing smoothness through a B-spline or Daubechies wavelet basis forces the posterior estimates of \(\rho\) to approach unity and the state variance \(\eta^2\) to shrink, irrespective of the true parameter values. Similarly, the random walk assumption in {\tt CY} acts as a form of implicit smoothing.}

The right panels of Figure~\ref{fig:simaccuracy} report computational speed on a log-minute scale. {{\tt TMB} is built specifically for computational efficiency and demonstrates the fastest performance across all scenarios. However, {\tt VB}, with or without smoothing, remains substantially more computationally efficient than both {\tt MCMC} and {\tt CY}.} The computational cost savings hold irrespective of the true persistence \(\rho\), with \texttt{VB} achieving 8-10x speedups for high persistence and 2× speedups for moderate persistence.

{This creates a three-way trade-off: \texttt{TMB} offers speed, \texttt{MCMC} provides benchmark accuracy, and our \texttt{VB} approach balances accuracy with computational efficiency. Higher accuracy translates into more reliable volatility forecasts for portfolio implementation, while computational efficiency enables real-time recursive estimation required for dynamic portfolio strategies.}

\begin{figure}[!ht]
\hspace{-1em}
	\subfigure[$t \in (1,10)$, \(\rho=0.98\)]{\includegraphics[width=.32\textwidth]{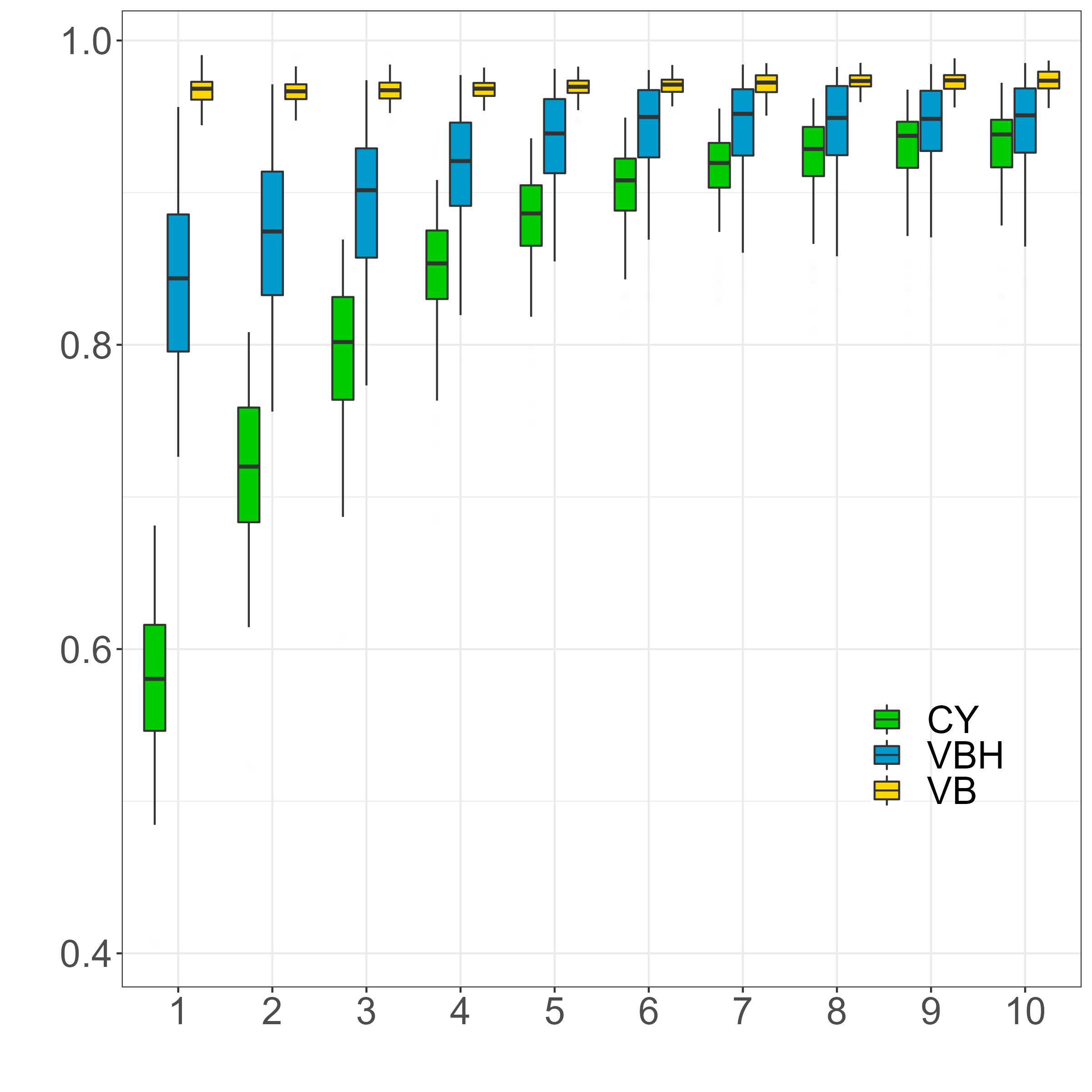}}
	\subfigure[$t \in (301,310)$, \(\rho=0.98\)]{\includegraphics[width=.32\textwidth]{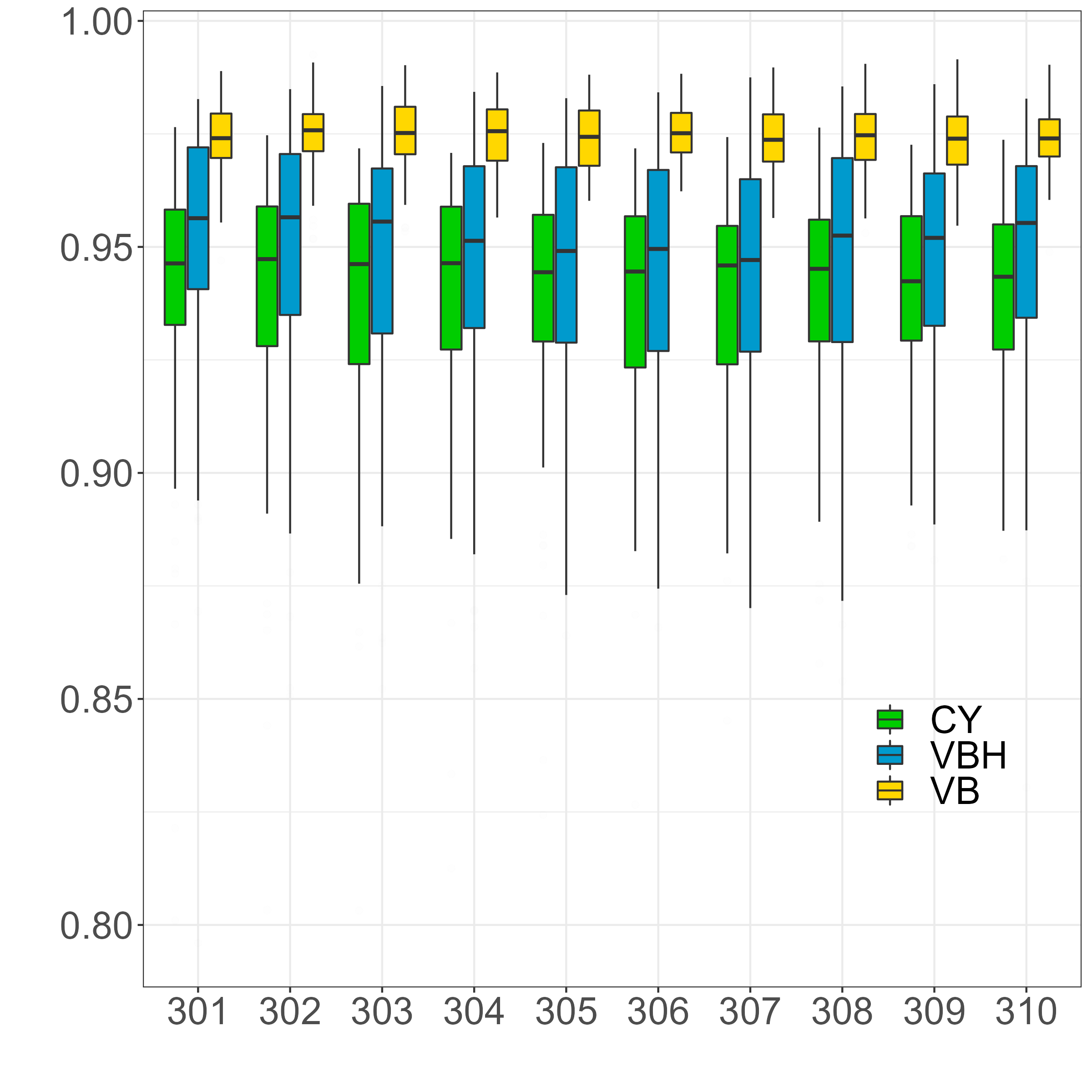}}
	\subfigure[$t \in (591,600)$, \(\rho=0.98\)]{\includegraphics[width=.32\textwidth]{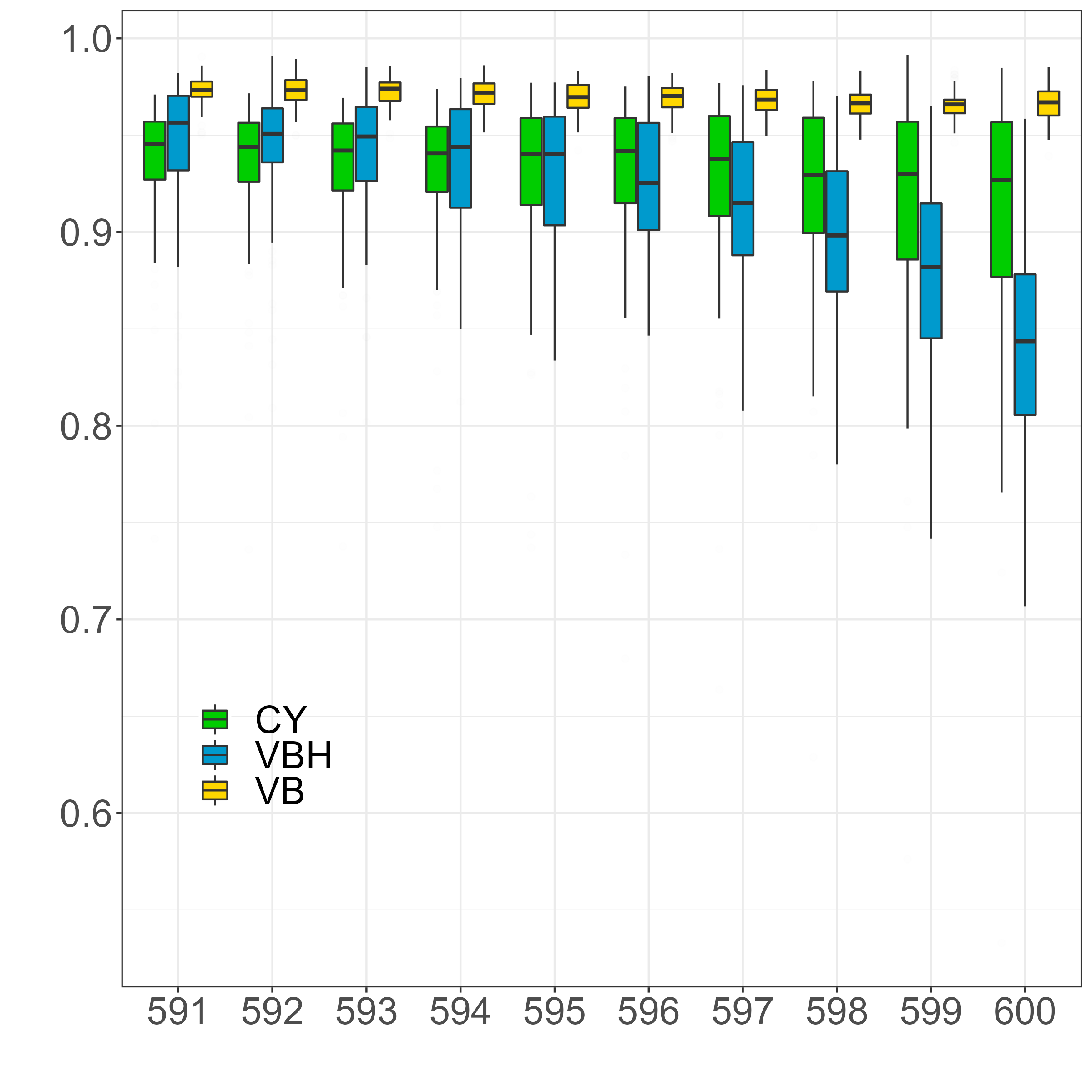}} \\
\vspace{1em}\hspace{-1em}	
	\subfigure[$t \in (1,10)$, \(\rho=0.70\)]{\includegraphics[width=.32\textwidth]{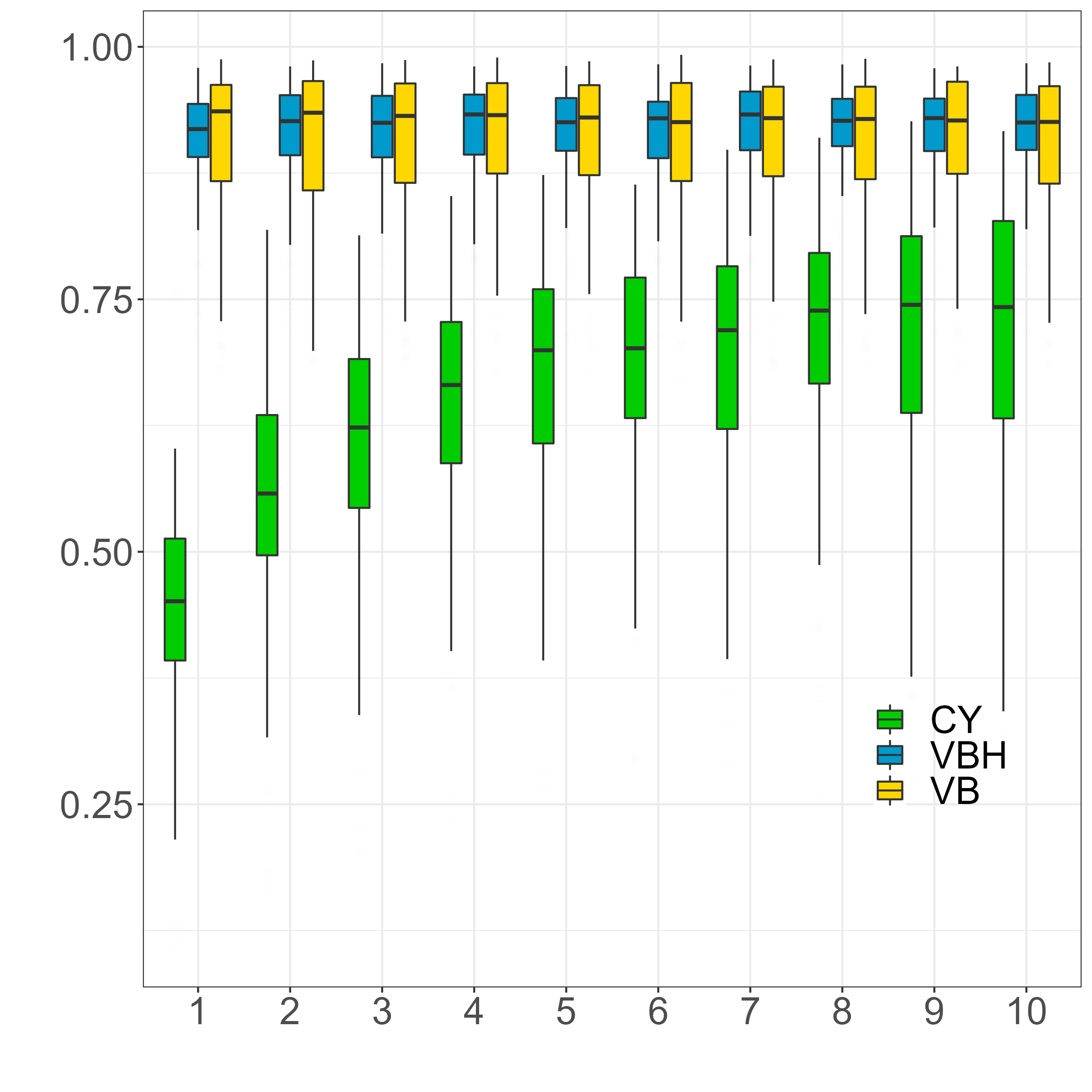}}
	\subfigure[$t \in (301,310)$, \(\rho=0.70\)]{\includegraphics[width=.32\textwidth]{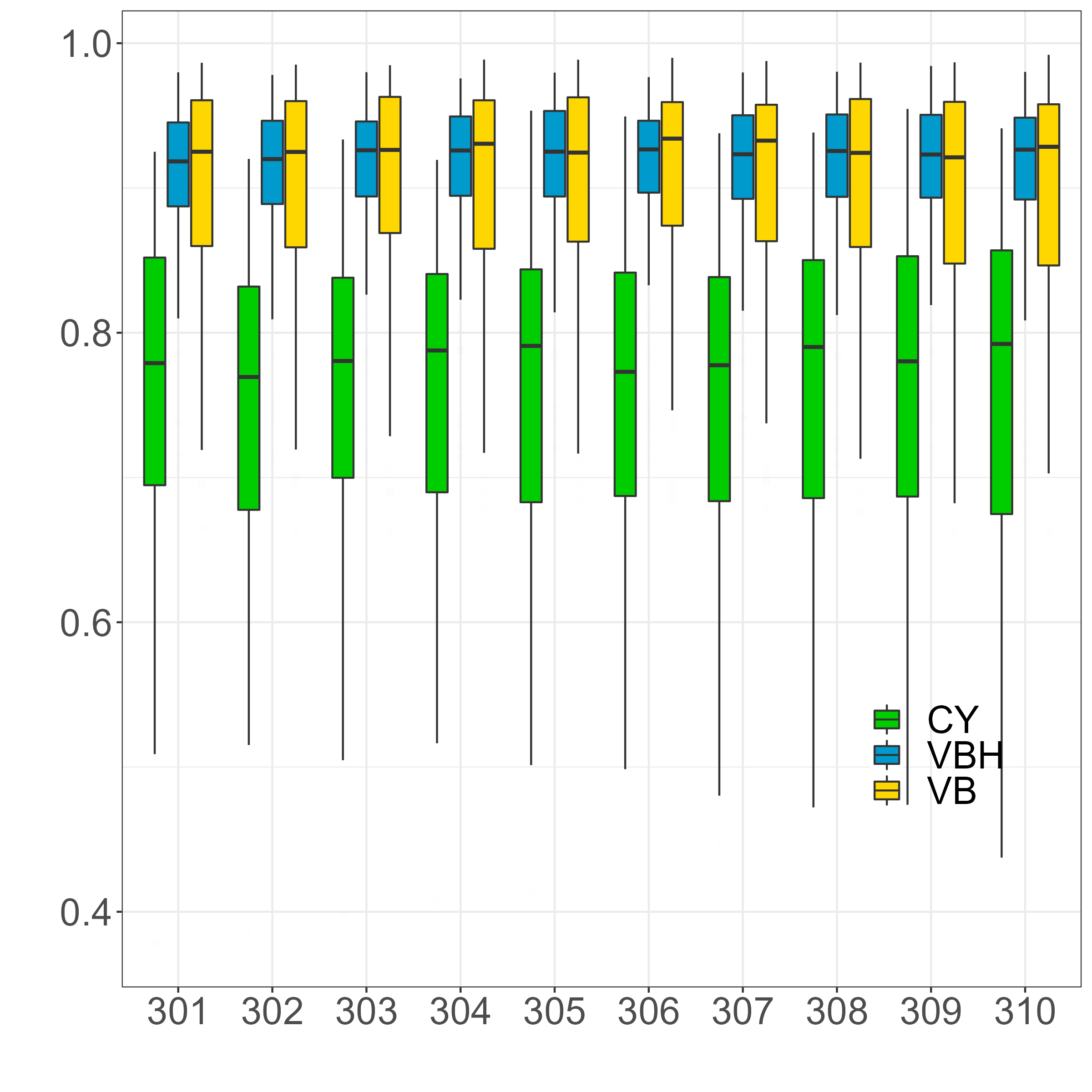}}
	\subfigure[$t \in (591,600)$, \(\rho=0.70\)]{\includegraphics[width=.32\textwidth]{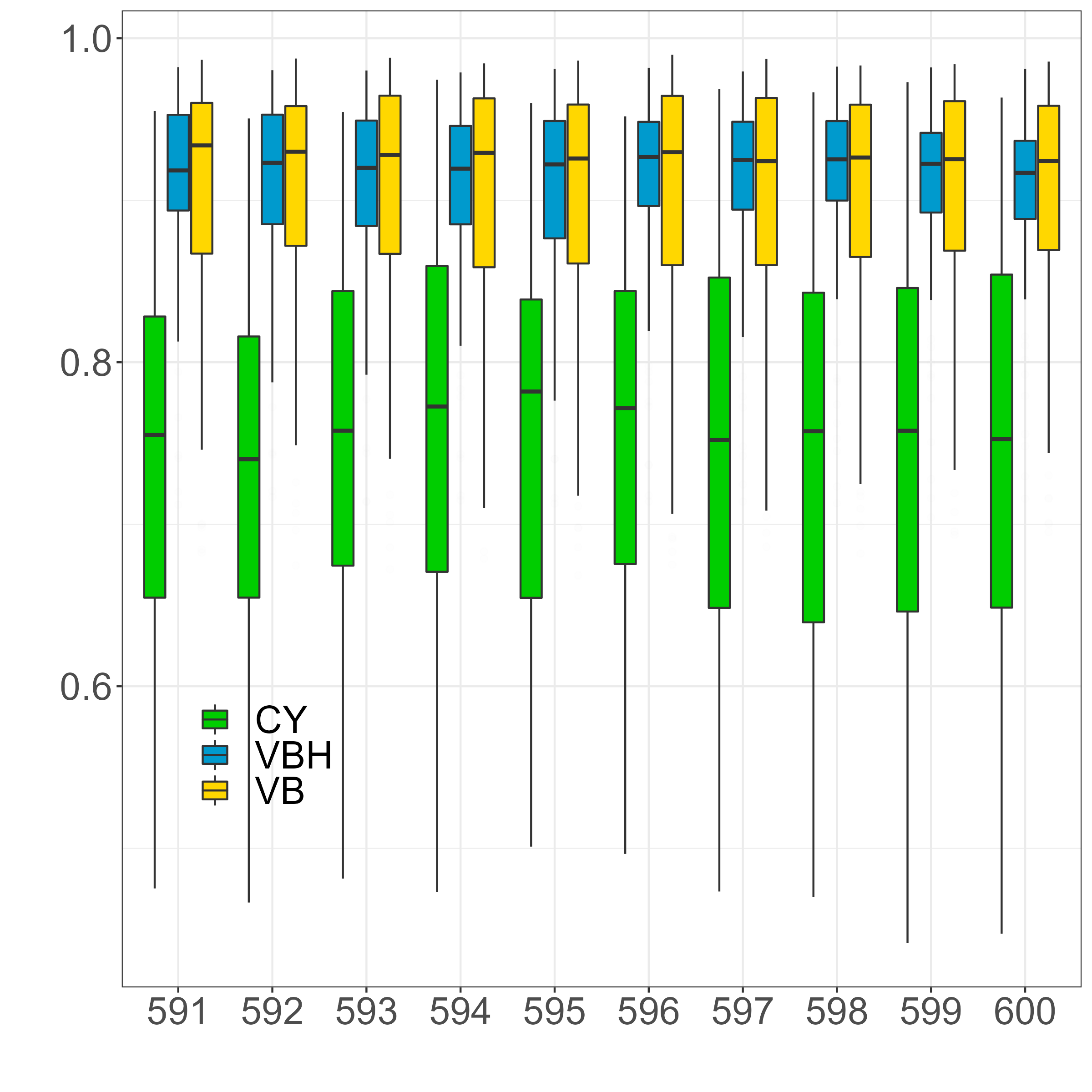}} 
    \caption{\footnotesize \bf Accuracy for Different Parts of the Sample\normalfont. This figure reports the accuracy of {\tt VB} and {\tt VBH} against {\tt CY}. The top (bottom) panels show results for \(\rho=0.98\) (\(\rho=0.7\)). The time steps \(t \in (1,10)\) are shown in the left panel, \(t \in (301,310)\) in the middle panel, and \(t \in (591,600)\) in the right panel. The benchmark is MCMC as in \citet{stochvol_package}.}
	\label{fig:sim_acc_time}
\end{figure}

Recall that besides the flexibility introduced by incorporating stochastic volatility estimates via \(\mathbf{W}\), our {\tt VB} approach relaxes the assumption that the initial distribution \(q(h_0)\) is independent of the remaining trajectory \(q(\mathbf{h}_1)\). This means we do not assume \(q(\mathbf{h}) = q(h_0)q(\mathbf{h}_1)\), unlike \citet{chan_yu2022}, which relies on the independence between \(q(h_0)\) and \(q(\mathbf{h}_1)\) for estimation purposes. Figure~\ref{fig:sim_acc_time} demonstrates that this assumption has important implications for the accuracy of the variational approximation, especially at the extremes of the estimation sample. The simulation results are reported for \(t \in (1,10)\) in the left panel, \(t \in (301,310)\) in the middle panel, and \(t \in (591,600)\) in the right panel, with the {exact} MCMC approach of \citet{stochvol_package} serving as the benchmark.

When the persistence of the latent state is high (top panels), our generalization significantly improves the global approximation accuracy relative to {\tt MCMC}, especially at the beginning and end of the sample. The results also confirm that an unconstrained estimate of \(\mathbf{\Sigma}_{q(h)}\) substantially improves approximation accuracy compared to the more restrictive \(\mathbf{\Sigma}_{q(h)} = \tau^2 \mathbf{\Gamma}^{-1}\) (labeled as {\tt VBH}). The bottom panels in Figure~\ref{fig:sim_acc_time} confirm that the global approximation accuracy of {\tt CY} deteriorates when the latent state is less persistent (\(\rho=0.7\)). In contrast, our {\tt VB} approach maintains superior approximation accuracy across the full sample.

\section{Volatility-managed portfolios}
\label{sec:appl}

Let \( y_t \) denote the return on a given equity strategy in month \( t \). {We assume a constant mean, \(\mu = \mathbf{x}_t^\intercal \beta\), where \(\mathbf{x}_t\) is an \( n \)-dimensional vector of ones in Eq.~\eqref{eq:svm1b}. This assumption ensures direct comparability with the relevant literature \citep{barroso2015momentum,moreira2017volatility,cederburg2020performance,barroso2021limits}.} The return on a real-time volatility-managed portfolio, \( y^\sigma_{t} \), is defined as:
\begin{align}
y^\sigma_{t+1} & = \frac{\gamma_{t}}{\mathbb{E}_{t}\left[\sigma_{t+1}^2\right]} y_{t+1}, \label{eq:volmanaged1}
\end{align}
{where \(\gamma_{t}\) is chosen to ensure that the variance of \( y^\sigma_{t} \) matches the variance of the original portfolio returns \( y_{t} \). Both the variance forecast \(\mathbb{E}_{t}\left[\sigma_{t+1}^2\right]\) and $\gamma_t$ are calculated in real-time using only information available up to the previous month \( t \), reflecting institutional constraints. This approach allows us to isolate the value of volatility timing from differences in average risk exposure. Without this normalization, performance differences could simply reflect different average leverage levels rather than superior volatility forecasting.\footnote{Alternative approaches requiring forward-looking volatility targets would not be feasible in practice, as investors cannot know future volatility levels when setting target parameters.}}

The dynamic adjustment \(\omega_{t} = \gamma_{t}/\mathbb{E}_{t}\left[\sigma_{t+1}^2\right]\) reflects the leverage (or deleverage) required to maintain a consistent level of risk. This is often referred to as volatility targeting \citep{harvey2018impact,bongaerts2020conditional}, as the portfolio adjusts to maintain a target level of volatility rather than a fixed notional capital exposure.

It is important to clarify why univariate volatility models suffice for this application. While correlations matter for portfolio allocation across multiple assets, volatility management focuses on timing the aggregate exposure to each individual factor. The economic value comes from tracking time-varying compensation for systematic risk factors rather than exploiting cross-asset correlation structures. This distinction explains why volatility-managed portfolios address a time-series adjustment to capital exposure to individual strategies, focusing on "the time series of many aggregate priced factors" rather than cross-sectional asset allocation where multivariate models excel \citep{moreira2017volatility}.

A simple yet popular approach to obtain \(\mathbb{E}_{t}\left[\sigma_{t+1}^2\right]\) is to use the previous month's realized variance ({\tt RV}), calculated based on daily portfolio returns \citep{moreira2017volatility,barroso2015momentum,cederburg2020performance,barroso2021limits,wang2021downside}.\footnote{Following this literature, we calculate the one-month realized variance based on daily returns as:
\begin{align}
    \mathbb{E}_{t}\left[\sigma_{t+1}^2\right] & = \frac{22}{n}\sum_{\tau=1}^{n}y_{\tau,t}^2,\label{eq:rv}
\end{align}
where \(y_{\tau,t}\) denotes the demeaned excess returns on a given portfolio on day \(\tau = 1, \ldots, n\) for month \(t\).} In addition to the realized variance, we compare our smooth stochastic volatility ({\tt SSV}) against a wealth of competing approaches. 

First, we follow \citet{moreira2017volatility} and replace the realized variance with the expected variance from a simple AR(1) forecast ({\tt RV AR}). Next, we use a six-month rolling estimate of the realized variance ({\tt RV6}), as proposed by \citet{barroso2015momentum,barroso2021limits}, using the daily data. We also include the heterogeneous autoregressive volatility forecast from \citet{corsi2009simple} ({\tt HAR}) {and a standard GARCH(1,1) model ({\tt Garch}) based on monthly returns}, which \citet{hansen2005forecast} identified as a challenging benchmark in volatility forecasting. Finally, we consider a conventional AR(1) latent stochastic volatility model ({\tt SV}), as described by \citet{taylor1994modeling}, {\it without} smoothing via \(\mathbf{W}\). This choice is motivated by the fact that the AR(1) structure of the latent stochastic volatility model inherently provides a relatively smoother estimate compared to realized variance-based methods.

\subsection{Data}
\label{subsec:data}

Alongside the return on the market portfolio in excess of the 30-day TBill rate (MktRf), we collect daily -- to construct realized variance measures -- and monthly returns for the size (SMB), value (HML), profitability (RMW), and investment (CMA) equity factors from \citet{fama2015five}. Additionally, we include the returns on the betting-against-beta (BAB) strategy from \citet{frazzini2014betting}, the profitability (ROE) and investment (IA) factors from \citet{hou2015digesting}, the short-term reversal (St Rev) of \citet{jegadeesh1990evidence}, and the long-term reversal (Lt Rev) of \citet{de1985does}.

We augment this initial set of equity factors with a broader collection of characteristic-based portfolios. {Specifically, we include 13 investment ``themes'' spanning accruals, debt issuance, investment, low leverage, low risk, momentum, profit growth, profitability, quality, seasonality, size, reversal, and value. These themes represent equal-weighted aggregations of a much larger pool of 153 characteristic portfolios, with clustering based on a k-means algorithm. For further details on portfolio construction, we refer the interested reader to \citet{jensen2021there}.}\footnote{Data on the \citet{fama2015five} factors, \citet{jegadeesh1993returns} momentum, and short- and long-term reversal factors are available on Kenneth French's website at \url{http://mba.tuck.dartmouth.edu/pages/faculty/ken.french/data_library.html}. Data on the betting-against-beta factor are available on the AQR website at \url{https://www.aqr.com/Insights/Datasets/Betting-Against-Beta-Original-Paper-Data}. Lu Zhang kindly provides data on profitability and investment factors at \url{http://global-q.org/index.html}. Data on the 13 themes are kindly provided by Bryan Kelly at \url{https://jkpfactors.com}.}

{In total, our empirical analysis considers 24 equity portfolios spanning almost a century of monthly data (see Table~\ref{tab:sample_periods_detailed} in Supplementary Material \ref{app:samples}).}

\subsection{Main empirical results}
\label{subsec:main results}

\paragraph{Weight statistics and turnover.} We start our analysis by comparing the sample statistics of the portfolio weights \(\omega_t = \gamma_{t}/\mathbb{E}_{t}\left[\sigma_{t+1}^2\right]\) computed in real-time. Recall that for {\tt SSV}, {\tt SV}, {\tt HAR}, and {\tt Garch}, forecasts are generated using an expanding window with a five-year burn-in period of monthly returns. The scaling constant \(\gamma_t\) is recalibrated each time a forecast is produced based on portfolio returns available up to time \(t\).

\begin{figure}[!h]
\centering
\begin{flushleft}
\bf Panel A\normalfont: Risk Factors
\end{flushleft}\medskip
\includegraphics[width=1\textwidth]{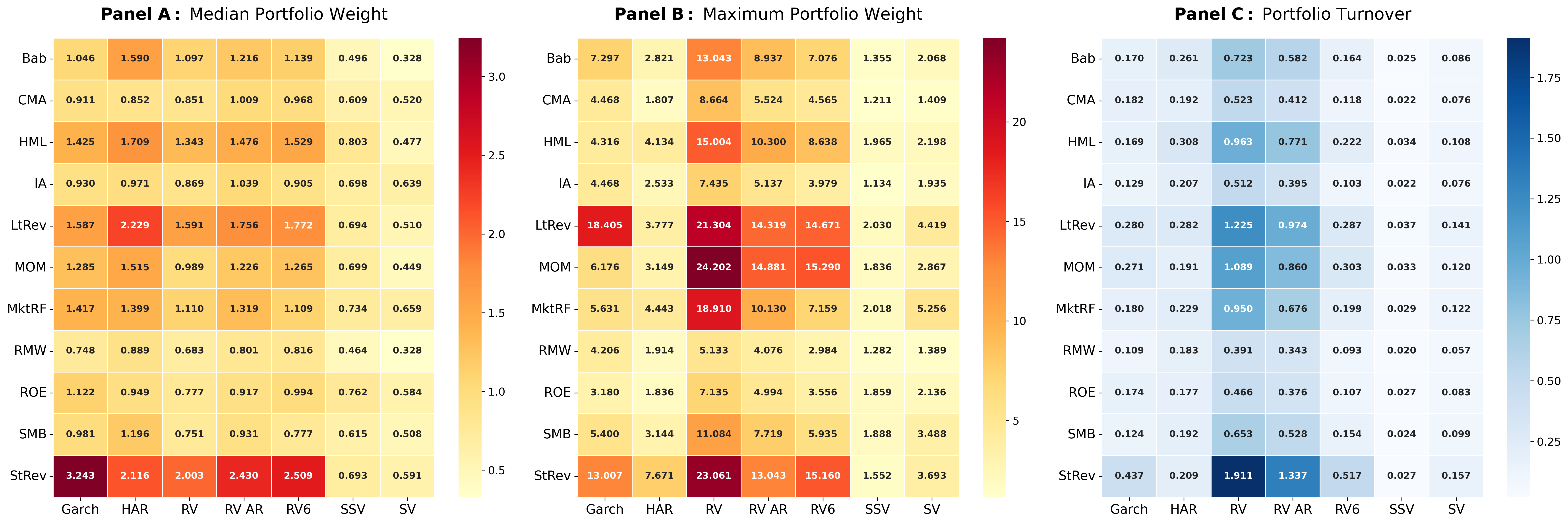}
\begin{flushleft}
\bf Panel B\normalfont: Characteristic-Based Themes
\end{flushleft}\medskip
\includegraphics[width=1\textwidth]{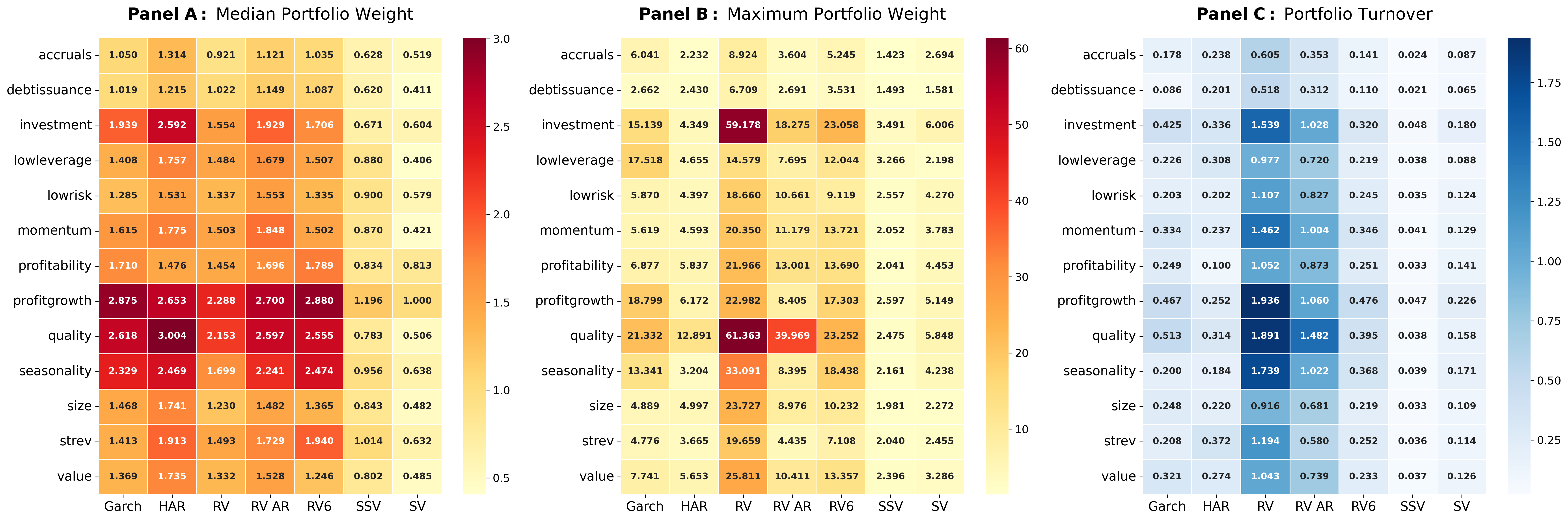}    
\caption{\footnotesize \bf Weight Statistics and Turnover\normalfont. This figure reports the median (left panel) and maximum (middle panel) of the portfolio allocation \(\omega_{t} = \frac{\gamma_{t}}{\mathbb{E}_{t}\left[\sigma_{t+1}^2\right]}\) calculated for each equity risk factor and each volatility timing approach. The right panel reports the average turnover calculated as $E\left[|\Delta \omega|\right]$.}    
	\label{fig:Leverage heatmaps}
\end{figure}

{Figure \ref{fig:Leverage heatmaps} demonstrates stark implementation differences across forecasting methods through three-panel heatmaps comparing portfolio weights and turnover. The analysis reveals a systematic hierarchy in implementation characteristics across models. RV-based approaches (\texttt{RV}, \texttt{RV AR}) generate the most aggressive median leverage (Panel A), often exceeding 2x leverage for momentum portfolios. \texttt{GARCH} and \texttt{HAR} occupy the middle ground with moderate allocations around 1x--1.5x median leverage, while our \texttt{SSV} produces the most conservative median weights, typically ranging 0.4x--1.2x median leverage. \texttt{RV6} demonstrates intermediate behavior, suggesting that longer estimation windows moderate \texttt{RV}'s aggressiveness.}

{Maximum weight patterns (Panel B) expose critical differences in leverage control. \texttt{RV} and \texttt{RV AR} generate extreme maximum positions exceeding 20 for multiple strategies, with some reaching 60. \texttt{GARCH} and \texttt{HAR} show improved but still concerning maximum leverage of 10x--25x. Our SSV demonstrates the most disciplined behavior with maximum positions consistently around 2x--2.5x maximum leverage across all strategies.}

{Turnover analysis (Panel C) reveals a clear efficiency ranking. \texttt{RV} methods exhibit the highest turnover (1.0--1.9), followed by \texttt{RV AR} with moderate reduction. \texttt{GARCH}, \texttt{HAR}, and \texttt{RV6} achieve meaningful turnover improvements, typically ranging 0.2--0.5. Our \texttt{SSV} consistently produces the lowest turnover across all strategies, typically below 0.2.}

{The results demonstrate that while traditional \texttt{RV} methods tend to produce higher average leverage, greater instability, and extreme values -- thus raising concerns about their practical implementation under liquidity constraints \citep{harvey2018impact,bongaerts2020conditional,patton2020you,schwarz2025performance} -- our \texttt{SSV} achieves superior signal extraction. The intermediate methods (\texttt{GARCH}, \texttt{HAR}, \texttt{RV6}) offer partial improvements but cannot match the controlled allocations and minimal rebalancing costs of our smoothed volatility model.}

{The dramatic differences in implementation characteristics stem from fundamentally different volatility dynamics across methods. Figure \ref{fig:volatility comparison} in Supplementary Material \ref{subsec:forecasting accuracy} provides compelling visual evidence for the market portfolio, showing that during crisis periods (Great Depression, GFC, COVID-19), \texttt{RV} exhibits extreme spikes reaching 20-25\% while \texttt{SSV} maintains stability around 5-10\%.}

\begin{figure}[!h]
\centering
\begin{flushleft}
\bf Panel A\normalfont: Risk Factors
\end{flushleft}\medskip
\includegraphics[width=1\textwidth]{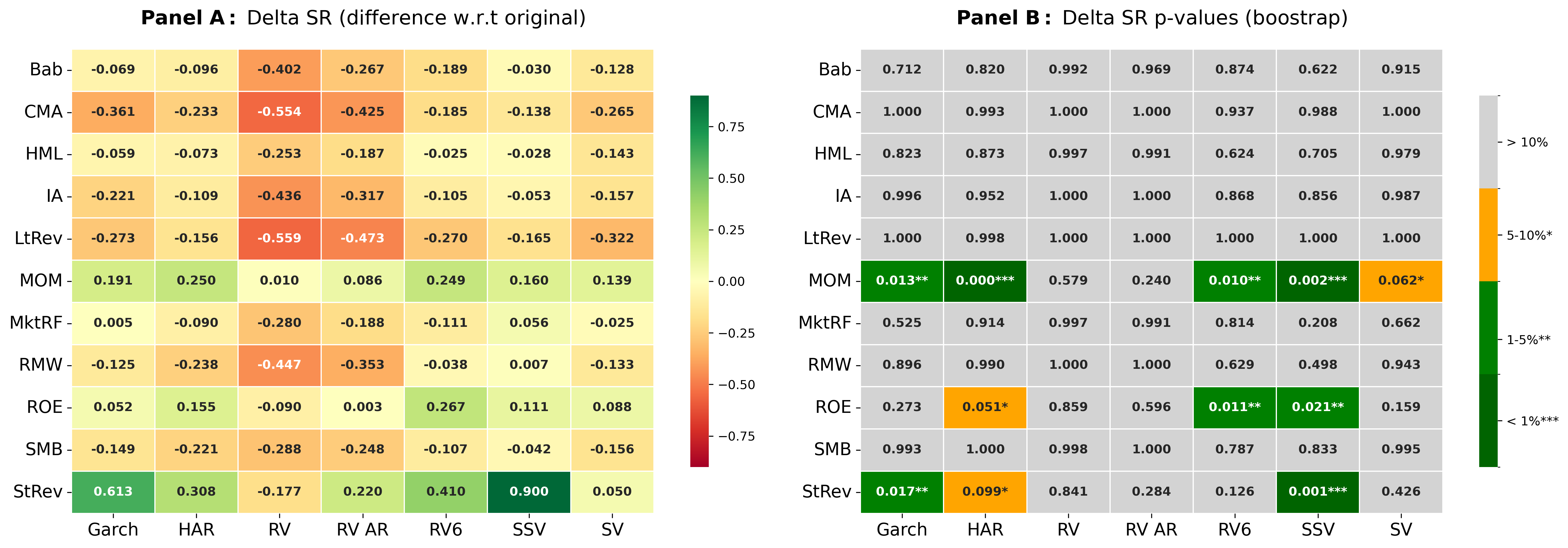}
\begin{flushleft}
\bf Panel B\normalfont: Characteristic-Based Themes
\end{flushleft}\medskip
\includegraphics[width=1\textwidth]{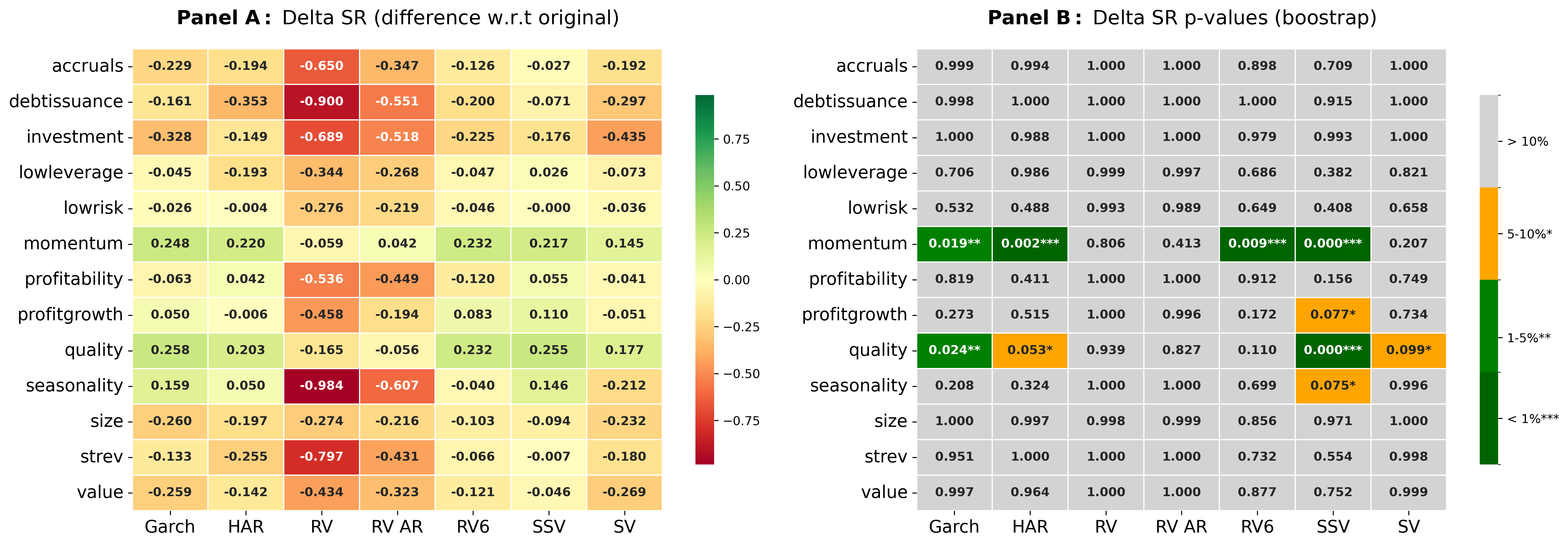}    
\caption{\footnotesize \bf Sharpe Ratio Differentials and Significance. \normalfont The left panel reports the annualized Sharpe ratio (SR) differential, defined as \(\Delta SR_{\tt \mathcal{M}} = SR_{\tt \mathcal{M}} - SR_{\tt U}\), where \(SR_{\tt \mathcal{M}}\) represents the Sharpe ratio of a given volatility-managed portfolio and \(SR_{\tt U}\) corresponds to the original portfolio. Transaction costs of 50 basis points are added as in \citet{demiguel2009optimal}. The right panel presents bootstrap p-values for the null hypothesis \(\mathcal{H}_0: \Delta SR_{\tt \mathcal{M}} = 0\), following the procedure outlined by \citet{ledoit2008robust}.}    
\label{fig:Sharpe ratios}
\end{figure}

\paragraph{Risk-adjusted returns net of transaction costs.} Figure \ref{fig:Sharpe ratios} reports the annualized Sharpe ratio (SR) differential, defined as \(\Delta SR_{\tt \mathcal{M}} = SR_{\tt \mathcal{M}} - SR_{\tt U}\), where \(SR_{\tt \mathcal{M}}\) represents the Sharpe ratio of a given volatility-managed portfolio and \(SR_{\tt U}\) corresponds to the original portfolio. To test the null hypothesis \(\mathcal{H}_0: \Delta SR_{\tt \mathcal{M}} = 0\), we follow \citet{cederburg2020performance} and implement a block bootstrap procedure as in \citet{ledoit2008robust}. 

We follow \citet{wang2021downside} and assume a transaction cost of 50 basis points (bps) to implement volatility targeting.\footnote{{We note that this represents a non-trivial execution cost for trading an ``ETF-like'' factor portfolio. The actual cost of constructing the equity factors, which are long-short portfolios built on individual asset characteristics, is arguably higher. However, these costs precede rebalancing costs due to volatility targeting, making them irrelevant for relative comparisons between volatility forecasts.}} Specifically, transaction costs (\(TC\)) are proportional to monthly changes in leverage, calculated as \(TC \times |\omega_t - \omega^+_{t-1}|\) where $\omega^+_{t-1}=\omega_{t-1}\times(1+y_t^\sigma)$. The wealth dynamics for a given volatility forecast can thus be expressed as \(W_t = W_{t-1}\left(1 + y_t^\sigma\right)\left(1 - TC \times |\omega_t - \omega^+_{t-1}|\right)\) for \(t = 1, \ldots, n,\). where the return net of transaction costs is given by \(y_{t,TC}^\sigma = \frac{W_t}{W_{t-1}} - 1\) as described in \citet{demiguel2009optimal}.

{The out-of-sample performance reveals the economic payoff from \texttt{SSV}'s superior implementation characteristics documented earlier. Panel A shows that \texttt{SSV} delivers positive Sharpe ratio improvements across 10 of 24 strategies, with particularly strong performance for momentum (+0.217), short-term reversal (+0.900), and quality (+0.255) strategies. In stark contrast, traditional \texttt{RV} methods show predominantly negative SR differentials, with particularly poor performance for investment themes (-0.900 for debt issuance, -0.984 for seasonality, -0.797 for reversal). The deep red coloring for \texttt{RV} across investment themes illustrates how implementation costs can completely erode economic value. Conversely, \texttt{SSV}'s disciplined leverage control (maximum $<3x$) and minimal turnover ($<0.1$) preserve the economic benefits of volatility timing.}

{The statistical significance patterns (Panel B) further validate \texttt{SSV}'s advantage. \texttt{SSV} achieves statistically significant improvements for 5 strategies (marked with green shading and significance stars), including momentum ($p<0.01$), short-term reversal ($p<0.001$), and quality ($p<0.001$). Traditional \texttt{RV}-based methods achieve zero significant improvements, reflecting their poor net-of-costs performance. The intermediate methods (\texttt{GARCH}, \texttt{HAR}, \texttt{RV6}) show mixed results---some positive improvements but limited statistical significance.}

\paragraph{Systematic performance analysis.} 

{To test whether \texttt{SSV}'s improvements represent genuine systematic enhancement rather than data mining, we implement a comprehensive testing framework following best practices from the anomaly meta-analysis literature \citep{cederburg2020performance,hou2020replicating}.} 

{We first examine the complete performance distribution to ensure that \texttt{SSV}'s benefits do not come at the cost of increased extreme negative outcomes. Using a threshold of $|\Delta SR| > 0.2$ for economically meaningful improvements, we classify each portfolio-method combination as exhibiting significant outperformance, significant underperformance, or neutral performance. Let $\Delta SR_{i,\mathcal{M}} = SR_{i,\mathcal{M}} - SR_{i,U}$ represent the Sharpe ratio improvement for portfolio $i$ using method $\mathcal{M}$ relative to the unmanaged portfolio $U$. We define Net Success as: $\text{Net Success}_{\mathcal{M}} = \sum_{i=1}^N \text{Success}_i - \sum_{i=1}^N \text{Failure}_i$ where $\text{Success}_i = \mathbb{I}(\Delta SR_{i,\mathcal{M}} > 0.2)$ and $\text{Failure}_i = \mathbb{I}(\Delta SR_{i,\mathcal{M}} < -0.2)$.}

{We also conduct two complementary joint tests to evaluate statistical significance. First, we test whether the success rate significantly differs from the 50\% null hypothesis of no systematic benefit using $Z_{\mathcal{M}} = \frac{X_{\mathcal{M}} - N \cdot 0.5}{\sqrt{N \cdot 0.25}}$, where $X_{\mathcal{M}}$ is the number of portfolios with positive improvements and $N$ is the total number of portfolios. Second, we test whether the mean improvement across portfolios significantly differs from zero using $t_{\mathcal{M}} = \frac{\overline{\Delta SR}_{\mathcal{M}}}{s_{\mathcal{M}}/\sqrt{N}}$, where $\overline{\Delta SR}_{\mathcal{M}}$ is the sample mean improvement and $s_{\mathcal{M}}$ is the sample standard deviation.}

{Finally, we implement left-tail analysis to address concerns that superior average performance might mask extreme negative outcomes. We examine the worst-performing portfolios for each method and count cases of extreme underperformance (SR $< -0.1$ annualized).}

\begin{table}[h!]
\centering
\resizebox{0.9\textwidth}{!}{
\begin{tabular}{l@{\hspace{0.3cm}}rrr@{\hspace{0.5cm}}rr@{\hspace{0.5cm}}rr}
\midrule 
& \multicolumn{3}{c}{Distribution Analysis} & \multicolumn{2}{c}{Joint Tests} & \multicolumn{2}{c}{Left Tail Analysis} \\
\cmidrule(lr){2-4} \cmidrule(lr){5-6} \cmidrule(lr){7-8}
Method & Sig.Out & Sig.Under & Net Success & Mean $\Delta$SR & Success Rate & Extreme Neg & Worst Decile \\
\midrule
\texttt{RV}     & 0  & 19 & -19 & -0.418$^{***}$ & 4.2\%  & 11 & -0.611 \\
\texttt{RV6}    & 5  & 2  & +3  & -0.027    & 25.0\% & 0  & -0.060 \\
\texttt{RV AR}  & 1  & 16 & -15 & -0.262$^{***}$ & 16.7\% & 8  & -0.311 \\
\texttt{HAR}    & 4  & 5  & -1  & -0.062    & 29.2\% & 1  & -0.098 \\
\texttt{GARCH}  & 3  & 7  & -4  & -0.049    & 33.3\% & 0  & -0.070 \\
\texttt{SV}     & 0  & 7  & -7  & -0.114$^{**}$  & 20.8\% & 2  & -0.148 \\
\texttt{SSV}    & 3  & 0  & +3  & +0.049    & 45.8\% & 0  & +0.015 \\
\midrule 
\end{tabular}}
\caption{\footnotesize \bf Systematic Performance Analysis\normalfont. This table presents a comprehensive systematic performance analysis across 24 portfolios using three complementary approaches. Distribution Analysis: "Sig.Out/Under" indicates portfolios with |$\Delta$SR| > 0.2; Net Success = Sig.Out - Sig.Under. Joint Tests: Mean $\Delta$SR tests average improvement across portfolios; Success Rate shows proportion with positive improvements. Left Tail: "Extreme Neg" counts portfolios with $\Delta$SR < -0.1; "Worst Decile" shows average performance of bottom 10\% portfolios. ***, **, * denote significance at 1\%, 5\%, 10\% levels for mean tests.}
\label{tab:systematic_performance}
\end{table}

{Table~\ref{tab:systematic_performance} reveals that \texttt{SSV} uniquely achieves positive performance across all three analyses. The distribution analysis shows \texttt{SSV} achieves positive net success (+3) with zero failures, while traditional \texttt{RV} methods exhibit severe deterioration: \texttt{RV} achieves zero significant outperformances against 19 underperformances (net success = -19), and \texttt{RV AR} produces only 1 outperformance against 16 underperformances (net success = -15).} 

{Joint tests confirm \texttt{SSV}'s 45.8\% success rate approaches the theoretical 50\% breakeven point and delivers 0.049 mean improvement -- the only positive average performance across all portfolios. Left-tail analysis shows \texttt{SSV} maintains positive performance even in worst-case scenarios (0.015 worst-decile average) with zero extreme negative outcomes, while \texttt{RV} methods show 8-11 cases of severe underperformance. This demonstrates that \texttt{SSV} provides systematic enhancement through consistent positive performance without the downside risk that characterizes alternative volatility forecasting approaches.}

\section{Robustness analysis and economic foundations}
\label{sec:additional_results}

\begin{figure}[!h]
\centering
\begin{flushleft}
\bf Panel A\normalfont: Risk Factors
\end{flushleft}\medskip
\includegraphics[width=1\textwidth]{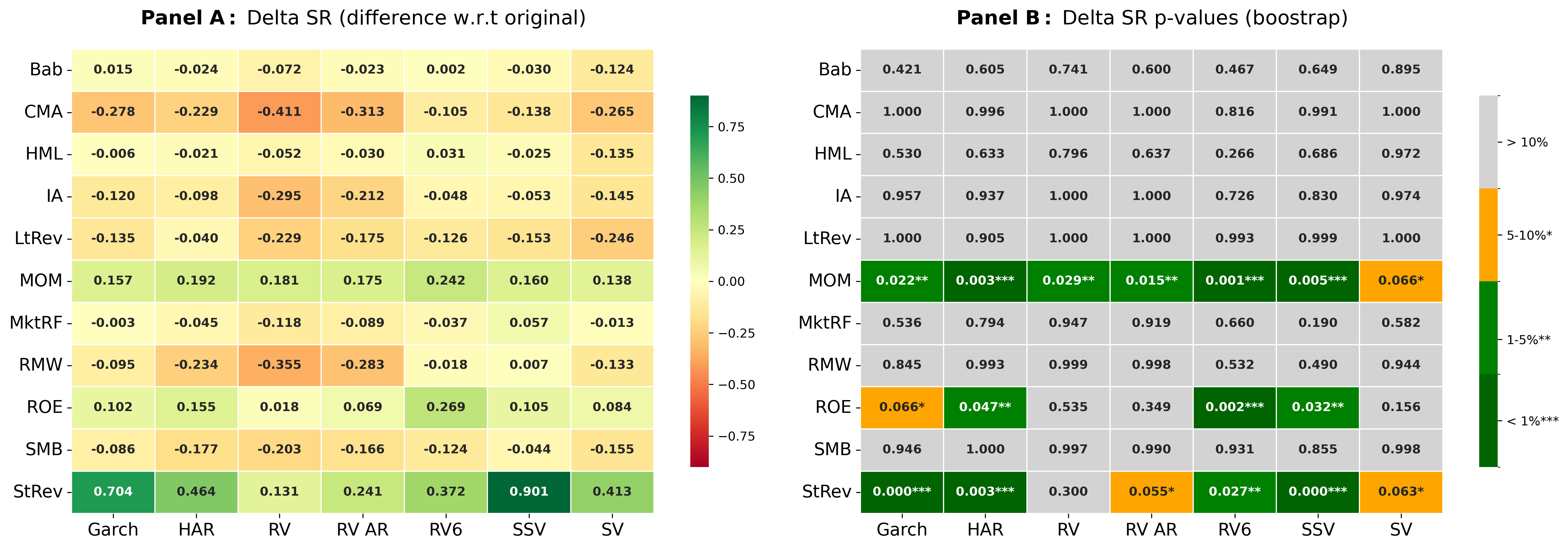}
\begin{flushleft}
\bf Panel B\normalfont: Characteristic-Based Themes
\end{flushleft}\medskip
\includegraphics[width=1\textwidth]{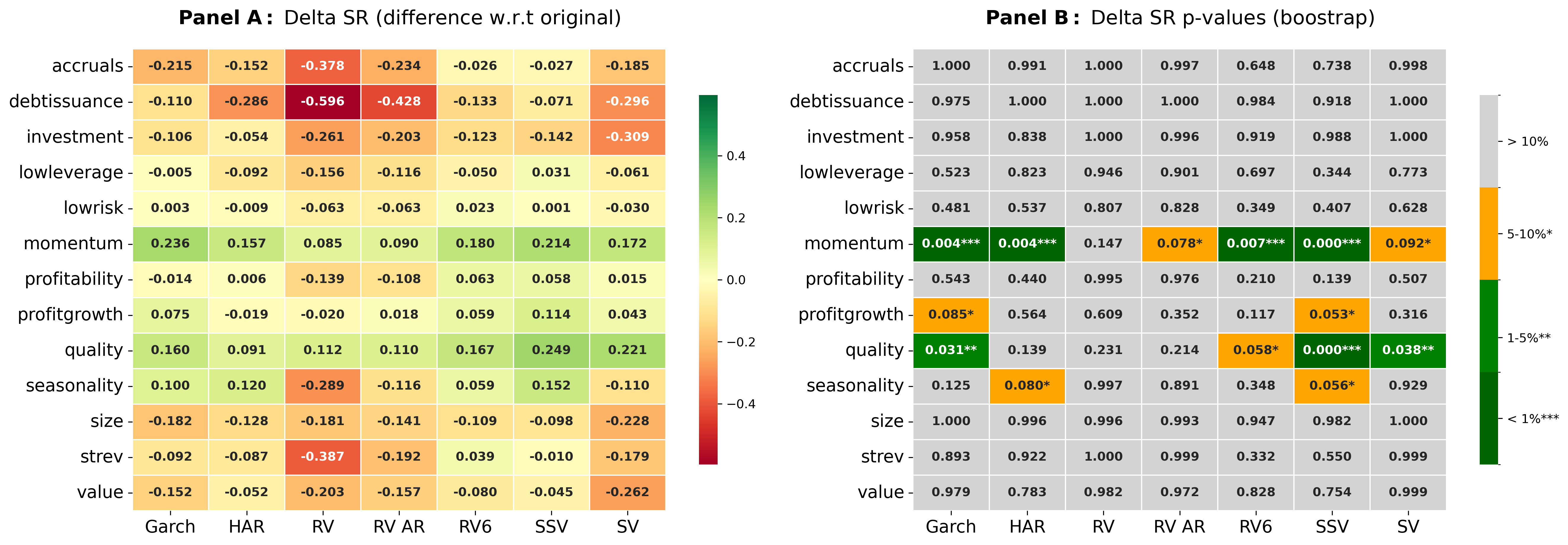}    
\caption{\footnotesize \bf Sharpe Ratio Differentials and Significance with 1.5x Leverage Cap. \normalfont The left panel reports the annualized Sharpe ratio (SR) differential, defined as \(\Delta SR_{\tt \mathcal{M}} = SR_{\tt \mathcal{M}} - SR_{\tt U}\), where \(SR_{\tt \mathcal{M}}\) represents the Sharpe ratio of a given volatility-managed portfolio and \(SR_{\tt U}\) corresponds to the original portfolio. Transaction costs of 50 basis points are added as in \citet{demiguel2009optimal}. The leverage is capped at 50\% (1.5x) such that $\omega_{t}=\min\left(\gamma_{t}/\mathbb{E}_{t}\left[\sigma_{t+1}^2\right],1.5\right)$. The right panel presents bootstrap p-values for the null hypothesis \(\mathcal{H}_0: \Delta SR_{\tt \mathcal{M}} = 0\), following the procedure outlined by \citet{ledoit2008robust}.}    
\label{fig:Sharpe ratios with constraints}
\end{figure}

\paragraph{Leverage constraints.} As shown in Figure~\ref{fig:Leverage heatmaps}, implementing volatility-managed portfolios often requires high leverage, which entails significant liquidity needs. A common approach to mitigate this issue is to impose a cap on leverage through arbitrary constraints. Figure~\ref{fig:Sharpe ratios with constraints} extends the out-of-sample SRs by evaluating the performance of different methods when leverage is capped at 50\%, such that \(\omega_{t} = \min\left(\gamma_{t}/\mathbb{E}_{t}\left[\sigma_{t+1}^2\right], 1.5\right)\). 

Under leverage constraints, \texttt{SSV} retains superiority across both factor and theme portfolios. \texttt{SSV} achieves positive improvements for momentum strategies, with high statistical significance ($p < 0.01$). This is consistent with existing literature \citep{barroso2021limits,schwarz2025performance}. For short-term reversal, \texttt{SSV} delivers exceptional performance with 0.901 improvement ($p < 0.001$), representing the strongest performance across all methods. Additionally, \texttt{SSV} shows significant positive improvements for ROE (0.105, $p < 0.05$), quality (0.249, $p < 0.001$), and seasonality (0.152, $p < 0.10$). 

The leverage constraint severely hampers traditional \texttt{RV}-based methods, which rely on extreme leverage to generate risk-adjusted returns. \texttt{RV} shows predominantly negative improvements, with particularly poor performance for CMA (-0.411), RMW (-0.355), and several investment themes. \texttt{RV AR} exhibits similar deterioration, with large negative improvements for CMA (-0.313) and RMW (-0.283). Overall, the leverage-constrained results provide compelling evidence in favour of \texttt{SSV}'s economic utility. 

In a set of unreported results, a less restrictive leverage cap of 500\%, such that \(\omega_{t} = \min\left(\gamma_{t}/\mathbb{E}_{t}\left[\sigma_{t+1}^2\right], 5\right)\), provides risk-adjusted performance which is largely similar. While the cap reduces both average returns and volatility, the Sharpe ratio exhibits only marginal changes compared to the unconstrained weights. 

\paragraph{Utility gains.} We follow \citet{cederburg2020performance} and further evaluate the economic value of different volatility methods using a conventional mean-variance allocation in {real-time using a 20-year rolling window} as,
\begin{equation}
\left[x_{\sigma,t},\ x_t\right]^\prime = \frac{1}{\delta}\widehat{\Sigma}_t^{-1}\widehat{\mu}_t,
\label{eq:mean variance}
\end{equation}
where $\widehat{\Sigma}_t$ and $\widehat{\mu}_t$ are the recursively estimated $2\times 2$ covariance structure of the original and volatility-managed portfolio returns and the $2\times 1$ vector of average returns, respectively. This strategy dynamically combines the original portfolio and the real-time volatility-scaled portfolio, generating the investment rule \(z_t = x_{\sigma,t}\omega_t + x_t.\) The returns of the combined strategy are given by \(y^*_{t+1} = z_t \cdot y_{t+1}\). 

\begin{figure}[!h]
\centering
\includegraphics[width=1\textwidth]{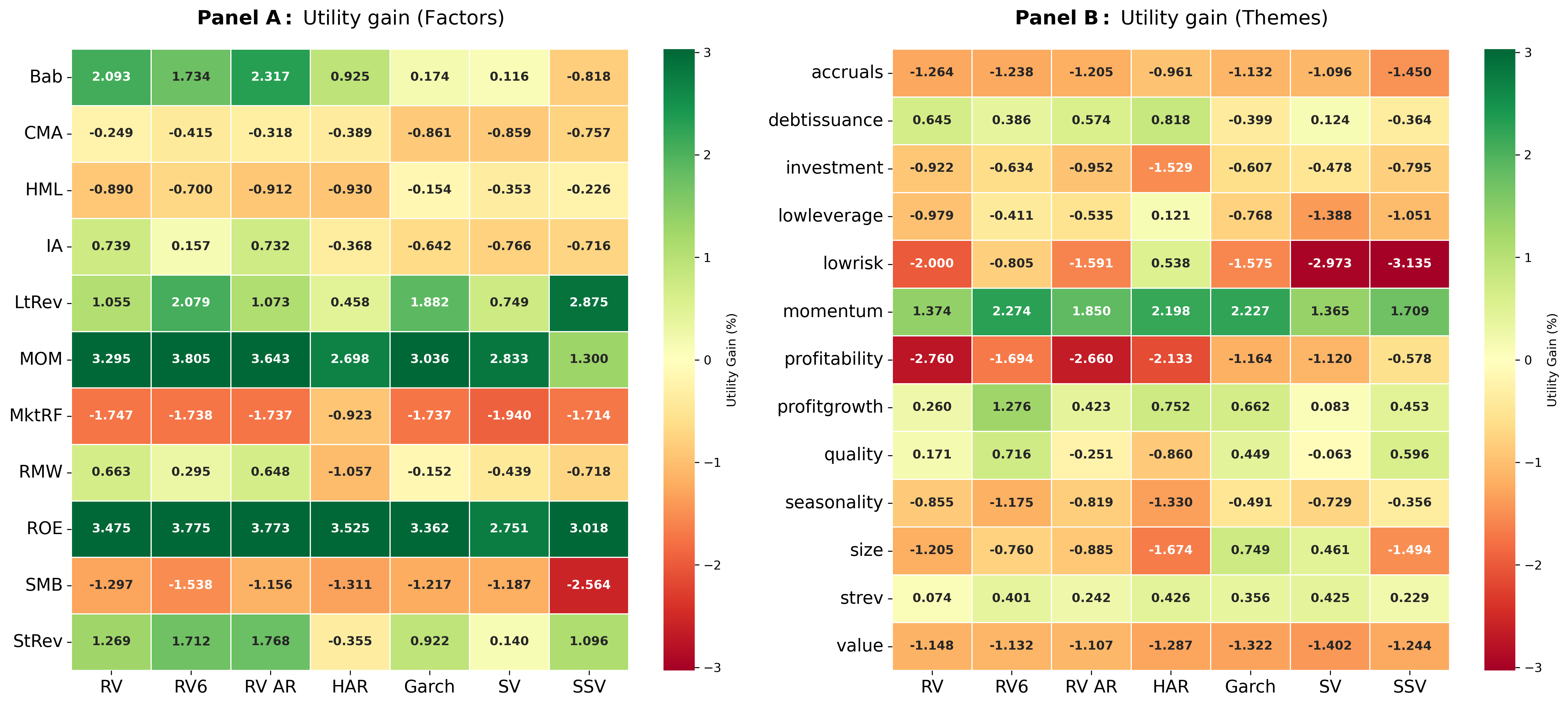}
\caption{\footnotesize \bf Out-of-Sample Utility Gains \normalfont This figure reports the utility gain $\Delta\text{CER} = \frac{\text{SR}\left(z^*_t\right) - \text{SR}\left(y_t\right)}{2\delta}$ where $\text{SR}\left(y_t\right)$ is the Sharpe ratio of the original equity portfolio and $\text{SR}\left(z^*_t\right)$ is the Sharpe ratio of the combined strategy $z_t=x_{\sigma,t}\omega_t + x_t$. The left panel reports the results for the factor portfolios whereas the right panel reports the results for the investment themes.}    
\label{fig:utility gains}
\end{figure}

Figure~\ref{fig:utility gains} reports the utility gain, calculated as \(\Delta \text{CER} = \frac{\text{SR}(z_t) - \text{SR}(y_t)}{2\delta},\) where \(\text{SR}(y_t)\) and \(\text{SR}(z_t)\) denote the Sharpe ratios of the original portfolio and the combined strategy, respectively. Following \citet{barroso2021limits}, we assume a risk aversion of \(\delta = 5\).

Our analysis reveals that the \texttt{SSV} method demonstrates comparable performance to alternative volatility estimation approaches across multiple dimensions. For factor portfolios (left panel), \texttt{SSV} achieves positive utility gains across key strategies including Bab ($-0.8\%$), LtRev ($2.9\%$), MOM ($1.3\%$), and ROE ($3.0\%$). Notably, \texttt{SSV} demonstrates a consistent performance which is similar to realized variance-based methods.

Compared to GARCH, \texttt{SSV} achieves higher utility gains for factor portfolios such as LtRev (2.9\% vs 1.9\%) and ROE (3.0\% vs 2.8\%). Relative to \texttt{HAR}, \texttt{SSV} demonstrates more consistent positive performance across the factor and theme portfolios universe, with fewer instances of substantial negative utility gains.

\paragraph{Spanning regressions.} 

The economic implication of Eq.\eqref{eq:mean variance} is that a volatility-managed portfolio expands the mean-variance frontier compared to a stand-alone investment in the original portfolio \citep[e.g.,][]{gibbons1989test}. This conclusion follows from the direct link between spanning tests and the mean-variance efficiency of portfolio allocation.

\begin{figure}[!h]
\centering
\begin{flushleft}
\bf Panel A\normalfont: Risk Factors
\end{flushleft}\medskip
\includegraphics[width=1\textwidth]{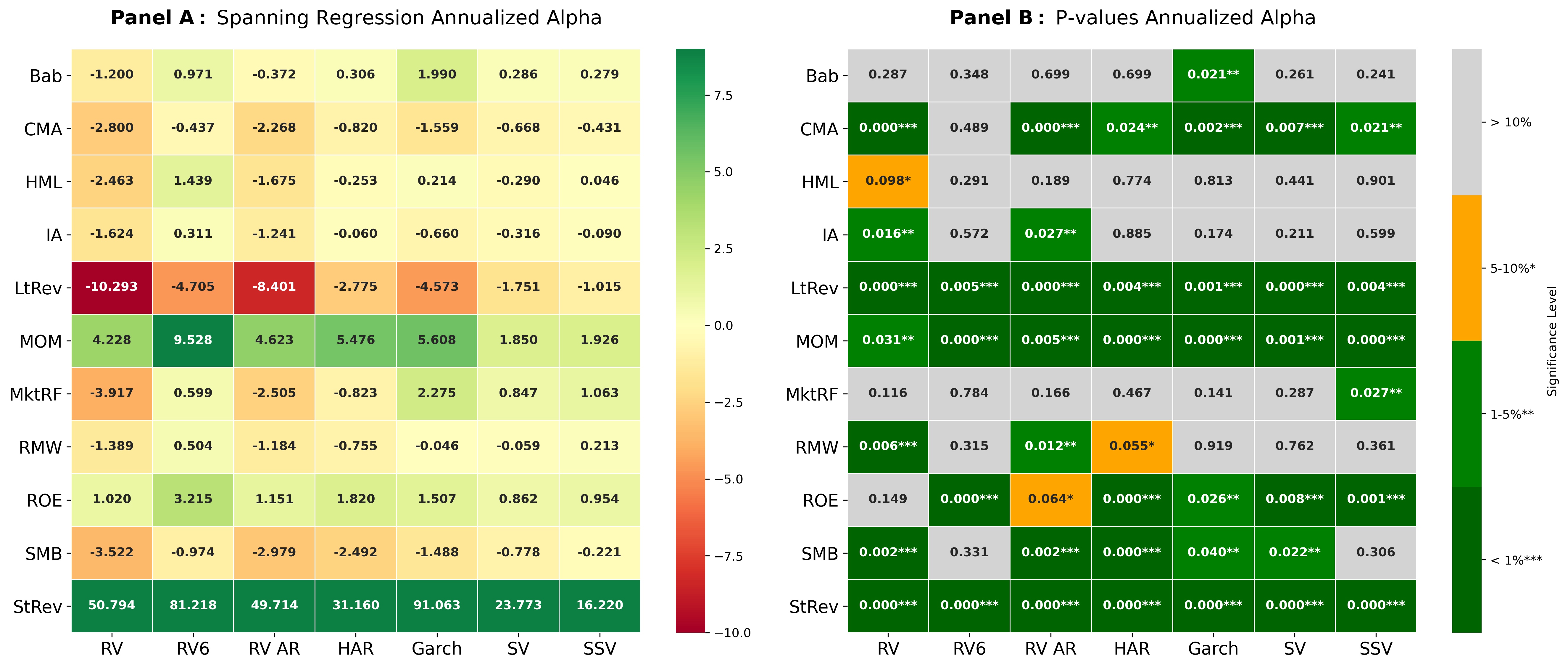}
\begin{flushleft}
\bf Panel B\normalfont: Characteristic-Based Themes
\end{flushleft}\medskip
\includegraphics[width=1\textwidth]{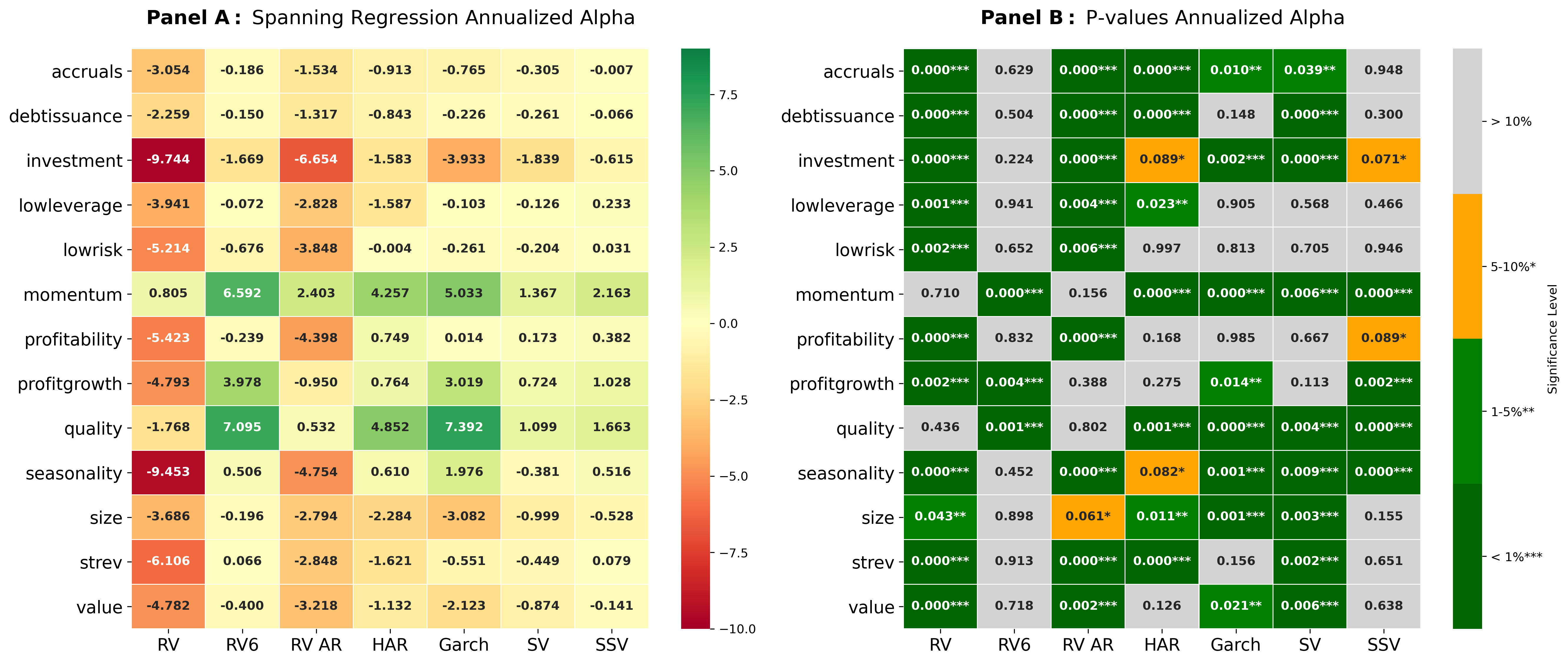} 
\caption{\footnotesize \bf In-Sample Spanning Regressions\normalfont. This figure reports the alphas and the corresponding p-values of a time-series regression of the form $y^\sigma_{t} = \alpha + \beta y_t + \epsilon_t$ where $y^\sigma_{t}$ is the return on the volatility-managed factor and $y_{t}$ the return on the original, unmanaged equity risk factor. A full description of the equity risk factors and volatility modelling methods is available in the main text.}    
\label{fig:alphas}
\end{figure}

Consider an investor who allocates her wealth between \(y^\sigma_{t}\) and \(y_t\) based on mean-variance preferences. Assume the volatility-managed return is an affine function of the original portfolio \(y^\sigma_{t} = \alpha + \beta y_t + \epsilon_t\). The in-sample mean-variance portfolio then takes the form,
\[
\left[x_\sigma, x\right]^\prime = \frac{1}{\delta} \widehat{\Sigma}^{-1} \widehat{\mu},
\]
where \( \widehat{\Sigma} = \widehat{\sigma}_f^2 \begin{bmatrix}
1 & \widehat{\rho} \\
\widehat{\rho} & 1
\end{bmatrix} \)
and
\(\widehat{\mu} = \left[\widehat{\alpha} + \widehat{\rho}\overline{y},\ \overline{y}\right]\),
with \(\widehat{\rho} = \text{Corr}(y^\sigma_{t}, y_t)\), \(\widehat{\sigma}_f^2 = \text{Var}(y_t)\), and \(\overline{y} = \text{E}[y_t]\). The optimal allocation to the volatility-managed portfolio is thus:
\[
x_\sigma = \frac{\widehat{\alpha}}{\delta \widehat{\sigma}_f^2 (1-\widehat{\rho})^2}.
\]
This implies that, for a given level of risk aversion \(\delta\), the optimal strategy assigns positive weight to the volatility-managed portfolio only if \(\alpha > 0\). Figure \ref{fig:alphas} reports the in-sample estimates of the \(\alpha\) from the in-sample regression $y^\sigma_{t} = \alpha + \beta y_t + \epsilon_t$.

\texttt{SSV} generates positive annualized alphas for several factor portfolios, including LtRev (16.22\%), MOM (1.93\%), and ROE (0.95\%). These compare favorably to RV, which produces negative alphas for multiple factor portfolios including SMB (-3.52\%), MktRF (-3.92\%), and CMA (-2.80\%). Among theme portfolios, \texttt{SSV} exhibits smaller negative alphas than RV across most strategies, with differences ranging from 1-3 percentage points.

{It is worth highlighting that while portfolios are indeed scaled to achieve similar volatilities, the lower alphas of \texttt{SSV} arise from a fundamental difference in how alpha scales with leverage: alpha generation depends critically on the \textit{magnitude} of deviations from the original portfolio, not just the volatility of those deviations. As \citet{moreira2017volatility} demonstrate, leverage-constrained volatility strategies produce lower alphas precisely because constraints limit the magnitude of portfolio adjustments. \texttt{SSV}'s lower leverage reflects its inherent efficiency in filtering volatility signals, requiring smaller position changes to capture the same economic benefits.}


\paragraph{Accuracy vs utility.} {It is important to highlight that our objective is not to improve volatility forecasting accuracy, but to optimize volatility estimates for economic utility in standard volatility-managed portfolio implementations. However, a fundamental question arises about the relationship between statistical forecasting accuracy and economic utility. To address this question, we calculate the out-of-sample $R_{oos}^2$ for each model as:
\begin{align}
R_{oos}^2 & = 1- \frac{\sum_{t=t_0}^{T-1}(RV_{t+1} - \sigma^2_{t+1|t}(\mathcal{M}))^2}{\sum_{t=t_0}^{T-1}(RV_{t+1} - \overline{RV_{t}})^2}\label{eq:R2oos}  
\end{align}
where the numerator is the sum-of-squared forecasting error from model $\sigma^2_{t+1|t}(\mathcal{M})$ and the denominator is the sum-of-squared residual of a naive forecast from the sample average of realized variance $\overline{RV_{t}}$. This provides a scale-invariant assessment of each model's ability to predict realized variance.}

{Supplementary Material \ref{subsec:forecasting accuracy} presents the results. The findings in Table \ref{tab:oos_r2_comparison} and Figure \ref{fig:tradeoff accuracy performance} reveal a fundamental disconnect between forecasting accuracy and portfolio performance. \texttt{RV AR} achieves the highest average $R^2_{oos}$ (0.380) but generates extreme leverage and turnover that erode net returns. In contrast, our \texttt{SSV} method achieves moderate forecasting accuracy (0.149 average $R^2_{oos}$) while delivering superior implementable portfolios. This supports our central thesis that economic utility often dominates pure forecasting precision in practical portfolio management.}

\paragraph{Temporal consistency analysis.} {To address concerns that performance gains might be concentrated in specific favorable periods rather than representing systematic improvement, we analyze \texttt{SSV}'s performance across distinct historical market regimes. Following \citet{cederburg2020performance}, we partition our sample into five major periods: Great Depression (1929-1939), Post-War Boom (1940-1969), Stagflation Era (1970-1982), Great Moderation (1983-2006), and Financial Crisis \& Recovery (2007-2020).}

{Figure~\ref{fig:performance per period} in Supplementary Material \ref{subsec:subsample} reports Sharpe ratio differences between each volatility-targeting strategy and the original portfolio across these periods for representative portfolios (MktRF, MOM, HML). The analysis reveals several key insights about the temporal robustness of our approach. \texttt{SSV} demonstrates positive SR differences across all five historical periods for momentum portfolios, with improvements ranging from 0.022 to 0.239. This contrasts sharply with traditional realized variance methods, which show highly volatile period-to-period performance, including extreme negative outcomes during certain regimes.}

{Importantly, unlike competing methods that show extreme negative performance in certain periods (e.g., \texttt{RV} achieving -0.553 for MktRF during the Financial Crisis), \texttt{SSV} exhibits more stable performance across different market regimes. Traditional volatility targeting methods (\texttt{RV}, \texttt{RV6}, \texttt{RV AR}) display period-dependent performance with large swings between positive and negative SR differences, suggesting their gains are concentrated in few periods.}

{The temporal analysis provides strong evidence that \texttt{SSV}'s outperformance represents systematic improvement in volatility forecasting rather than gains concentrated in a few favorable periods. While other methods alternate between strong and weak performance across regimes, \texttt{SSV}'s more consistent pattern indicates that its superior performance stems from fundamental methodological advantages rather than timing luck.}

\section{Conclusion}
\label{sec:conclu}

Volatility-managed portfolios promise superior risk-adjusted returns, yet implementation challenges often erode their theoretical benefits. This paper demonstrates that smoothing volatility forecasts through a novel variational Bayes approach resolves this implementation paradox. Our smoothed stochastic volatility (\texttt{SSV}) method achieves positive net success across all portfolios with zero implementation failures. Across 24 equity factors and themes spanning nearly a century, \texttt{SSV} combines superior risk-adjusted performance with controlled leverage and minimal turnover. When transaction costs are considered, \texttt{SSV} delivers systematic economic gains while traditional realized variance methods suffer severe deterioration.

\bibliographystyle{cas-model2-names}
\bibliography{cas-refs}

\newpage
\appendix

\pagenumbering{arabic}

\renewcommand\thetheorem{\thesection.\arabic{theorem}} 
\renewcommand\thelemma{\thesection.\arabic{lemma}} \renewcommand\thedefinition{\thesection.\arabic{definition}} 
\renewcommand\thecorollary{\thesection.\arabic{corollary}} 
\renewcommand\theproposition{\thesection.\arabic{proposition}} 
\renewcommand\thefigure{\thesection.\arabic{figure}} 
\renewcommand\thetable{\thesection.\arabic{table}}    
\numberwithin{equation}{section}
\numberwithin{figure}{section}
\numberwithin{table}{section}

\justifying
\noindent{\Large\bf Supplementary Material}\\
\vspace{0.2cm}\\
This supplementary material contains the complete derivation of the optimal variational densities outlined in the main text. The supplementary material also provides additional simulation and empirical results. All the additional results included in this supplementary material are referred to in the main text.\\

\section{Derivation of the Optimal Variational Densities}
\label{app:params}

The mean-field approach factorizes the joint variational distribution according to a partition $q(\boldsymbol{\vartheta})=\prod_{j=1}^M q(\boldsymbol{\vartheta}_j)$,
where each component $q(\boldsymbol{\vartheta}_j)$ can be computed as (see \citealp{wand_ormerod.2011}),
\begin{equation}
    q_j^\ast(\boldsymbol{\vartheta}_j) \propto \exp\left\{\mathbb{E}_{q(\boldsymbol{\vartheta}\setminus\boldsymbol{\vartheta}_j)}\Big[\log p(\mathbf{y},\boldsymbol{\vartheta})\Big] \right\},\qquad
q(\boldsymbol{\vartheta}\setminus\boldsymbol{\vartheta}_j)=\prod_{\substack{i=1\\i\neq j}}^d q_i(\boldsymbol{\vartheta}_i),\label{eq:mfvb_opt}
\end{equation}
where the expectation is taken with respect to the joint approximating density with the $j$-th element of the partition removed $q^\star(\boldsymbol{\vartheta}\setminus\boldsymbol{\vartheta}_j)$. A valid alternative to Eq.\eqref{eq:mfvb_opt} is given by:
\begin{equation}
    q_j^\ast(\boldsymbol{\vartheta}_j) \propto \exp\left\{\mathbb{E}_{q(\boldsymbol{\vartheta}\setminus\boldsymbol{\vartheta}_j)}\left[\log p(\boldsymbol{\vartheta}_j|\text{rest})\right]\right\},\label{eq:mfvb_opt2}
\end{equation}
where $p(\boldsymbol{\vartheta}_j|\text{rest})$ denotes the full conditional distribution of $\boldsymbol{\vartheta}_j$ (see \citealp{ormerod2010explaining}). The following derivations leverage Eq.\eqref{eq:mfvb_opt2} to specify the optimal variational densities for the model parameters and the latent volatility state. Assume that $\mathbf{y}$ is a $n$-dimensional vector, $\mathbf{X}$ a $p\times n$ matrix and $\boldsymbol{\mathbf{\beta}}$ a $p$-dimensional vector of parameters whose prior distribution is $\beta\sim \mathsf{N}_p\left(\mathbf{\mu}_\beta,\mathbf{\Sigma}_\beta\right)$. 
\begin{proposition}\label{eq:prop_beta}
The optimal variational density for $\mathbf{\beta}$ is $q^\ast(\mathbf{\beta})\equiv\mathsf{N}_p(\mathbf{\mu}_{q(\beta)},\mathbf{\Sigma}_{q(\beta)})$, with
	\begin{align}\label{eq:up_beta}
		\mathbf{\Sigma}_{q(\beta)}= \left(\mathbf{X}^\intercal\mathbf{H}^{-1}\mathbf{X}+\mathbf{\Sigma}_\beta^{-1}\right)^{-1} \qquad
		\mathbf{\mu}_{q(\beta)}=\mathbf{\Sigma}_{q(\beta)}\left(\mathbf{X}^\intercal\mathbf{H}^{-1}\mathbf{y}+\mathbf{\Sigma}_\beta^{-1}\mathbf{\mu}_\beta\right),
	\end{align}
	Where $\mathbf{H}=\mathsf{Diag}\left(\mathbb{E}_q\left[\mathrm{e}^{\mathbf{h}}\right]\right)$ is a diagonal matrix with elements that depend on the optimal density for the latent log-volatilities $q\left(\mathbf{h}\right)$ (see Proposition \ref{eq:prop_h_het}). 
\end{proposition}
\begin{proof}
	The logarithm of the full conditional $p(\mathbf{\beta}|\text{rest})$ is proportional to:
	\begin{align*}
	    \log p(\mathbf{\beta}|\text{rest}) &\propto -\frac{1}{2}\left(\mathbf{y}-\mathbf{X}\mathbf{\beta}\right)^\intercal\mathsf{diag}\left(\mathrm{e}^{\mathbf{h}}\right)\left(\mathbf{y}-\mathbf{X}\mathbf{\beta}\right) -\frac{1}{2}\left(\mathbf{\beta}-\mathbf{\mu}_\beta\right)^\intercal\mathbf{\Sigma}_\beta^{-1}\left(\mathbf{\beta}-\mathbf{\mu}_\beta\right) \\
	    &\propto -\frac{1}{2}\left(\mathbf{\beta}^\intercal\mathbf{X}^\intercal\mathsf{diag}\left(\mathrm{e}^{\mathbf{h}}\right)\mathbf{X}\mathbf{\beta}-2\mathbf{\beta}^\intercal\mathbf{X}^\intercal\mathsf{diag}\left(\mathrm{e}^{\mathbf{h}}\right)\mathbf{y}\right) -\frac{1}{2}\left(\mathbf{\beta}^\intercal\mathbf{\Sigma}_\beta^{-1}\mathbf{\beta}-2\mathbf{\beta}^\intercal\mathbf{\Sigma}_\beta^{-1}\mathbf{\mu}_\beta\right).
	\end{align*}
	Compute the optimal variational density as $ \log q^\ast(\mathbf{\beta})=\mathbb{E}_{q(\boldsymbol{\vartheta}\setminus\mathbf{\beta})}\left[\log p(\mathbf{\beta}|\text{rest})\right]$:
	\begin{align*}
	    q^\ast(\mathbf{\beta}) &\propto -\frac{1}{2}\left(\mathbf{\beta}^\intercal\mathbf{X}^\intercal\mathsf{diag}\left(\mathbb{E}_h\left[\mathrm{e}^{\mathbf{h}_1}\right]\right)\mathbf{X}\mathbf{\beta}-2\mathbf{\beta}^\intercal\mathbf{X}^\intercal\left(\mathbb{E}_h\left[\mathrm{e}^{\mathbf{h}}\right]\right)\mathbf{y}\right) \\ &\qquad-\frac{1}{2}\left(\mathbf{\beta}^\intercal\mathbf{\Sigma}_\beta^{-1}\mathbf{\beta}-2\mathbf{\beta}^\intercal\mathbf{\Sigma}_\beta^{-1}\mathbf{\mu}_\beta\right)\\
	    &= -\frac{1}{2}\left(\mathbf{\beta}^\intercal(\mathbf{X}^\intercal\mathbf{H}^{-1}\mathbf{X}+\mathbf{\Sigma}_\beta^{-1})\mathbf{\beta}-2\mathbf{\beta}^\intercal(\mathbf{X}^\intercal\mathbf{H}^{-1}\mathbf{y}+\mathbf{\Sigma}_\beta^{-1}\mathbf{\mu}_\beta)\right),
	\end{align*}
	where $\mathbf{H}^{-1}=\mathsf{diag}\left(\mathbb{E}_h\left[\mathrm{e}^{\mathbf{h}}\right]\right)$. Take the exponential and you obtain the kernel of a multivariate Gaussian distribution with mean $ 	\mathbf{\mu}_{q(\beta)}=\mathbf{\Sigma}_{q(\beta)}\left(\mathbf{X}^\intercal\mathbf{H}^{-1}\mathbf{y}+\mathbf{\Sigma}_\beta^{-1}\mathbf{\mu}_\beta\right)$ and covariance $\mathbf{\Sigma}_{q(\beta)}= \left(\mathbf{X}^\intercal\mathbf{H}^{-1}\mathbf{X}+\mathbf{\Sigma}_\beta^{-1}\right)^{-1}$.  
\end{proof}

We assume a standard Gaussian prior $c\sim\mathsf{N}\left(\mu_c,\sigma^2\right)$ for the unconditional mean of the latent state. The corresponding optimal variational density is derived by leveraging a Gaussian Markov Random Field (GMRF) representation of the latent state $\mathbf{h}\sim\mathsf{N}_{n+1}(c\boldsymbol{\iota}_{n+1},\eta^2\mathbf{Q}^{-1})$ that preserves the time dependence structure of the autoregressive process. Notice the matrix $\mathbf{Q}$ is a tridiagonal precision matrix with diagonal elements $q_{1,1}=q_{n+1,n+1}=1$ and $q_{i,i}=1+\rho^2$ for $i=2,\ldots,n$, and off-diagonal elements $q_{i,j}=-\rho$ if $|i-j|=1$ and $0$ elsewhere (see \citealp{rue_held.2005}).

\begin{proposition}\label{eq:prop_c}
    The optimal variational density for $c$ is $q^\ast(c)\equiv\mathsf{N}(\mu_{q(c)},\sigma^2_{q(c)})$ where:
	\begin{equation}\begin{aligned}\label{eq:up_c_app}
		\sigma^2_{q(c)}&= (\mu_{q(1/\eta^2)}\boldsymbol{\iota}_{n+1}^\intercal\mathbf{\mu}_{q(\mathbf{Q})}\boldsymbol{\iota}_{n+1}+1/\sigma^2_c)^{-1} \\
		\mu_{q(c)}&=\sigma^2_{q(c)}(\mu_{q(1/\eta^2)}\boldsymbol{\iota}_{n+1}^\intercal\mathbf{\mu}_{q(\mathbf{Q})}\mathbf{\mu}_{q(\mathbf{h})}+\mu_c/\sigma^2_c).
	\end{aligned}\end{equation}
	where is the element-wise expectation of the tridiagonal matrix $\mathbf{Q}$,
	\begin{equation*}
		\mathbf{\mu}_{q(\mathbf{Q})} = 
		\begin{bmatrix}
			1      & -\mu_{q(\rho)}    &\dots & 0         & 0 \\
			-\mu_{q(\rho)}  & 1+\mu_{q(\rho^2)} &\dots & 0         & 0\\
			\vdots & \vdots   &\ddots &\vdots     &\vdots\\
			0      & 0        &\dots  & 1+\mu_{q(\rho^2)}  & -\mu_{q(\rho)} \\
			0      & 0        &\dots  & -\mu_{q(\rho)}     & 1
		\end{bmatrix}.
	\end{equation*}
\end{proposition}
\begin{proof}
	The logarithm of the full conditional $p(c|\text{rest})$ is proportional to:
	\begin{align*}
	    \log p(c|\text{rest}) &\propto -\frac{1}{2\eta^2}(\mathbf{h}-c\boldsymbol{\iota}_{n+1})^\intercal\mathbf{Q}(\mathbf{h}-c\boldsymbol{\iota}_{n+1}) -\frac{1}{2\sigma^2_c}(c-\mu_c)^2\\
	    &\propto -\frac{1}{2\eta^2}(c^2\boldsymbol{\iota}_{n+1}^\intercal\mathbf{Q}\boldsymbol{\iota}_{n+1}-2c\boldsymbol{\iota}_{n+1}^\intercal\mathbf{Q}\mathbf{h}) -\frac{1}{2\sigma^2_c}(c^2-2c\mu_c).
	\end{align*}
	Compute the optimal variational density as $\log q^\ast(c)=\mathbb{E}_{q(\boldsymbol{\vartheta}\setminus c)}\left[\log p(c|\text{rest})\right]$:
	\begin{align*}
	    \log q^\ast(c) &\propto -\frac{1}{2}\mathbb{E}_{\eta^2}[1/\eta^2](c^2\boldsymbol{\iota}_{n+1}^\intercal\mathbb{E}_\rho[\mathbf{Q}]\boldsymbol{\iota}_{n+1}-2c\boldsymbol{\iota}_{n+1}^\intercal\mathbb{E}_\rho[\mathbf{Q}]\mathbb{E}_h[\mathbf{h}]) -\frac{1}{2\sigma^2_c}(c^2-2c\mu_c)\\
	    &=-\frac{1}{2}\mu_{q(1/\eta^2)}(c^2\boldsymbol{\iota}_{n+1}^\intercal\mathbf{\mu}_{q(\mathbf{Q})} \boldsymbol{\iota}_{n+1}-2c\boldsymbol{\iota}_{n+1}^\intercal\mathbf{\mu}_{q(\mathbf{Q})} \mathbf{\mu}_{q(h)}) -\frac{1}{2\sigma^2_c}(c^2-2c\mu_c)\\
	    &=-\frac{1}{2}\left(c^2(\mu_{q(1/\eta^2)}\boldsymbol{\iota}_{n+1}^\intercal\mathbf{\mu}_{q(\mathbf{Q})} \boldsymbol{\iota}_{n+1}+1/\sigma^2_c)-2c(\boldsymbol{\iota}_{n+1}^\intercal\mathbf{\mu}_{q(\mathbf{Q})}\mathbf{\mu}_{q(h)}+\mu_c/\sigma^2_c) \right),
	\end{align*}
	where $\mathbf{\mu}_{q(\mathbf{Q})}$ denotes the element-wise expectation of the matrix $\mathbf{Q}$. Take the exponential, and you obtain the kernel of a univariate Gaussian distribution with parameters with mean $\mu_{q(c)}=\sigma^2_{q(c)}(\mu_{q(1/\eta^2)}\boldsymbol{\iota}_{n+1}^\intercal\mathbf{\mu}_{q(\mathbf{Q})}\mathbf{\mu}_{q(\mathbf{h})}+\mu_c/\sigma^2_c)$ and variance $\sigma^2_{q(c)} = (\mu_{q(1/\eta^2)}\boldsymbol{\iota}_{n+1}^\intercal\mathbf{\mu}_{q(\mathbf{Q})}\boldsymbol{\iota}_{n+1}+1/\sigma^2_c)^{-1}$. 
 \end{proof}

\begin{proposition}\label{eq:prop_rho}
	The optimal variational density for $\rho$ has the following form:
	\begin{equation}
		\log q(\rho)\propto\frac{1}{2}\log(1-\rho^2)-\frac{1}{2}\mu_{q(1/\eta^2)}\left(\rho^2\sum_{t=1}^{n-1}a_t-2\rho\sum_{t=0}^{n-1}b_t\right), \quad \rho\in(-1,1)
	\end{equation}
	with
	\begin{align}\label{eq:ab_eq1}
		a_t &= \mathbb{E}_q\left[(h_t-c)^2\right] = (\mu_{q(h_t)}-\mu_{q(c)})^2+\sigma^2_{q(h_t)}+\sigma^2_{q(c)} \\
		b_t &= \mathbb{E}_q\left[(h_t-c)(h_{t+1}-c)\right] = (\mu_{q(h_t)}-\mu_{q(c)})(\mu_{q(h_{t+1})}-\mu_{q(c)})+\sigma_{q(h_t,h_{t+1})}+\sigma^2_{q(c)},\label{eq:ab_eq2}
	\end{align}
	where $\sigma_{q(h_t,h_{t+1})}$ denotes the covariance between $h_t$ and $h_{t+1}$ under the approximating density $q$. Notice that $\log q(\rho)$ can be written as:
	\begin{equation}\label{eq:up_rho}
		\log q(\rho)\propto\frac{1}{2}\log(1-\rho^2)-\frac{1}{2}\mu_{q(1/\eta^2)}\left(\sum_{t=1}^{n-1}a_t\right)\left(\rho-\frac{\sum_{t=0}^{n-1}b_t}{\sum_{t=1}^{n-1}a_t}\right)^2, \quad \rho\in(-1,1)
	\end{equation}
	thus the normalizing constant and the first two moments can be found by Monte Carlo methods by sampling from a univariate Gaussian distribution with mean $\frac{\sum_{t=0}^{n-1}b_t}{\sum_{t=1}^{n-1}a_t}$ and precision $\mu_{q(1/\eta^2)}\left(\sum_{t=1}^{n-1}a_t\right)$. 
\end{proposition}
\begin{proof}
	The logarithm of the full conditional $p(\rho|\text{rest})$ is proportional to:
	\begin{align*}
	    \log p(\rho|\text{rest}) &\propto \frac{1}{2}\log|\mathbf{Q}|-\frac{1}{2\eta^2}(\mathbf{h}-c\boldsymbol{\iota}_{n+1})^\intercal\mathbf{Q}(\mathbf{h}-c\boldsymbol{\iota}_{n+1}) \\
	    &\propto \frac{1}{2}\log(1-\rho^2)-\frac{1}{2\eta^2}\left(\rho^2\sum_{t=1}^{n-1}(h_t-c)^2-2\rho\sum_{t=0}^{n-1}(h_t-c)(h_{t+1}-c)\right),
	\end{align*}
	for $\rho\in(-1,1)$.
	Compute the optimal variational density as $\log q^\ast(\rho)=\mathbb{E}_{q(\boldsymbol{\vartheta}\setminus\rho)}\left[\log p(\rho|\text{rest})\right]$:
	\begin{align*}
	    \log q^\ast(\rho) &\propto \frac{1}{2}\log(1-\rho^2)-\frac{1}{2}\mathbb{E}_q\left[1/\eta^2\right]\left(\rho^2\sum_{t=1}^{n-1}\mathbb{E}_q\left[(h_t-c)^2\right]-2\rho\sum_{t=0}^{n-1}\mathbb{E}_q\left[(h_t-c)(h_{t+1}-c)\right]\right) \\
	    &=\frac{1}{2}\log(1-\rho^2)-\frac{1}{2}\mu_{q(1/\eta^2)}\left(\rho^2\sum_{t=1}^{n-1}a_t-2\rho\sum_{t=0}^{n-1}b_t\right), \quad \rho\in(-1,1),
	\end{align*}
	where $a_t$ and $b_t$ are as in Eq.\eqref{eq:ab_eq1}-Eq.\eqref{eq:ab_eq2}. Take the exponential and obtain:
	\begin{align*}
	    q^\ast(\rho) &\propto \sqrt{1-\rho^2}\,\mathbb{I}_{\rho\in(-1,1)}\,\phi\left(\rho;\frac{\sum_{t=0}^{n-1}b_t}{\sum_{t=1}^{n-1}a_t},\frac{1}{\mu_{q(1/\eta^2)}\sum_{t=1}^{n-1}a_t}\right),
	\end{align*}
	where $\phi(x;m,s^2)$ denotes the probability density function of an univariate Gaussian distribution with mean $m$ and variance $s^2$.
\end{proof}

\begin{proposition}\label{eq:prop_eta2}
	The optimal variational density for $\eta^2$ is an Inverse-Gamma distribution $q^\ast(\eta^2)\equiv\mathsf{IG}(A_{q(\eta^2)},B_{q(\eta^2)})$, where:
	\begin{align}\label{eq:up_eta21}
		A_{q(\eta^2)} &=  A + \frac{n+1}{2}\\
		B_{q(\eta^2)} &=  B + \frac{1}{2}(\mathbf{\mu}_{q(\mathbf{h})}-\mu_{q(c)}\boldsymbol{\iota}_{n+1})^\intercal\mathbf{\mu}_{q(\mathbf{Q})}(\mathbf{\mu}_{q(\mathbf{h})}-\mu_{q(c)}\boldsymbol{\iota}_{n+1}) \\\label{eq:up_eta22}
		&\qquad\quad +\frac{1}{2}\left(
		\mathsf{tr}\left\{\mathbf{\Sigma}_{q(\mathbf{h})}\mathbf{\mu}_{q(\mathbf{Q})}\right\} + \sigma^2_{q(c)}\boldsymbol{\iota}_{n+1}^\intercal\mathbf{\mu}_{q(\mathbf{Q})}\boldsymbol{\iota}_{n+1}\right),\notag 
	\end{align}
	and recall that $\mu_{q(1/\eta^2)}=A_{q(\eta^2)}/B_{q(\eta^2)}$.
\end{proposition}
\begin{proof}
	The logarithm of the full conditional $p(\eta^2|\text{rest})$ is proportional to:
	\begin{align*}
	    \log p(\eta^2|\text{rest}) &\propto -\frac{n+1}{2}\log\eta^2-\frac{1}{2\eta^2}(\mathbf{h}-c\boldsymbol{\iota}_{n+1})^\intercal\mathbf{Q}(\mathbf{h}-c\boldsymbol{\iota}_{n+1}) -(A+1)\log\eta^2-B/\eta^2\\
	    &\propto -\left(A+\frac{n+1}{2}+1\right)\log\eta^2-\frac{1}{\eta^2}\left(B+\frac{1}{2}(\mathbf{h}-c\boldsymbol{\iota}_{n+1})^\intercal\mathbf{Q}(\mathbf{h}-c\boldsymbol{\iota}_{n+1})\right).
	\end{align*}
	Compute the optimal variational density as $\log q^\ast(\eta^2)=\mathbb{E}_{q(\boldsymbol{\vartheta}\setminus\rho)}\left[\log p(\eta^2|\text{rest})\right]$:
	\begin{align*}
	    \log q^\ast(\eta^2) &\propto -\left(A+\frac{n+1}{2}+1\right)\log\eta^2-\frac{1}{\eta^2}\left(B+\frac{1}{2}\mathbb{E}_{c,\rho,h}\left[(\mathbf{h}-c\boldsymbol{\iota}_{n+1})^\intercal\mathbf{Q}(\mathbf{h}-c\boldsymbol{\iota}_{n+1})\right]\right),
	\end{align*}
	where
	\begin{align*}
	    \mathbb{E}_{c,\rho,h}\left[(\mathbf{h}-c\boldsymbol{\iota}_{n+1})^\intercal\mathbf{Q}(\mathbf{h}-c\boldsymbol{\iota}_{n+1})\right] &= \mathbb{E}_{c,\rho,h}\left[\mathbf{h}^\intercal\mathbf{Q}\mathbf{h}-2c\mathbf{h}^\intercal\mathbf{Q}\boldsymbol{\iota}_{n+1}+c^2\boldsymbol{\iota}_{n+1}^\intercal\mathbf{Q}\boldsymbol{\iota}_{n+1}\right] \\
	    &= \mathbb{E}_{h}\left[\mathbf{h}^\intercal\mathbf{\mu}_{q(\mathbf{Q})}\mathbf{h}\right]+\mathbb{E}_c[c^2]\boldsymbol{\iota}_{n+1}^\intercal\mathbf{\mu}_{q(\mathbf{Q})}\boldsymbol{\iota}_{n+1} \\
	    &\qquad -2\mu_{q(c)}\mathbf{\mu}_{q(h)}^\intercal\mathbf{\mu}_{q(\mathbf{Q})}\boldsymbol{\iota}_{n+1} \\
	    &= \mathsf{tr}\left\{\mathbb{E}_{h}[\mathbf{h}\mathbf{h}^\intercal]\mathbf{\mu}_{q(\mathbf{Q})}\right\}+(\mu_{q(c)}^2+\sigma^2_{q(c)})\boldsymbol{\iota}_{n+1}^\intercal\mathbf{\mu}_{q(\mathbf{Q})}\boldsymbol{\iota}_{n+1} \\
	    &\qquad -2\mu_{q(c)}\mathbf{\mu}_{q(h)}^\intercal\mathbf{\mu}_{q(\mathbf{Q})}\boldsymbol{\iota}_{n+1} \\
	    &= \mathsf{tr}\left\{\left(\mathbf{\mu}_{q(h)}\mathbf{\mu}_{q(h)}^\intercal+\mathbf{\Sigma}_{q(h)}\right)\mathbf{\mu}_{q(\mathbf{Q})}\right\} \\
	    &\qquad +(\mu_{q(c)}^2+\sigma^2_{q(c)})\boldsymbol{\iota}_{n+1}^\intercal\mathbf{\mu}_{q(\mathbf{Q})}\boldsymbol{\iota}_{n+1} \\
	    &\qquad -2\mu_{q(c)}\mathbf{\mu}_{q(h)}^\intercal\mathbf{\mu}_{q(\mathbf{Q})}\boldsymbol{\iota}_{n+1} \\
	    &= \mathbf{\mu}_{q(h)}^\intercal\mathbf{\mu}_{q(\mathbf{Q})}\mathbf{\mu}_{q(h)}  +\mu_{q(c)}^2\boldsymbol{\iota}_{n+1}^\intercal\mathbf{\mu}_{q(\mathbf{Q})}\boldsymbol{\iota}_{n+1} \\
	    &\qquad -2\mu_{q(c)}\mathbf{\mu}_{q(h)}^\intercal\mathbf{\mu}_{q(\mathbf{Q})}\boldsymbol{\iota}_{n+1}\\
	    &\qquad
	    +\mathsf{tr}\left\{\mathbf{\Sigma}_{q(h)}\mathbf{\mu}_{q(\mathbf{Q})}\right\} +\sigma^2_{q(c)}\boldsymbol{\iota}_{n+1}^\intercal\mathbf{\mu}_{q(\mathbf{Q})}\boldsymbol{\iota}_{n+1} \\
	    &= (\mathbf{\mu}_{q(h)}-\mu_{q(c)}\boldsymbol{\iota}_{n+1})^\intercal\mathbf{\mu}_{q(\mathbf{Q})}(\mathbf{\mu}_{q(h)}-\mu_{q(c)}\boldsymbol{\iota}_{n+1})\\
	    &\qquad
	    +\mathsf{tr}\left\{\mathbf{\Sigma}_{q(h)}\mathbf{\mu}_{q(\mathbf{Q})}\right\} +\sigma^2_{q(c)}\boldsymbol{\iota}_{n+1}^\intercal\mathbf{\mu}_{q(\mathbf{Q})}\boldsymbol{\iota}_{n+1}.
	\end{align*}
	Take the exponential of $\log q^\ast(\eta^2)$ and you obtain the kernel of an inverse-gamma distribution with scale and shape parameters as in Eq.\eqref{eq:up_eta21}-Eq.\eqref{eq:up_eta22}.
\end{proof}

\subsection{Latent Volatility State}

In the following, we provide the full derivation of optimal variational density for the latent stochastic volatility state. We start with the unconstrained case in which the variational density $q\left(\mathbf{h}\right)$ is assumed to take the form $\mathbf{h}\sim\mathsf{N}_{n+1}(\mathbf{\mu}_{q(h)},\mathbf{\Omega}_{q(h)}^{-1})$ with mean vector $\mathbf{\mu}_{q(h)}=\mathbf{W}\mathbf{f}_{q(h)}$ and unconstrained variance-covariance matrix $\mathbf{\Sigma}_{q(h)}=\mathbf{\Omega}_{q(h)}^{-1}$. This is the specification used in the main text (see Proposition \ref{eq:prop5}). 

\begin{proposition}\label{eq:prop_h_het}
Let $\mathbf{\mu}_{q(\mathbf{s})}=(\mu_{q(s_1)},\ldots,\mu_{q(s_n)})^\intercal$ with $\mu_{q(s_t)} = (y_t-\mathbf{x}_t^\intercal\mathbf{\mu}_{q(\beta)})^2+\mathsf{tr}\left\{\mathbf{\Sigma}_{q(\beta)}\mathbf{x}_t\mathbf{x}_t^\intercal\right\}$, and $\mathbf{\mu}_{q(\beta)},\mathbf{\Sigma}_{q(\beta)}$ denote the variational mean and covariance from Proposition \ref{eq:prop_beta}. Assuming a representation $\mathbf{h}\sim\mathsf{N}_{n+1}(\mathbf{\mu}_{q(h)},\mathbf{\Omega}_{q(h)}^{-1})$, with mean vector $\mathbf{\mu}_{q(h)}=\mathbf{W}\mathbf{f}_{q(h)}$ and covariance matrix $\mathbf{\Sigma}_{q(h)}=\mathbf{\Omega}_{q(h)}^{-1}$, an iterative algorithm can be set as:
\begin{align}
	\mathbf{\Sigma}_{q(h)}^{new} &= \left[\nabla_{\boldsymbol{\mu}_{q(h)}\boldsymbol{\mu}_{q(h)}}^2 S(\mathbf{\mu}_{q(h)}^{old},\mathbf{\Sigma}_{q(h)}^{old})\right]^{-1}, \\
	\mathbf{f}_{q(h)}^{new} &= \mathbf{f}_{q(h)}^{old} + \mathbf{W}^+\,\mathbf{\Sigma}_{q(h)}^{new}\nabla_{\boldsymbol{\mu}_{q(h)}} S(\mathbf{\mu}_{q(h)}^{old},\mathbf{\Sigma}_{q(h)}^{old}),\\
	\mathbf{\mu}_{q(h)}^{new} &= \mathbf{W}\mathbf{f}_{q(h)}^{new},
\end{align}
with $\mathbf{W}^+=(\mathbf{W}^\intercal\mathbf{W})^{-1}\mathbf{W}^\intercal$ the left Moore–Penrose pseudo-inverse of $\mathbf{W}$, and $S(\mathbf{\mu}_{q(h)},\mathbf{\Sigma}_{q(h)})$ equal to $\mathbb{E}_q(\log p(\mathbf{h},\mathbf{y}))$ such that,
\begin{align}
	\nabla_{\boldsymbol{\mu}_{q(h)}} S(\mathbf{\mu}_{q(h)},\mathbf{\Sigma}_{q(h)}) &= -\frac{1}{2}[0,\boldsymbol{\iota}_n^\intercal]^\intercal+\frac{1}{2}[0,\mathbf{\mu}_{q(\mathbf{s})}^\intercal]^\intercal\odot\mathrm{e}^{-\boldsymbol{\mu}_{q(h)}+\frac{1}{2}\mathsf{diag}\left(\boldsymbol{\Sigma}_{q(\mathbf{h})}\right)} \\
	&\qquad-\mu_{q(1/\eta^2)}\mathbf{\mu}_{q(\mathbf{Q})}(\mathbf{\mu}_{q(h)}-\mu_{q(c)}\boldsymbol{\iota}_{n+1}),\\
	\nabla_{\boldsymbol{\mu}_{q(h)}\boldsymbol{\mu}_{q(h)}}^2 S(\mathbf{\mu}_{q(h)},\mathbf{\Sigma}_{q(h)}) &=-\frac{1}{2}\mathsf{Diag}\Bigg[[0,\mathbf{\mu}_{q(\mathbf{s})}^\intercal]^\intercal\odot\mathrm{e}^{-\boldsymbol{\mu}_{q(h)}+\frac{1}{2}\mathsf{diag}\left(\boldsymbol{\Sigma}_{q(\mathbf{h})}\right)}\Bigg]
	-\mu_{q(1/\eta^2)}\mathbf{\mu}_{q(\mathbf{Q})},
\end{align}
where $\boldsymbol{\iota}_n$ is an n-dimensional vector of ones, $\mu_{q\left(1/\eta^2\right)}$ is the variational mean of $1/\eta^2$, $\mathbf{\mu}_{q(\mathbf{Q})}$ is the element-wise variational mean of $\mathbf{Q}$, and $\odot$ denotes the Hadamard product.
\end{proposition}
\begin{proof}
To find the optimal parameters of the approximating density $(\mathbf{f}_{q(h)},\mathbf{\Sigma}_{q(h)})$, we have to solve the following optimization problem:
\begin{equation}
	\widehat{\boldsymbol{\xi}}=\arg\max_\xi \psi(\mathbf{f}_{q(h)},\mathbf{\Sigma}_{q(h)}),
\end{equation}
where $\psi(\mathbf{f}_{q(h)},\mathbf{\Sigma}_{q(h)})=\mathbb{E}_q(\log p(\mathbf{h},\mathbf{y}))-\mathbb{E}_q(\log q(\mathbf{h}))$ is called variational lower bound (ELBO).
To this aim, we can exploit a result provided by \cite{rohde_wand2016} valid when the approximating density is a multivariate Gaussian distribution. The latter states an iterative update scheme for the variational parameters:
\begin{align}
	\mathbf{\Sigma}^{new} &= \left[\nabla_{\mathbf{\mu},\mathbf{\mu}}^2 S(\mathbf{\mu}^{old},\mathbf{\Sigma}^{old})\right]^{-1} \\
	\mathbf{\mu}^{new} &= \mathbf{\mu}^{old} + \mathbf{\Sigma}^{new}\nabla_{\mathbf{\mu}} S(\mathbf{\mu}^{old},\mathbf{\Sigma}^{old}),
\end{align}
where $\nabla_{\mathbf{\mu}} S(\mathbf{\mu}^{old},\mathbf{\Sigma}^{old})$ and $\nabla_{\mathbf{\mu},\mathbf{\mu}}^2 S(\mathbf{\mu}^{old},\mathbf{\Sigma}^{old})$ denote the first and second derivative of $S(\mathbf{\mu},\mathbf{\Sigma})$ with respect to $\mathbf{\mu}$ and evaluated at $(\mathbf{\mu}^{old},\mathbf{\Sigma}^{old})$. The function $S$ is the so-called \textit{non-entropy function} which is given by $\mathbb{E}_q(\log p(\mathbf{h},\mathbf{y}))$. In our scenario, we have that
\begin{align}\label{eq:seq_appendix}
	S(\mathbf{\mu}_{q(h)},\mathbf{\Sigma}_{q(h)})&=-\frac{1}{2}[0,\boldsymbol{\iota}_n^\intercal]\mathbf{\mu}_{q(h)}-\frac{1}{2}[0,\mathbf{\mu}_{q(\mathbf{s})}^\intercal]\mathrm{e}^{-\mathbf{\mu}_{q(h)}+\frac{1}{2}\boldsymbol{\sigma}^2_{q(\mathbf{h})}}-\frac{1}{2}\mu_{q(1/\eta^2)}\mathsf{tr}(\mathbf{\Sigma}_{q(\mathbf{h})}\mathbf{\mu}_{q(\mathbf{Q})})\nonumber\\
	&\qquad
	-\frac{1}{2}\mu_{q(1/\eta^2)}(\mathbf{\mu}_{q(h)}-\mu_{q(c)}\boldsymbol{\iota}_{n+1})^\intercal\mathbf{\mu}_{q(\mathbf{Q})}(\mathbf{\mu}_{q(h)}-\mu_{q(c)}\boldsymbol{\iota}_{n+1}),
\end{align}
where $\boldsymbol{\sigma}^2_{q(h)}=\mathsf{diag}(\mathbf{\Sigma}_{q(h)})$ is the vector of variances and the $\mathsf{diag}$ operator extracts the diagonal vector from the input matrix. Moreover, we obtain:
\begin{align}
	\nabla_{\boldsymbol{\mu}_{q(h)}} S(\mathbf{\mu}_{q(h)},\mathbf{\Sigma}_{q(h)}) &= -\frac{1}{2}[0,\boldsymbol{\iota}_n^\intercal]^\intercal+\frac{1}{2}[0,\mathbf{\mu}_{q(\mathbf{s})}^\intercal]^\intercal\odot\mathrm{e}^{-\boldsymbol{\mu}_{q(h)}+\frac{1}{2}\boldsymbol{\sigma}^2_{q(h)}} \nonumber\\
	&\qquad-\mu_{q(1/\eta^2)}\mathbf{\mu}_{q(\mathbf{Q})}(\mathbf{\mu}_{q(h)}-\mu_{q(c)}\boldsymbol{\iota}_{n+1}),\\
	\nabla_{\boldsymbol{\mu}_{q(h)}\boldsymbol{\mu}_{q(h)}}^2 S(\mathbf{\mu}_{q(h)},\mathbf{\Sigma}_{q(h)}) &=-\frac{1}{2}\mathsf{Diag}\Bigg[[0,\mathbf{\mu}_{q(\mathbf{s})}^\intercal]^\intercal\odot\mathrm{e}^{-\boldsymbol{\mu}_{q(h)}+\frac{1}{2}\boldsymbol{\sigma}^2_{q(h)}}\Bigg]
	-\mu_{q(1/\eta^2)}\mathbf{\mu}_{q(\mathbf{Q})},
\end{align}
where $\boldsymbol{\iota}_n$ is an n-dimensional vector of ones, $\mu_{q\left(1/\eta^2\right)}$ is the variational mean of $1/\eta^2$, $\mathbf{\mu}_{q(\mathbf{Q})}$ is the element-wise variational mean of $\mathbf{Q}$, and $\odot$ denotes the Hadamard product.
Then, the updating scheme becomes:
\begin{align}
	\mathbf{\Sigma}_{q(h)}^{new} &= \left[\nabla_{\boldsymbol{\mu}_{q(h)}\boldsymbol{\mu}_{q(h)}}^2 S(\mathbf{\mu}_{q(h)}^{old},\mathbf{\Sigma}_{q(h)}^{old})\right]^{-1}, \\
	\mathbf{f}_{q(h)}^{new} &= \mathbf{f}_{q(h)}^{old} + \mathbf{W}^+\,\mathbf{\Sigma}_{q(h)}^{new}\nabla_{\boldsymbol{\mu}_{q(h)}} S(\mathbf{\mu}_{q(h)}^{old},\mathbf{\Sigma}_{q(h)}^{old}),\\
	\mathbf{\mu}_{q(h)}^{new} &= \mathbf{W}\mathbf{f}_{q(h)}^{new},
\end{align}
with $\mathbf{W}^+=(\mathbf{W}^\intercal\mathbf{W})^{-1}\mathbf{W}^\intercal$ the left Moore–Penrose pseudo-inverse of $\mathbf{W}$.
\end{proof}

Notice that under the multivariate gaussian approximation of $q(\mathbf{h})$ with mean vector $\mathbf{\mu}_{q(h)}$ and covariance matrix $\mathbf{\Sigma}_{q(h)}$, the optimal density of the vector of variances $\boldsymbol{\sigma}^2=\exp\{\mathbf{h}\}$, namely $q(\boldsymbol{\sigma}^2)$, is a multivariate log-normal distribution such that:
\begin{align}
    \mathbb{E}_q[\sigma_t^2] &= \exp\{\mu_{q(h_t)}+1/2\sigma^2_{q(h_t)}\},\\
    \mathsf{Var}_q[\sigma_t^2] &= \exp\{2\mu_{q(h_t)}+\sigma^2_{q(h_t)}\}(\exp\{\sigma^2_{q(h_t)}\}-1), \\
    \mathsf{Cov}_q[\sigma_t^2,\sigma_{t+1}^2] &= \exp\{\mu_{q(h_t)}+\mu_{q(h_{t+1})}+1/2(\sigma^2_{q(h_t)}+\sigma^2_{q(h_{t+1})})\}(\exp\{\mathsf{Cov}_q[h_t,h_{t+1}]\}-1).
\end{align}

\subsection{Homoschedastic Approximation of $q^\ast\left(\mathbf{h}\right)$}
\label{app:h_homo}
Next, we discuss the optimal variational density for the latent state under the assumption that $\mathbf{h}\sim\mathsf{N}_{n+1}(\mathbf{W}\mathbf{f},\tau^2\mathbf{\Gamma}^{-1})$ where $\mathbf{\mu}_{q(h)}=\mathbf{W}\mathbf{f}$ is the mean vector and $\mathbf{\Sigma}_{q(h)}=\tau^2\mathbf{\Gamma}^{-1}$ is the variance-covariance matrix. $\Gamma$ is a tridiagonal precision matrix with diagonal elements $\Gamma_{1,1}=\Gamma_{n+1,n+1}=1$ and $\Gamma_{i,i}=1+\gamma^2$ for $i=2,\ldots,n$, and off-diagonal elements $\Gamma_{i,j}=-\gamma$ if $|i-j|=1$ and $0$ elsewhere (see \citealp{rue_held.2005}). Under this setting, the density function of the approximate distribution is given by
\begin{equation}
\log q(\mathbf{h})\propto-\frac{n+1}{2}\log(\tau^2)-\frac{n}{2}\log(1-\gamma^2)-\frac{1}{2\tau^2}(\mathbf{h}-\mathbf{W}\mathbf{f})^\intercal\mathbf{\Gamma}(\mathbf{h}-\mathbf{W}\mathbf{f}).
\end{equation}
Recall that $\mathbf{h}^\intercal=(h_0,h_1,\ldots,h_n)$ admits a Gaussian Markov Random Field (GMRF) representation $\mathbf{h}\sim\mathsf{N}_{n+1}(c\boldsymbol{\iota}_{n+1},\eta^2\mathbf{Q}^{-1})$ that preserves the time dependence structure of the autoregressive process. The matrix $\mathbf{Q}$ is a tridiagonal precision matrix with diagonal elements $q_{1,1}=q_{n+1,n+1}=1$ and $q_{i,i}=1+\rho^2$ for $i=2,\ldots,n$, and off-diagonal elements $q_{i,j}=-\rho$ if $|i-j|=1$ and $0$ elsewhere (see \citealp{rue_held.2005}). As a result, the joint distribution of the latent states and the observations is given by:
\begin{align}
	\log p(\mathbf{h},\mathbf{y}|\mathbf{X})&\propto\log p(\mathbf{y}\vert\mathbf{h},\mathbf{X})+\log p(\mathbf{h})\nonumber\\
	&=-\frac{1}{2}\boldsymbol{\iota}_n^\intercal\mathbf{h}-\frac{1}{2}\mathbf{s}^\intercal\mathrm{e}^{-\mathbf{h}}-\frac{1}{2\eta^2}(\mathbf{h}-c\boldsymbol{\iota}_{n+1})^\intercal\mathbf{Q}(\mathbf{h}-c\boldsymbol{\iota}_{n+1}),
\end{align}
where $\mathbf{s}=(s_1,\ldots,s_n)^\intercal$ with $s_t = (y_t-\mathbf{x}_t^\intercal\mathbf{\beta})^2$, $\mathbf{h}=(h_1,\dots,h_n)^\intercal$ and $\mathrm{e}^{\mathbf{h}}=(\mathrm{e}^{h_1},\dots,\mathrm{e}^{h_n})^\intercal$.
Define the variational lower bound (ELBO) as: 
\begin{align}
\psi(\mathbf{f},\tau^2,\gamma)&=\mathbb{E}_q(\log p(\mathbf{h},\mathbf{y}))-\mathbb{E}_q(\log q(\mathbf{h}))\nonumber\\
&\propto-\frac{1}{2}\boldsymbol{\iota}_n^\intercal\mathbf{W}_1\mathbf{f}-\frac{1}{2}\mathbf{\mu}_{q(\mathbf{s})}^\intercal\mathrm{e}^{-\mathbf{W}_1\mathbf{f}+\frac{1}{2}\tau^2\boldsymbol{\iota}_n}\nonumber\\
&\qquad
-\frac{1}{2}\mu_{q(1/\eta^2)}(\mathbf{W}\mathbf{f}-\mu_{q(c)}\boldsymbol{\iota}_{n+1})^\intercal\mathbf{\mu}_{q(\mathbf{Q})}(\mathbf{W}\mathbf{f}-\mu_{q(c)}\boldsymbol{\iota}_{n+1})\nonumber\\
&\qquad
-\frac{1}{2}\mu_{q(1/\eta^2)}\tau^2\mathsf{tr}(\mathbf{\Gamma}^{-1}\mathbf{\mu}_{q(\mathbf{Q})})\nonumber\\
&\qquad+\frac{n+1}{2}\log(\tau^2)+\frac{n}{2}\log(1-\gamma^2),
\end{align}
where $\mathbf{\mu}_{q(\mathbf{s})}=(\mu_{q(s_1)},\ldots,\mu_{q(s_n)})^\intercal$ with $\mu_{q(s_t)} = (y_t-\mathbf{x}_t^\intercal\mathbf{\mu}_{q(\beta)})^2+\mathsf{tr}\left\{\mathbf{\Sigma}_{q(\beta)}\mathbf{x}_t\mathbf{x}_t^\intercal\right\}$, and $\mathbf{W}_1\in\mathbb{R}^{n\times k}$ denotes the matrix obtained by deleting the first row of $\mathbf{W}$. Moreover
\begin{align*}
\mathsf{tr}(\mathbf{\Gamma}^{-1}\mathbf{\mu}_{q(\mathbf{Q})})&=2+(1+\mu_{q(\rho^2)})(n-1)-2n\gamma\mu_{q(\rho)}.
\end{align*}
Let $\boldsymbol{\xi}=(\mathbf{f},\tau^2,\gamma)$ be the collection of parameters. To find the optimal values $\widehat{\boldsymbol{\xi}}$ we need to solve the optimization $\widehat{\boldsymbol{\xi}}=\arg\max_\xi \psi(\mathbf{f},\tau^2,\gamma)$. The gradient and Hessian of the objective function $\psi(\mathbf{f},\tau^2,\gamma)$ can be found analytically. The gradient vector takes the form,
\begin{align*}
	\nabla_\xi \psi(\mathbf{f},\tau^2,\gamma)=\begin{bmatrix}
		\nabla_\mathbf{f} \psi(\mathbf{f},\tau^2,\gamma),\quad  
		\nabla_{\tau^2}\psi(\mathbf{f},\tau^2,\gamma),\quad  
		\nabla_{\gamma}\psi(\mathbf{f},\tau^2,\gamma)
	\end{bmatrix}^\prime,
\end{align*}
where
\begin{align}
\nabla_\mathbf{f} \psi(\mathbf{f},\tau^2,\gamma)&=-\frac{1}{2}\mathbf{W}^\intercal[0,\boldsymbol{\iota}_n^\intercal]^\intercal+\frac{1}{2}\mathbf{W}^\intercal\left([0,\mathbf{\mu}_{q(\mathbf{s})}^\intercal]^\intercal\odot\mathrm{e}^{-\mathbf{W}\mathbf{f}+\frac{1}{2}\tau^2\boldsymbol{\iota}_{n+1}}\right)\nonumber\\
&\qquad-\mu_{q(1/\eta^2)}\mathbf{W}^\intercal\mathbf{\mu}_{q(\mathbf{Q})}(\mathbf{W}\mathbf{f}-\mu_{q(c)}\boldsymbol{\iota}_{n+1}), \\
\nabla_{\tau^2}
\psi(\mathbf{f},\tau^2,\gamma)&=-\frac{1}{4}(\mathbf{\mu}_{q(\mathbf{s})}\odot\boldsymbol{\iota}_n)^\intercal\mathrm{e}^{-\mathbf{W}_1\mathbf{f}+\frac{1}{2}\tau^2\boldsymbol{\iota}_n}\nonumber\\
&\qquad-\frac{1}{2}\mu_{q(1/\eta^2)}(2+(1+\mu_{q(\rho^2)})(n-1)-2n\gamma\mu_{q(\rho)}) +\frac{n+1}{2\tau^2},\\
\nabla_{\gamma}
\psi(\mathbf{f},\tau^2,\gamma)&=n\tau^2\mu_{q(1/\eta^2)}\mu_{q(\rho)}-\frac{n\gamma}{1-\gamma^2},
\end{align}
and the Hessian matrix is equal to:
\begin{equation*}
	\mathcal{H}_\xi=\begin{bmatrix}
		\nabla_{\mathbf{f},\mathbf{f}}^2 \psi(\mathbf{f},\tau^2,\gamma)  & \nabla_{\mathbf{f},\tau^2}^2 \psi(\mathbf{f},\tau^2,\gamma) & \nabla_{\mathbf{f},\gamma}^2 \psi(\mathbf{f},\tau^2,\gamma)\\
		\nabla_{\mathbf{f},\tau^2}^2 \psi(\mathbf{f},\tau^2,\gamma)  & \nabla_{\tau^2,\tau^2}^2 \psi(\mathbf{f},\tau^2,\gamma) & \nabla_{\tau^2,\gamma}^2 \psi(\mathbf{f},\tau^2,\gamma)\\
		\nabla_{\mathbf{f},\gamma}^2 \psi(\mathbf{f},\tau^2,\gamma)  & \nabla_{\tau^2,\gamma}^2 \psi(\mathbf{f},\tau^2,\gamma) & \nabla_{\gamma,\gamma}^2 \psi(\mathbf{f},\tau^2,\gamma)\\
	\end{bmatrix},
\end{equation*}
with
\begin{align}
\nabla_{\mathbf{f},\mathbf{f}}^2 \psi(\mathbf{f},\tau^2,\gamma)&=-\frac{1}{2}\mathbf{W}^\intercal\left\{\mathsf{Diag}\Bigg[[0,\mathbf{\mu}_{q(\mathbf{s})}^\intercal]^\intercal\odot\mathrm{e}^{-\mathbf{W}\mathbf{f}+\frac{1}{2}\tau^2\boldsymbol{\iota}_{n+1}}\Bigg]
+\mu_{q(1/\eta^2)}\mathbf{\mu}_{q(\mathbf{Q})}\right\}\mathbf{W} \\
\nabla_{\tau^2,\tau^2}^2\psi(\mathbf{f},\tau^2,\gamma)&=-\frac{1}{8}(\mathbf{\mu}_{q(\mathbf{s})}\odot\boldsymbol{\iota}_n)^\intercal\mathrm{e}^{-\mathbf{W}_1\mathbf{f}+\frac{1}{2}\tau^2\boldsymbol{\iota}_n}-\frac{n+1}{2\tau^4}\\
\nabla_{\gamma,\gamma}^2
\psi(\mathbf{f},\tau^2,\gamma)&= -\frac{n(1+\gamma^2)}{(1-\gamma^2)^2}\\
\nabla_{\mathbf{f},\tau^2}^2\psi(\mathbf{f},\tau^2,\gamma)&=\frac{1}{4}\mathbf{W}^\intercal([0,\mathbf{\mu}_{q(\mathbf{s})}^\intercal]^\intercal\odot\mathrm{e}^{-\mathbf{W}\mathbf{f}+\frac{1}{2}\tau^2\boldsymbol{\iota}_{n+1}}) \\
\nabla_{\mathbf{f},\gamma}^2
\psi(\mathbf{f},\tau^2,\gamma)&= \mathbf{0}_k\\
\nabla_{\tau^2,\gamma}^2
\psi(\mathbf{f},\tau^2,\gamma)&= n\mu_{q(\rho)}\mu_{q(1/\eta^2)}
\end{align}
where $\mathbf{a}=\mathsf{diag}(\mathbf{A})$ returns the vector $\mathbf{a}\in\mathbb{R}^n$ of elements belonging to the main diagonal of the square matrix $\mathbf{A}\in\mathbb{R}^{n\times n}$, while $\mathbf{A}=\mathsf{Diag}(\mathbf{a})$ returns a diagonal square matrix $\mathbf{A}\in\mathbb{S}_+^{n}$ whose entries consist of the corresponding elements of the vector $\mathbf{a}\in\mathbb{R}^n$. 

\clearpage
\subsection{Pseudo-Code for the Algorithm Implementation}
\label{app:algorithm}
In this section, we present the pseudo-code for the implementation of the proposed algorithm. The algorithm is iterative and the convergence is achieved when the variation in the parameters' update $q^*(\boldsymbol{\vartheta},\mathbf{h})^{(\text{iter})}-q^*(\boldsymbol{\vartheta},\mathbf{h})^{(\text{iter}-1)}$ is smaller than a threshold $\Delta$.
\begin{table}[!ht]
    \centering
    \begin{algorithm}[H]
        \SetAlgoLined
        \kwInit{\(q^*(\boldsymbol{\vartheta})\), \(\mathbf{W}\), \(\Delta\)}
        \While{\(\big(\widehat{\Delta} > \Delta\big)\)}{
            Update \(q^*(\mathbf{\beta}) = \mathsf{N}_p(\mathbf{\mu}_{q(\beta)}, \mathbf{\Sigma}_{q(\beta)})\) as in Proposition~\ref{eq:prop1};\\
            \vspace{0.2cm}
            Update \(q^*(c) = \mathsf{N}(\mu_{q(c)}, \sigma^2_{q(c)})\) as in Proposition~\ref{eq:prop2};\\
            \vspace{0.2cm}
            Update \(q^*(\rho)\) as in Proposition~\ref{eq:prop3};\\
            \vspace{0.2cm}
            Update \(q^*(\eta^2) = \mathsf{IG}(A_{q(\eta^2)}, B_{q(\eta^2)})\) as in Proposition~\ref{eq:prop4};\\
            \vspace{0.2cm}
            Update \(q^*(\mathbf{h})\) as in Proposition~\ref{eq:prop5};\\
            \vspace{0.2cm}
            Compute \(\widehat{\Delta} = q^*(\boldsymbol{\vartheta})^{(\text{iter})} - q^*(\boldsymbol{\vartheta})^{(\text{iter}-1)}\);
        }
        \caption{Variational Bayes inference for smooth stochastic volatility.}
        \label{code:VBSV}
    \end{algorithm}
\end{table}

\clearpage
\section{Smoothing the Latent Volatility State}
\label{sec:smoothing}

\begin{figure}[!ht]
\centering
\includegraphics[width=.45\textwidth]{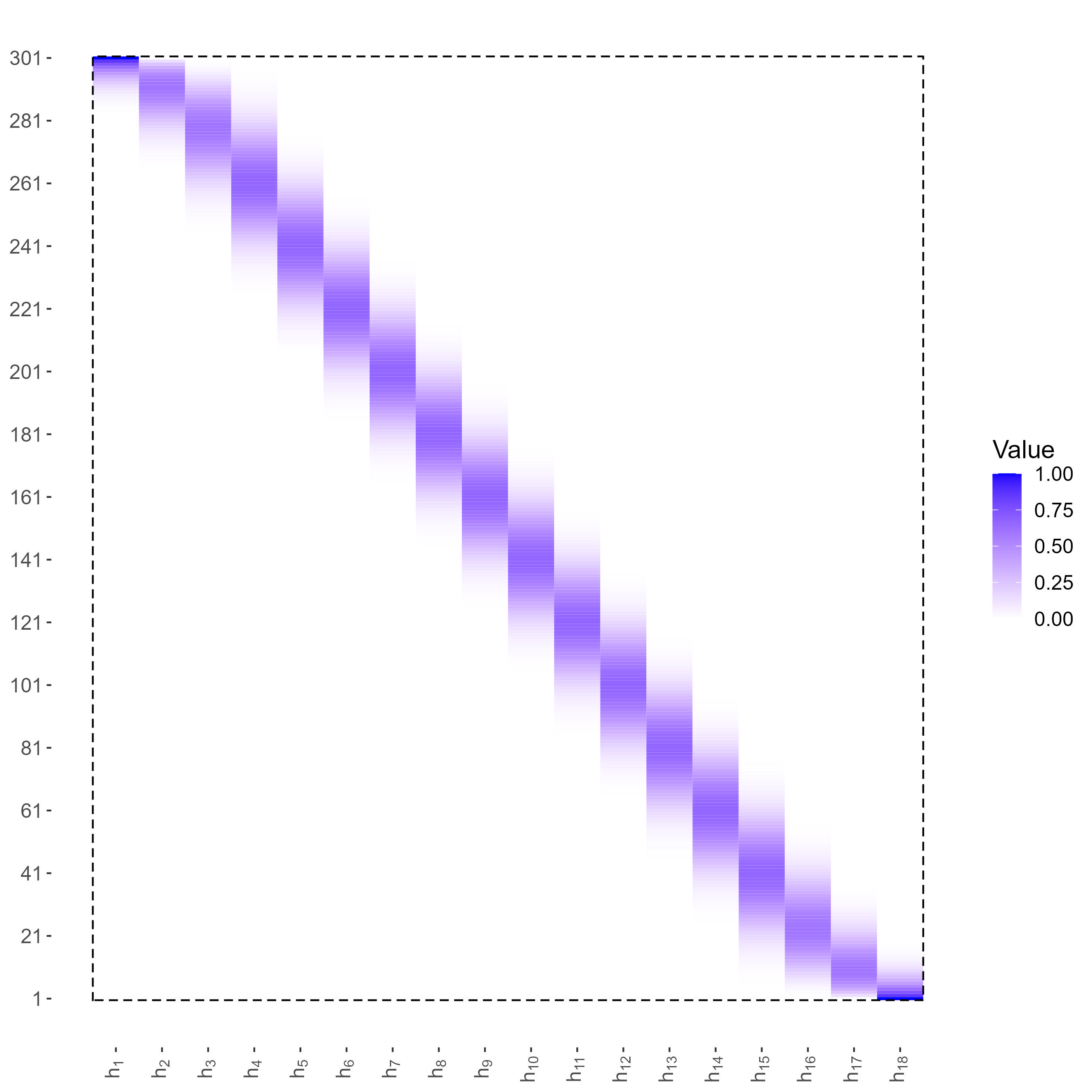}
\includegraphics[width=.42\textwidth]{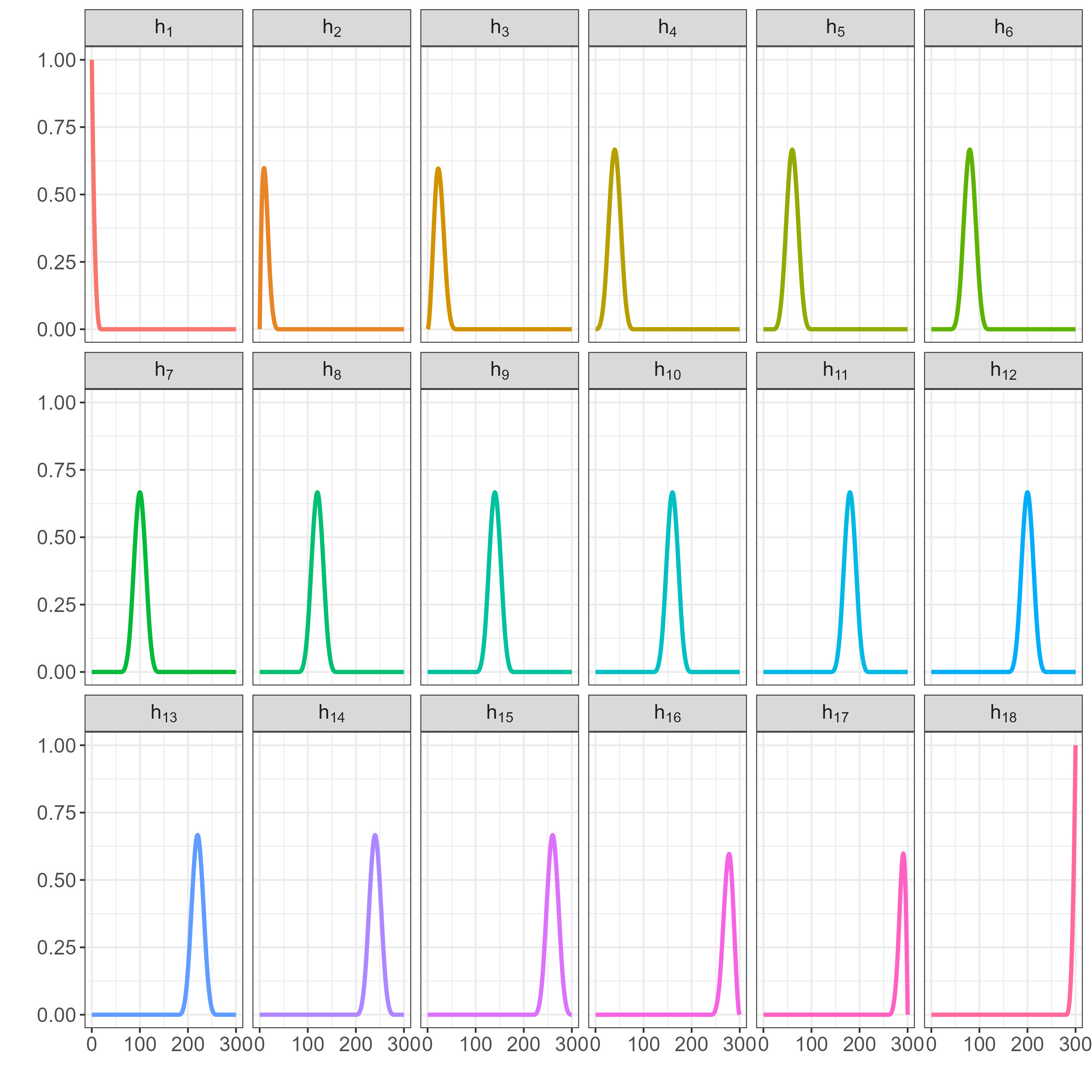}
\caption{\footnotesize \bf Smoothing Volatility Based on B-spline Basis\normalfont. The form of $\mathbf{W}$ for the case of a B-spline basis functions. The B-spline basis is a sequence of piecewise polynomial functions of a given degree, in this case, $dg=3$. The knots determine the locations of the pieces; here, we assume $kn=20$ equally spaced knots.}

	\label{fig:form_of_W_app}
\end{figure}

In this section, we further discuss the smoothing properties for different specifications of $\mathbf{W}$. Figure 1 in the main text depicts the form $\mathbf{W}$ taking, assuming Daubechies wavelet basis functions with level $l=5$ and resolution $R=16$. Alternative functional forms can be used for smoothing the distribution of the latent state. Figure \ref{fig:form_of_W_app} depicts the form of $\mathbf{W}$ when B-spline functions are used. The left panel shows the form of $\mathbf{W}$ in the case of a B-spline basis. The right panel corresponds to the columns of the matrix $\mathbf{W}$. The B-spline basis is a sequence of piecewise polynomials of a given degree, in this case, $dg=3$. The knots determine the locations of the pieces; here, we assume $kn=20$ equally spaced knots.

\begin{figure}[!ht]
\hspace{-2em}\subfigure[Identity matrix]{\includegraphics[width=.52\textwidth]{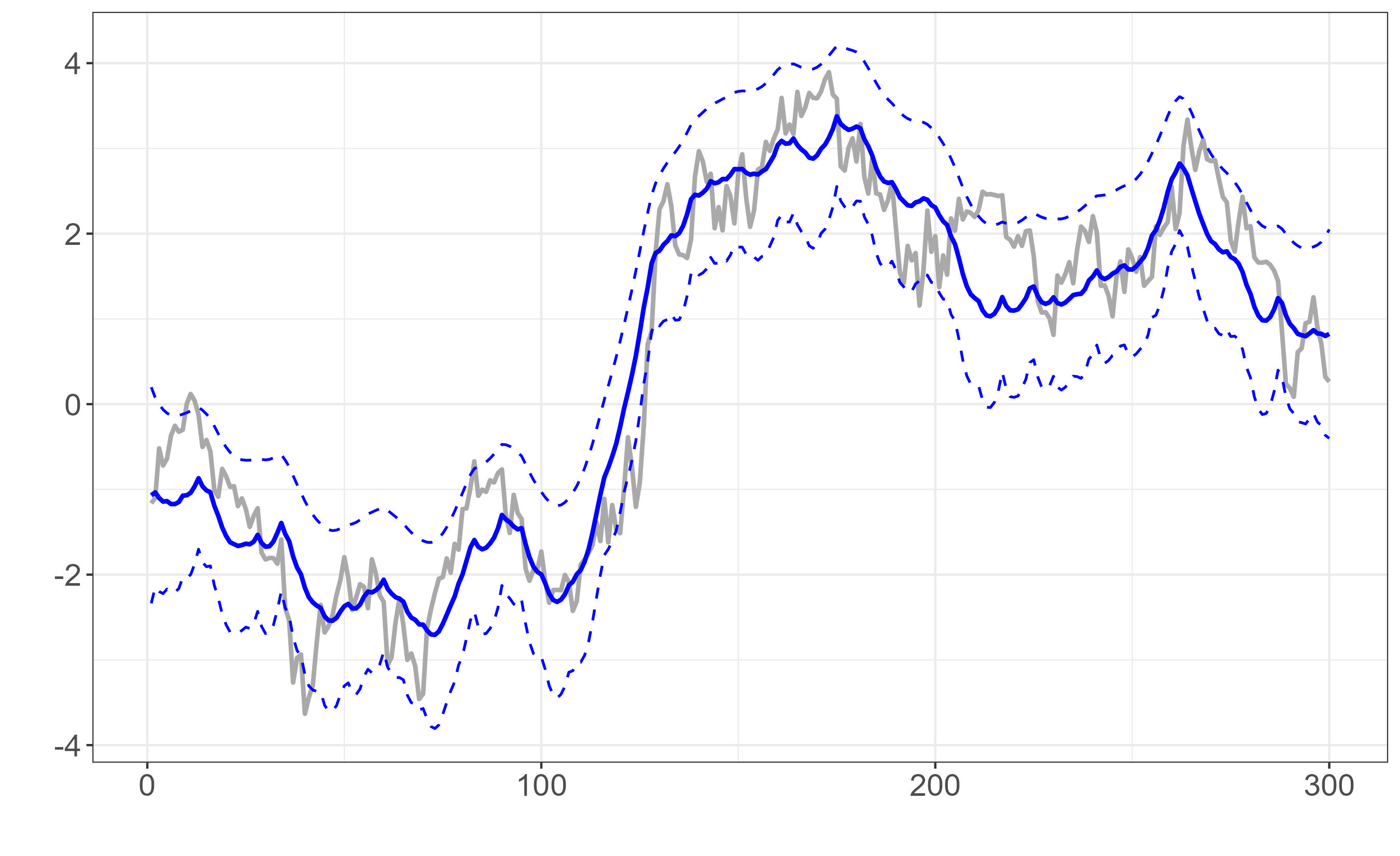}}\subfigure[Daubechies wavelet basis matrix with $l=4$]{\includegraphics[width=.52\textwidth]{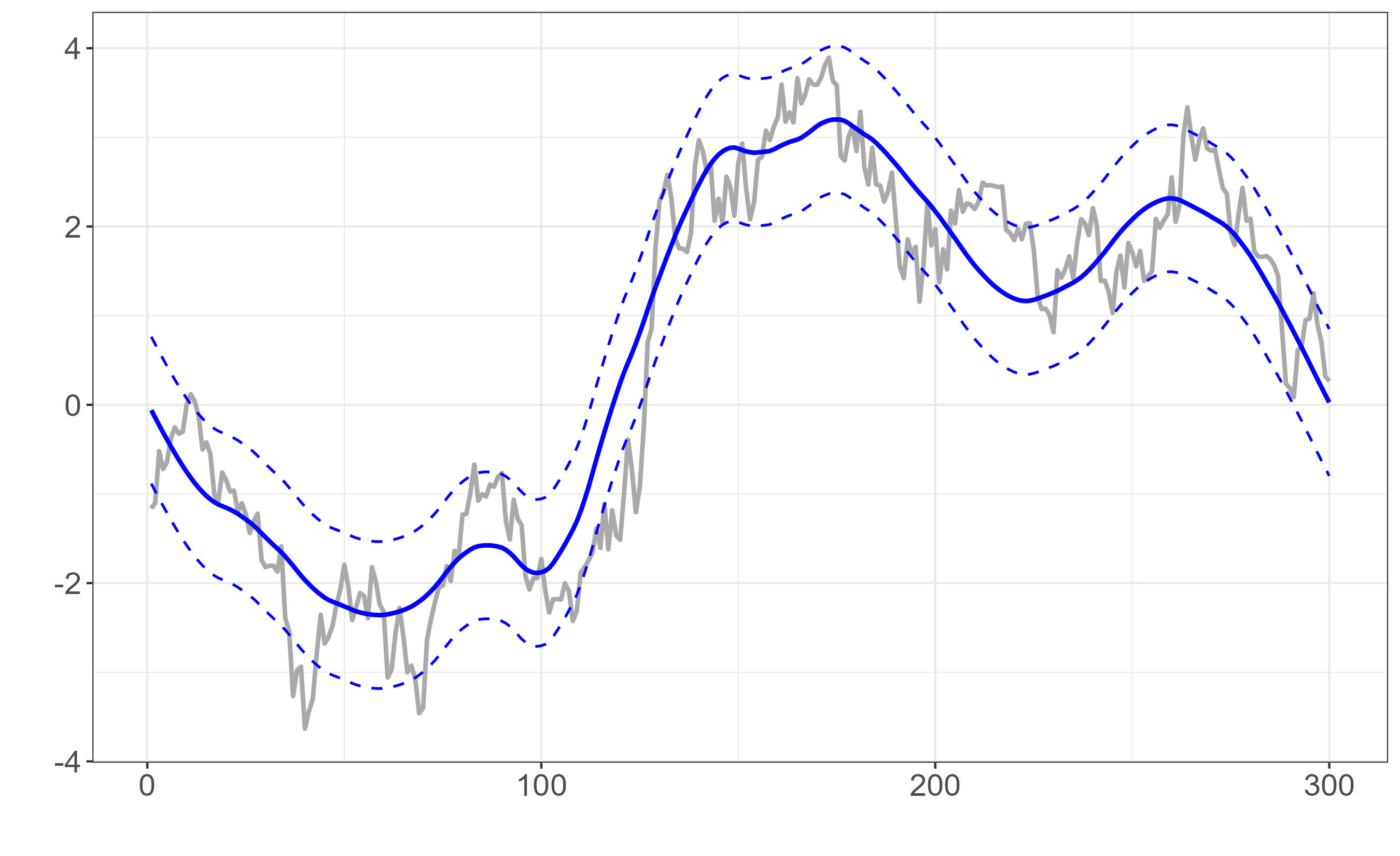}}
	\hspace{-2em}
	
\hspace{-2em}\subfigure[Identity $+$ Daubechies wavelet basis matrix]{\includegraphics[width=.52\textwidth]{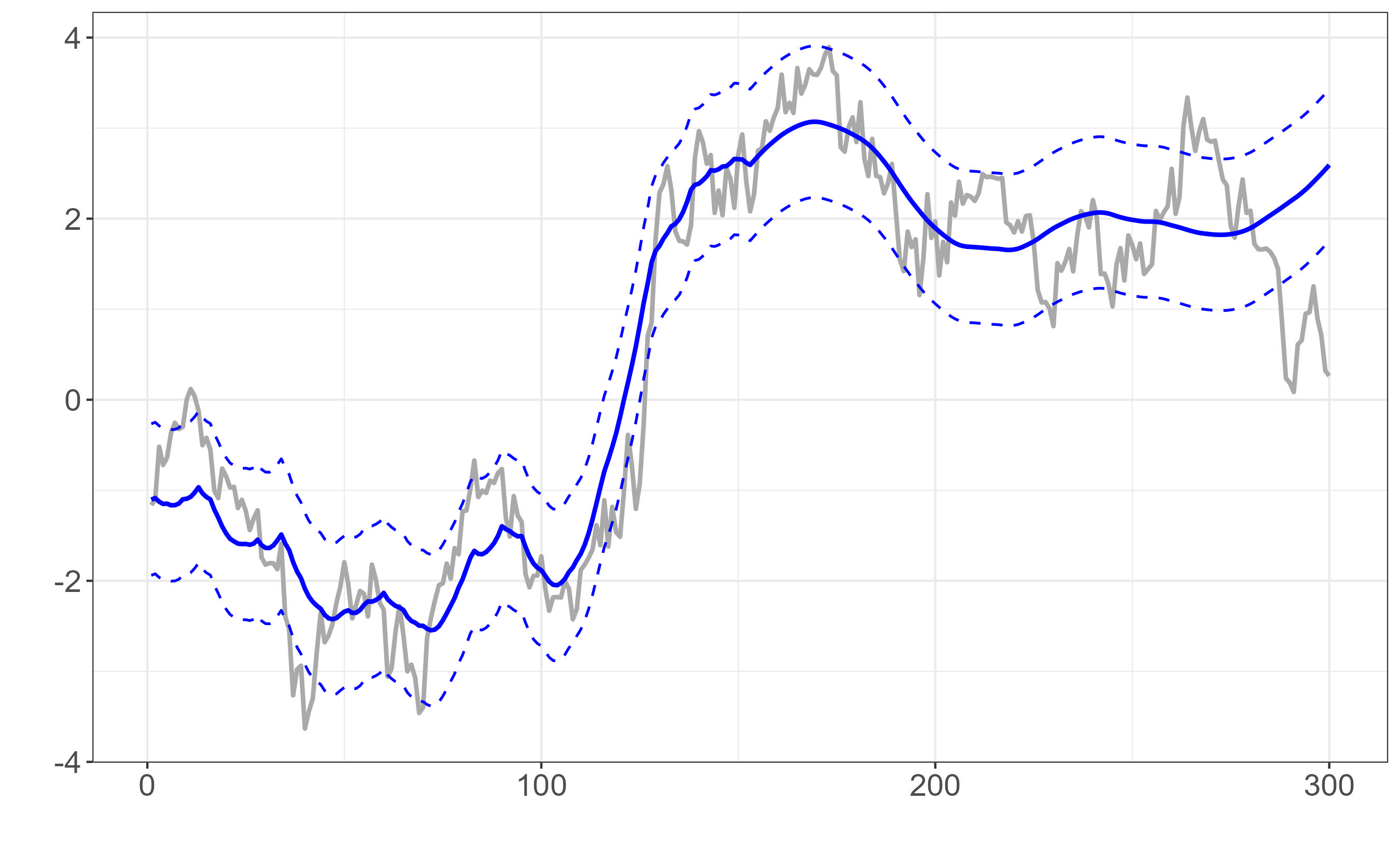}}\subfigure[B-spline basis matrix with $kn=20$ and $dg=3$]{\includegraphics[width=.52\textwidth]{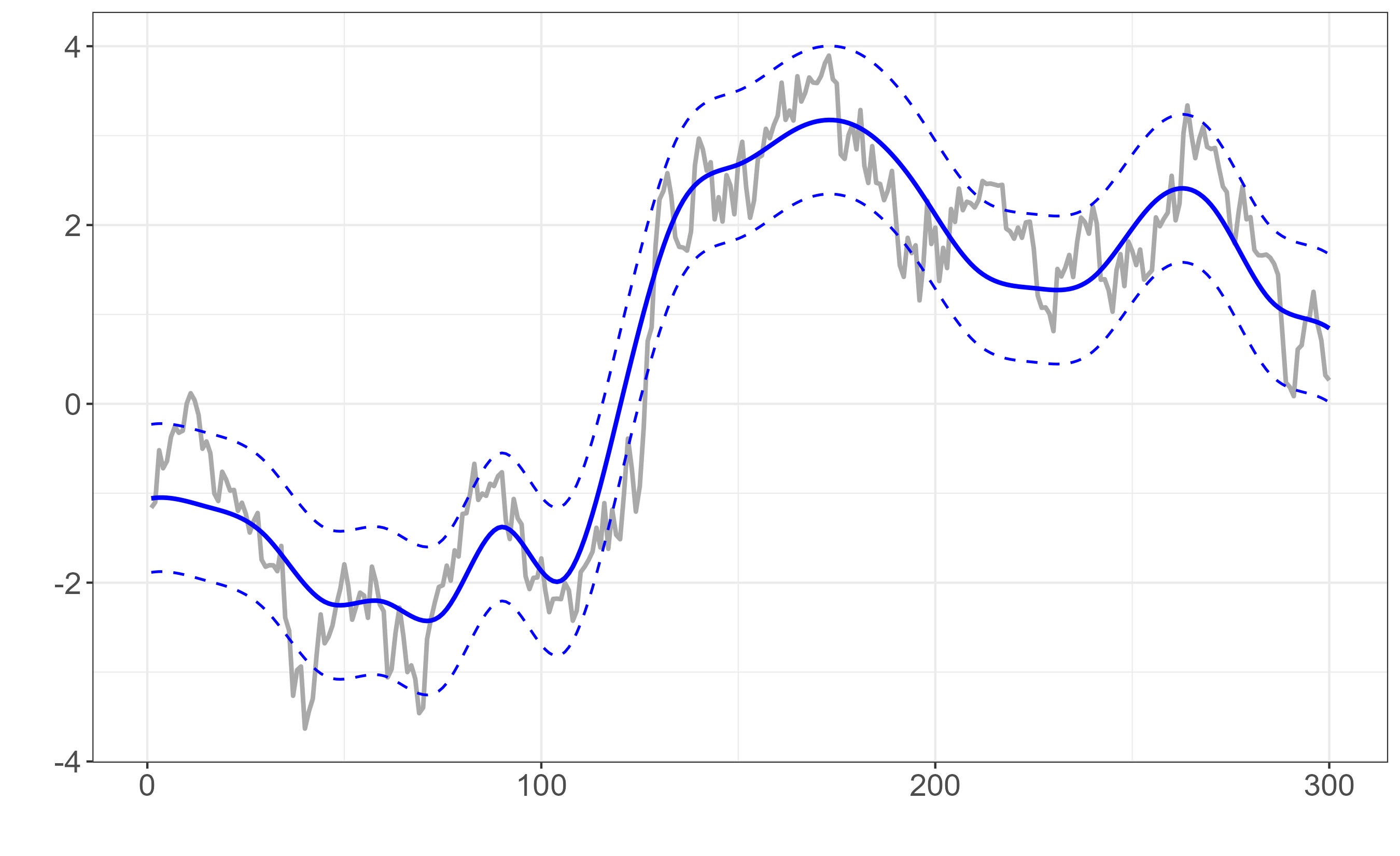}}\hspace{-2em}
\caption{\footnotesize \bf Shape of the Log-Volatility Estimates for Different $\mathbf{W}$\normalfont. This figure shows the posterior mean and credibility intervals for a single trajectory of simulated log-volatility based on different smoothing assumptions.}
\label{fig:choice_of_W}
\end{figure}

Figure \ref{fig:choice_of_W} shows examples of the shape of $\mathbf{\mu}_{q\left(\mathbf{h}\right)}=\mathbf{W}\mathbf{f}_{q(h)}$ for different choices of $\mathbf{W}$ (solid line), and the corresponding confidence intervals implied by $\mathbf{\Sigma}_{q(h)}$ (dashed line). The grey trajectory represents the true simulated value of the log-stochastic volatility $\mathbf{h}^\intercal=(h_0,h_1,\ldots,h_n)$ for $n=300$. The top-left panel reports the posterior estimates obtained by setting $\mathbf{W}=\mathbf{I}_{n+1}$, with $\mathbf{I}_{n+1}$ an identity matrix of dimension $n+1$. This represents a non-smooth estimate akin to the output of a standard MCMC estimation scheme \citep[see, e.g.,][]{stochvol_package}. 

The remaining panels of Figure \ref{fig:choice_of_W} highlight a key feature of our estimation strategy; that is, it allows us to customise the volatility forecasts without changing the underlying model assumptions. For instance, the top-right panel shows the posterior estimates of the latent volatility state with $\mathbf{W}$ a matrix of wavelet basis functions with a fixed degree of smoothness $l=4$ \citep[see][]{wand_ormerod.2011}. The fact that the matrix $\mathbf{W}$ enters both in the conditional mean and covariance of the optimal variational density $q^\ast\left(\mathbf{h}\right)$ allows for smoothing not only the conditional mean of the latent volatility state but also the corresponding confidence intervals. 

The bottom panels in Figure \ref{fig:choice_of_W} highlight the flexibility of our approach; the left panel shows that more than one smoothing assumption can coexist in the same optimal variational density. For instance, the shape of the posterior estimates assuming $\mathbf{W}=\mathbf{I}_{n+1}$ for the first half of the sample and $\mathbf{W}$ a wavelet basis function with $l=4$ for the second half of the sample. The bottom-right panel shows that a variety of smoothing functions can be adopted; for instance, the estimates of the latent stochastic volatility can be smoothed based on $\mathbf{W}$ equal to a B-spline basis matrix representing the family of piecewise polynomials with the pre-specified interior knots ($kn$), degree ($dg$), and boundary knots. 

This flexibility in choosing $\mathbf{W}$ allows for context-specific smoothing that can adapt to different market conditions or portfolio requirements. For volatility-managed portfolios, this translates to more stable portfolio weights during crisis periods while maintaining responsiveness to economically meaningful volatility changes, as demonstrated in the empirical analysis.

\clearpage
\section{Additional Results}

\subsection{Sample Sizes and Observations}
\label{app:samples}
Table \ref{tab:sample_periods_detailed} reports the sample start and end dates for each portfolio ised in the main empirical analysis. The sample spans almost a century of data, with the earliest observations starting in January 1928 and the most recent ending in March 2022. The shortest sample is Debtissuance (816 observations) and the longest samples have 1,128 observations. The varying start dates reflect data availability constraints for different portfolio construction methodologies, which is standard in the literature \citep{hou2020replicating,jensen2021there}.

\begin{table}[!htbp]
\centering
\renewcommand{\arraystretch}{1}
\resizebox{.6\textwidth}{!}{\begin{tabular}{lrrr}
\multicolumn{4}{l}{\textbf{Panel A}: Risk Factors} \\
\midrule 
Portfolio & Start Date & End Date & Observations \\
\midrule 
MktRF & July 1928 & March 2022 & 1,125 \\
SMB & July 1928 & March 2022 & 1,125 \\
HML & July 1928 & March 2022 & 1,125 \\
StRev & July 1928 & March 2022 & 1,125 \\
MOM & December 1928 & March 2022 & 1,120 \\
LtRev & April 1932 & March 2022 & 1,080 \\
Bab & December 1932 & February 2022 & 1,071 \\
RMW & July 1965 & March 2022 & 681 \\
CMA & July 1965 & March 2022 & 681 \\
ROE & January 1969 & December 2021 & 636 \\
IA & January 1969 & December 2021 & 636 \\
\midrule
\multicolumn{4}{l}{}\\
\multicolumn{4}{l}{\textbf{Panel B:} Characteristic-Based Themes} \\
\midrule 
Portfolio & Start Date & End Date & Observations \\
\midrule 
Size & January 1928 & December 2021 & 1,128 \\
Lowrisk & February 1928 & December 2021 & 1,127 \\
Lowleverage & February 1928 & December 2021 & 1,127 \\
Strev & February 1928 & December 2021 & 1,127 \\
Seasonality & February 1928 & December 2021 & 1,127 \\
Momentum & April 1928 & December 2021 & 1,125 \\
Profitability & April 1928 & December 2021 & 1,125 \\
Value & December 1928 & December 2021 & 1,117 \\
Quality & January 1929 & December 2021 & 1,116 \\
Profitgrowth & January 1929 & December 2021 & 1,116 \\
Investment & January 1933 & December 2021 & 1,068 \\
Accruals & January 1948 & December 2021 & 888 \\
Debtissuance & January 1954 & December 2021 & 816 \\
\midrule 
\end{tabular}}
\caption{\footnotesize \bf Sample Periods and Observations\normalfont.  This table reports the sample periods and number of monthly observations for the 24 portfolios used in the empirical analysis. Panel A shows 11 equity risk factors and Panel B shows 13 characteristic-based portfolio themes.}
\label{tab:sample_periods_detailed}
\end{table}

\subsection{Simulation Study}
\label{app:simulation}

Figure 2 in the main text suggests that the accuracy of our variational Bayes estimation framework deteriorates when smoothness on the latent state is imposed via the structure in $\mathbf{W}$. We now investigate in more detail why that is the case by looking at the posterior estimates of the parameters of interest $\{c,\eta^2,\rho\}$ for different specifications of $\mathbf{W}$.

\begin{figure}[!ht]
\centering
\vspace{1em}
    \subfigure[$\hat{c}$ when $\rho=0.98$]{\includegraphics[width=.32\textwidth]{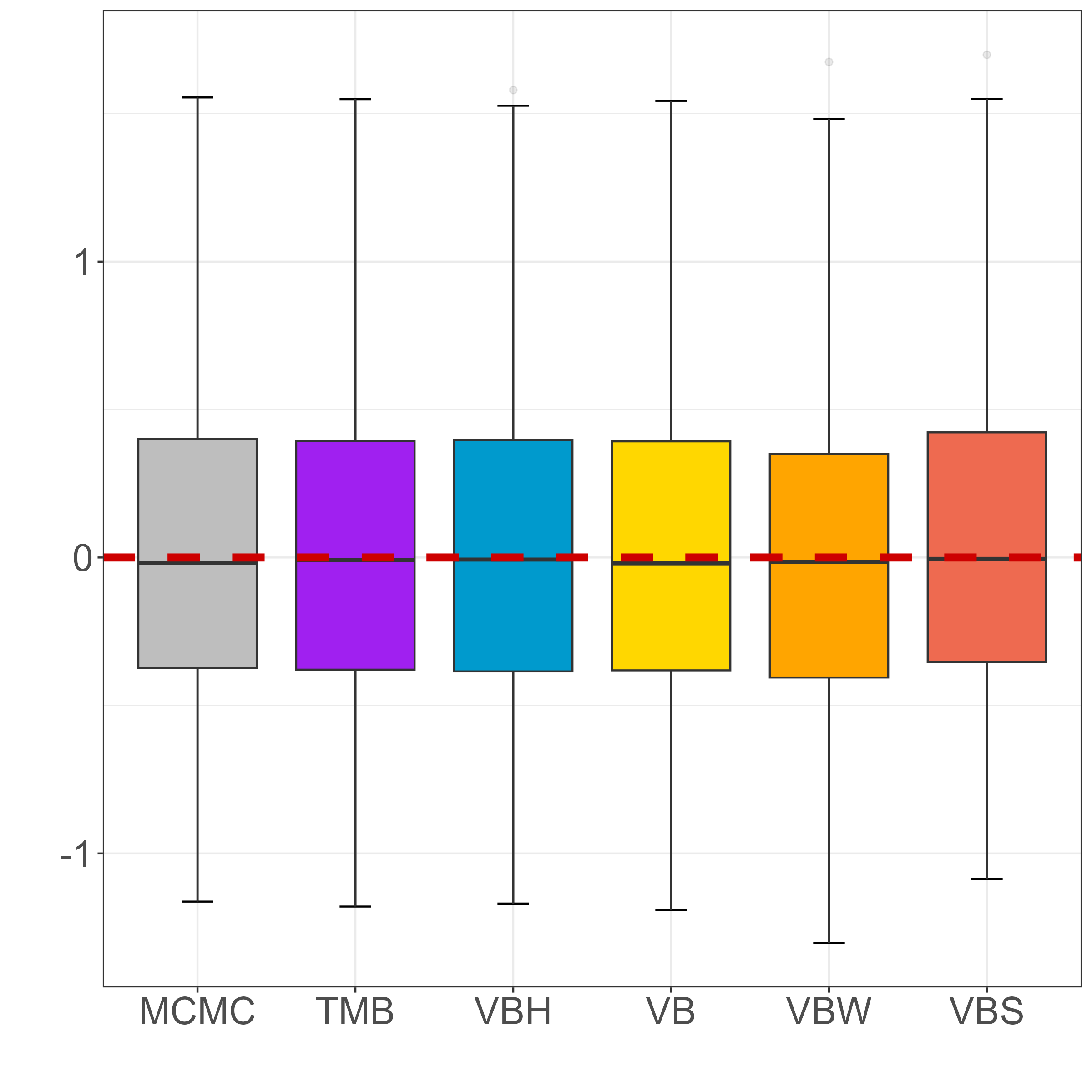}}
	\subfigure[$\hat{\eta}^2$ when $\rho=0.98$]{\includegraphics[width=.32\textwidth]{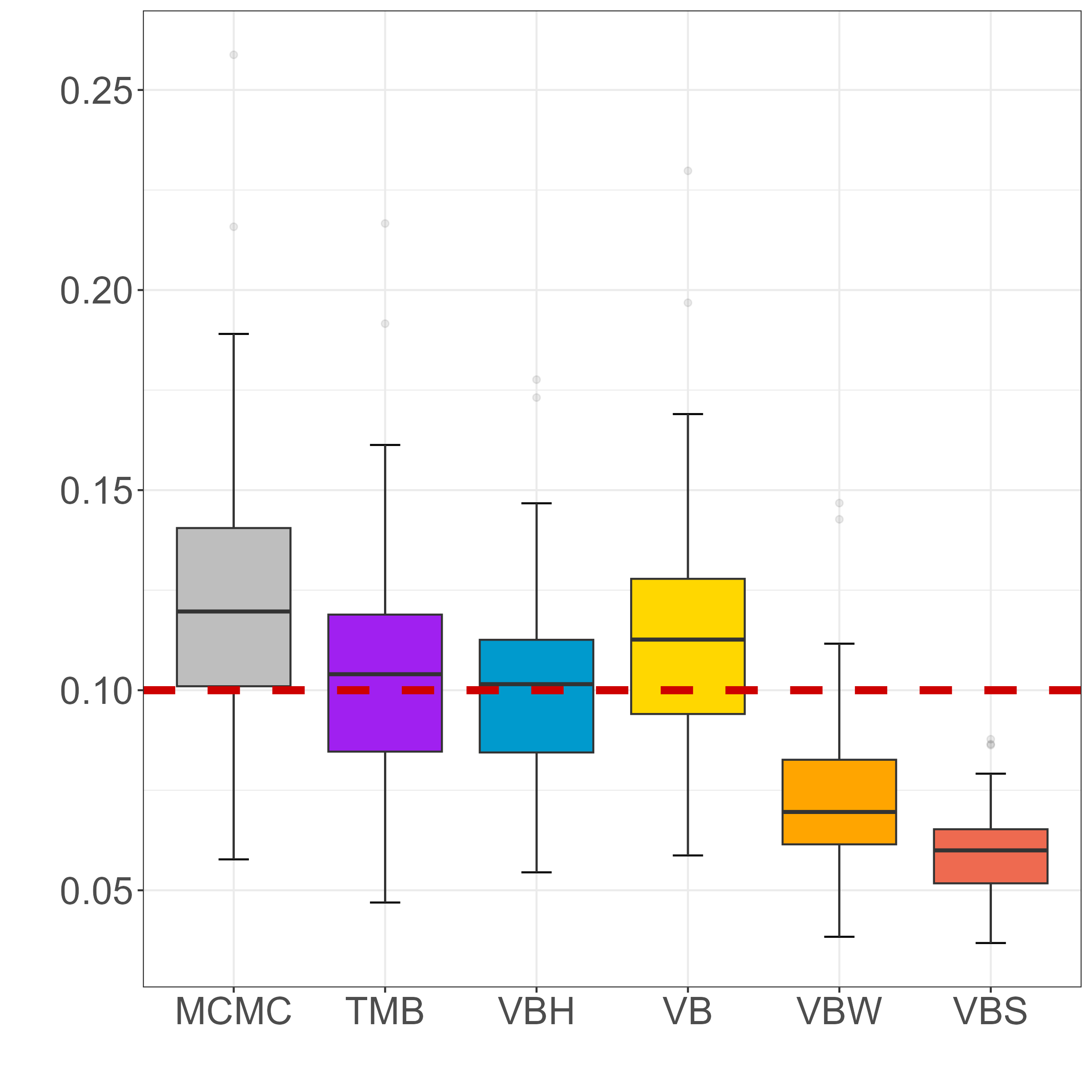}}
	\subfigure[$\hat{\rho}$ when $\rho=0.98$]{\includegraphics[width=.32\textwidth]{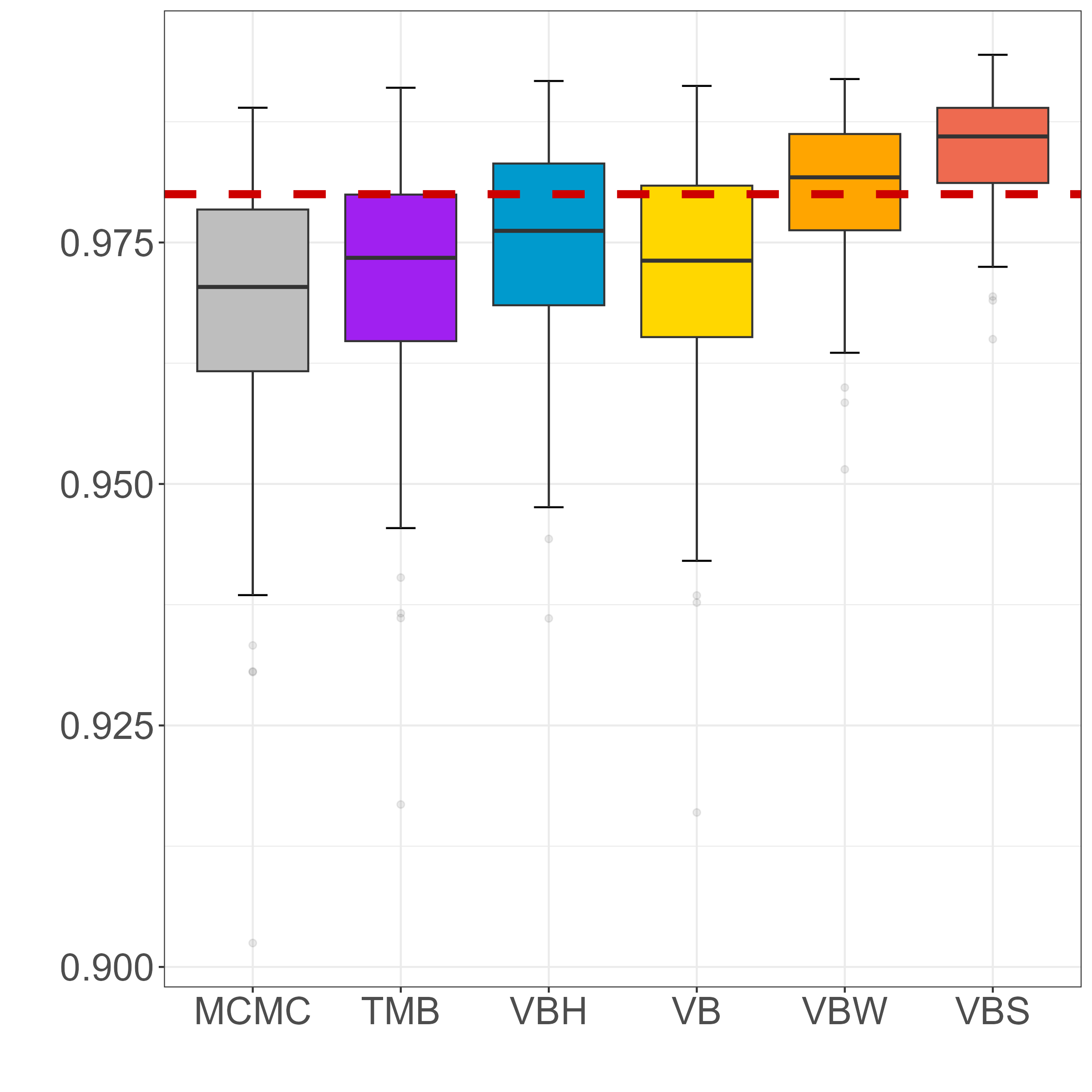}} \\
	\subfigure[$\hat{c}$ when $\rho=0.70$]{\includegraphics[width=.32\textwidth]{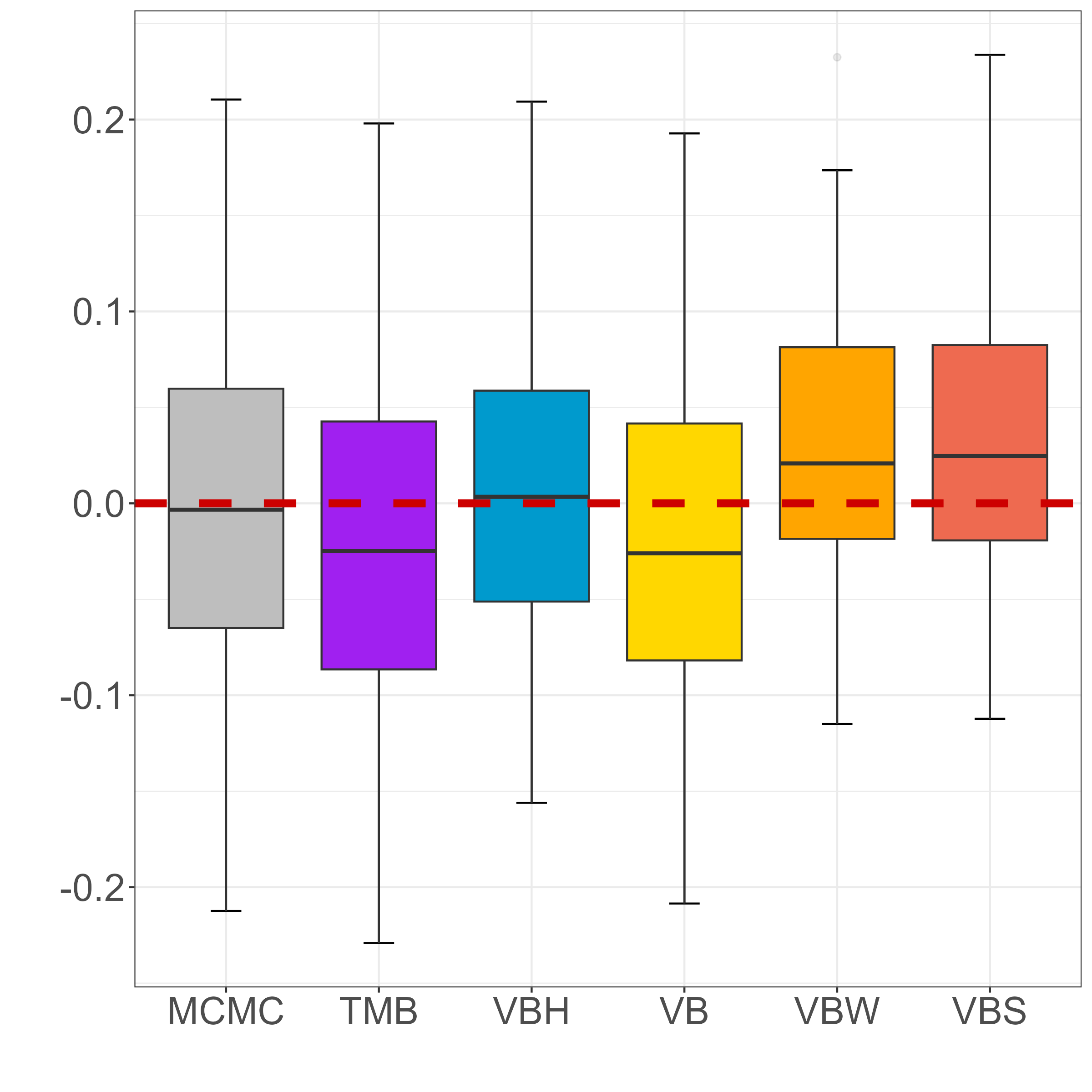}}
	\subfigure[$\hat{\eta}^2$ when $\rho=0.70$]{\includegraphics[width=.32\textwidth]{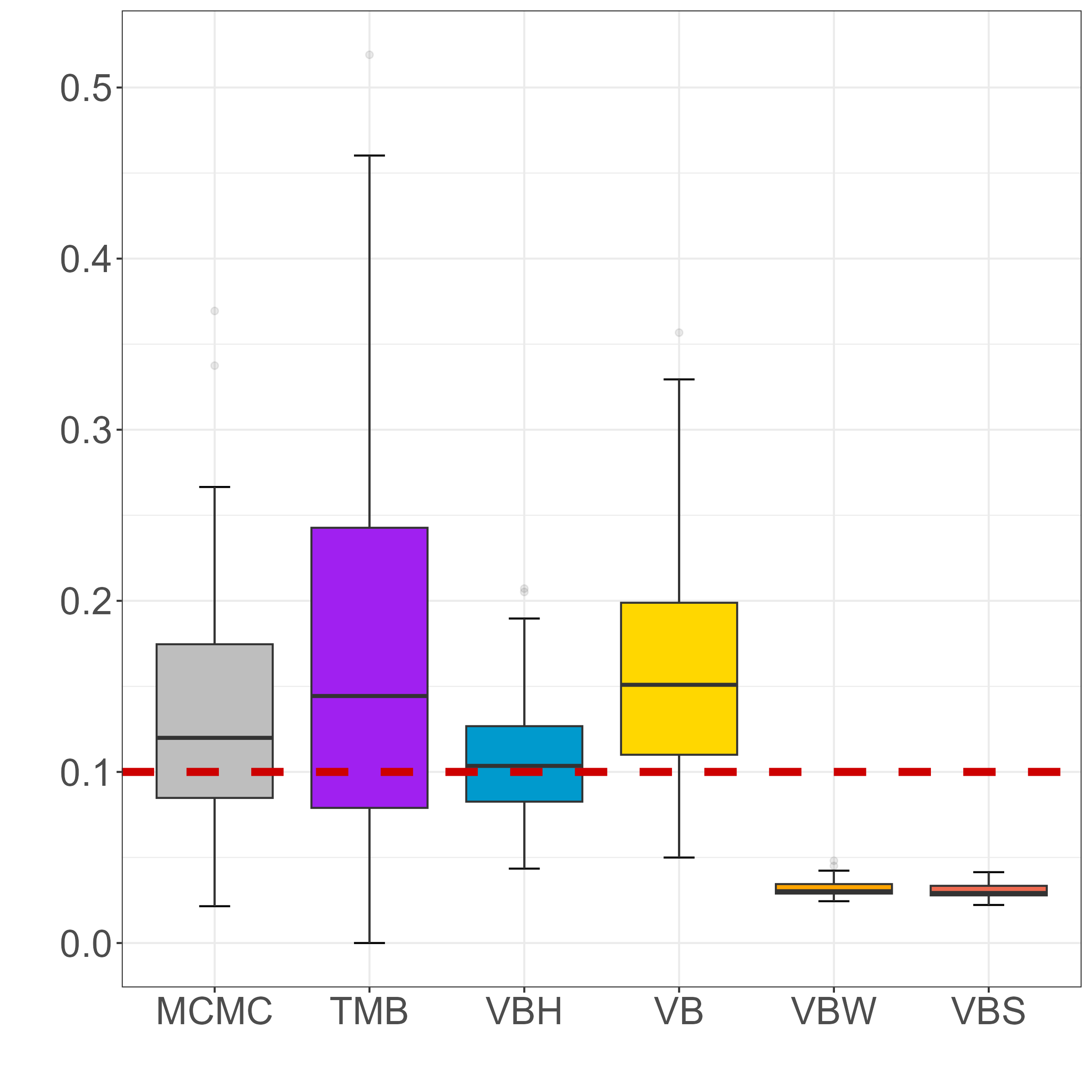}}
	\subfigure[$\hat{\rho}$ when $\rho=0.70$]{\includegraphics[width=.32\textwidth]{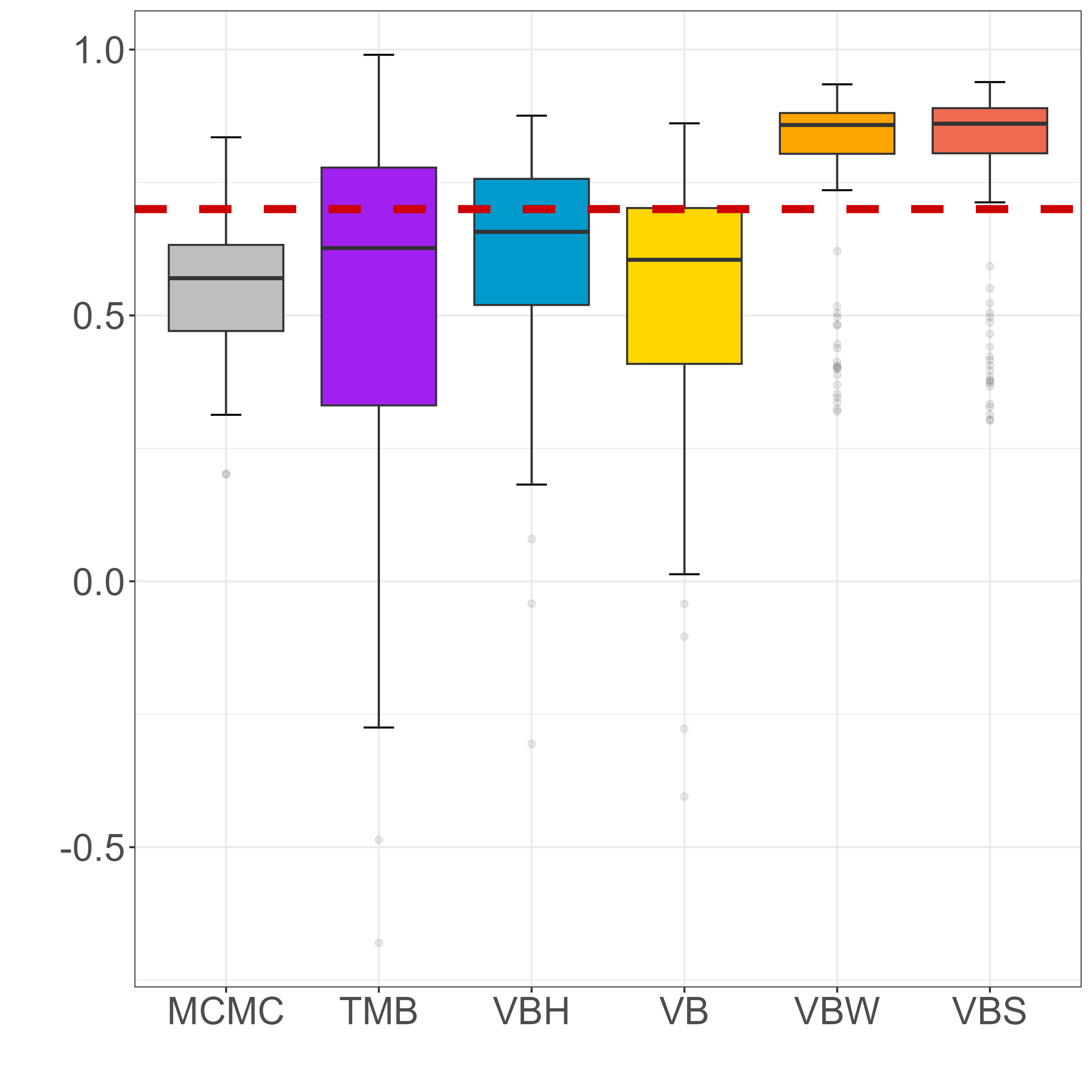}}
\caption{\footnotesize \bf Estimates of the Parameters of the Latent State\normalfont. This figure reports the posterior estimates of the parameters of interest for the stochastic volatility models across simulations and for different inference methods. We report the simulation results for $\rho=0.98$ (top panels) and $\rho=0.7$ (bottom panels). We compare our variational Bayes methods, with and without smoothing, against a standard MCMC and the TMB approach. The red dashed line denotes the true value of the parameter used in the simulations.}   
    \label{fig:sim_par_hat}
\end{figure}

Figure \ref{fig:sim_par_hat} shows that imposing smoothness in the form of either B-spline or a Daubechies wavelet basis forces the posterior estimates of $\rho$ to be close to one, irrespective of the actual level of persistence in the underlying latent process. Similarly, the estimates of the latent state variance $\eta^2$ are smaller for both {\tt VBS} and {\tt VBW} versus {\tt MCMC}'s, and even more so when $\rho=0.7$. This confirms the intuition that a lower accuracy of the posterior estimates of the latent state is due to tight regularization of the parameters implied by smoothing.

\subsection{Forecasting Accuracy vs Economic Utility}
\label{subsec:forecasting accuracy}

The main text demonstrates a fundamental disconnect between forecasting accuracy and portfolio performance using aggregate statistics. Here we provide additional insights by examining the complete distribution of results across all portfolios.

\begin{figure}[!ht]
\centering
	\subfigure[Themes portfolios]{\includegraphics[width=.45\textwidth]{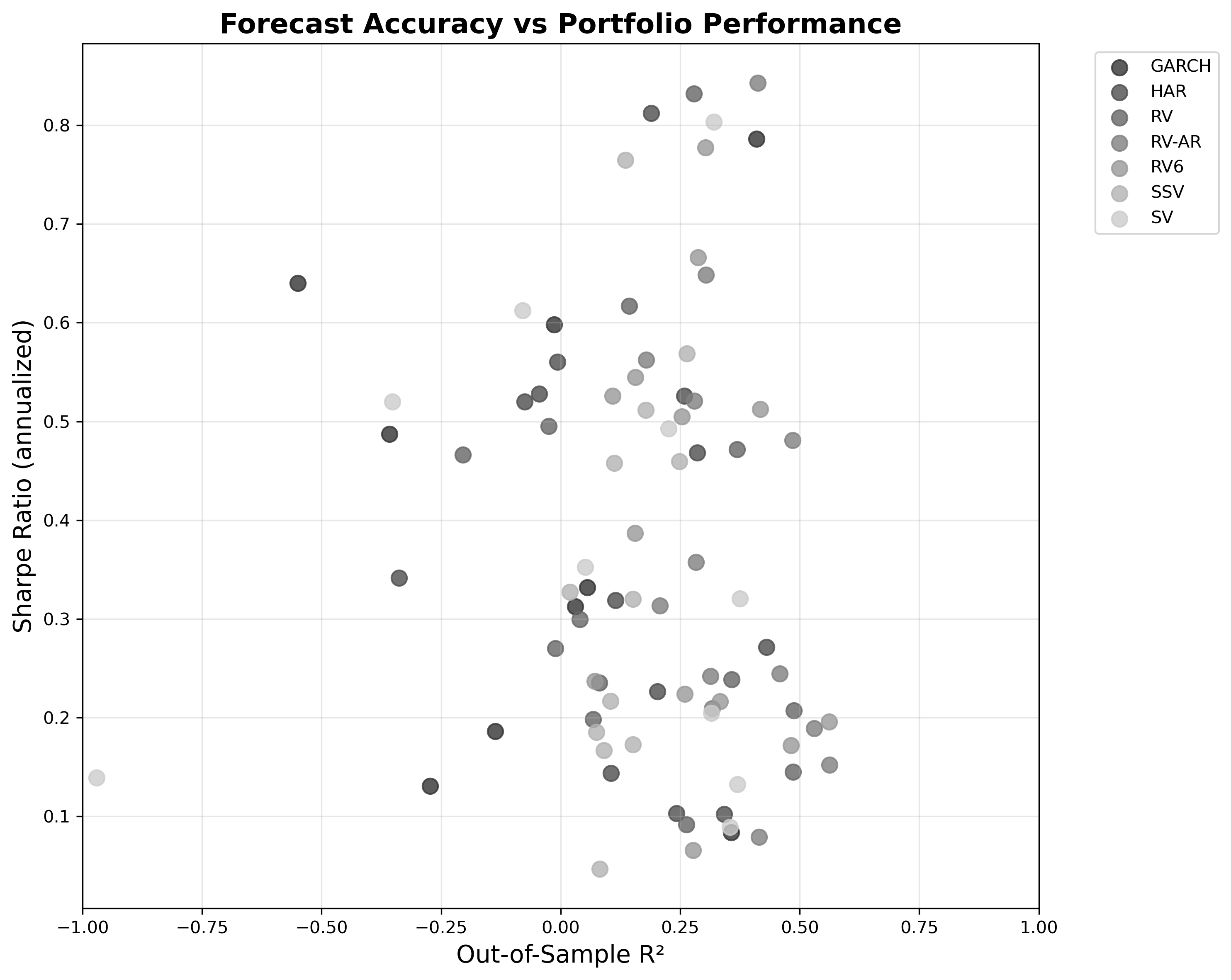}}
	\subfigure[Risk factors]{\includegraphics[width=.45\textwidth]{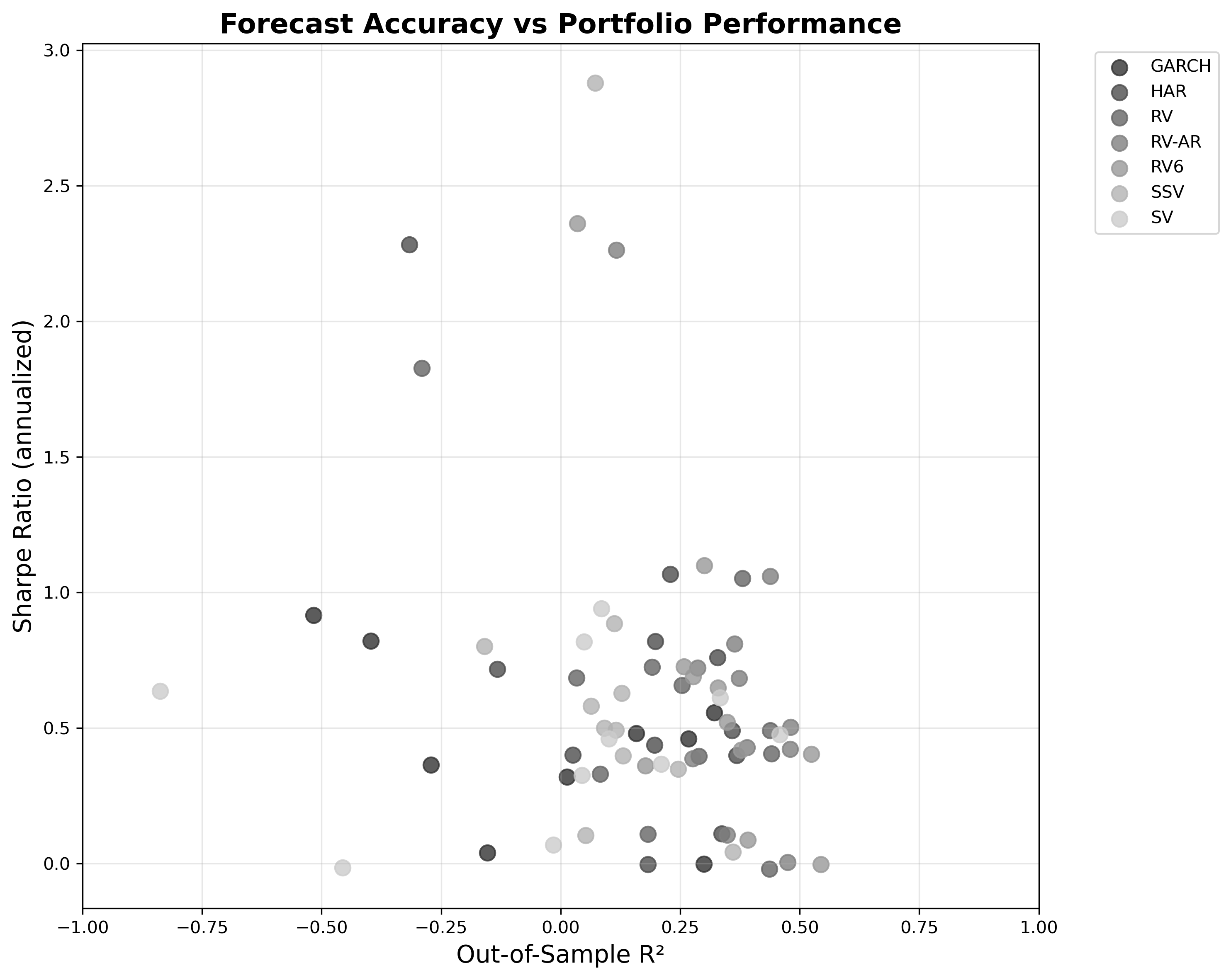}} 
\caption{\footnotesize \bf Forecasting Accuracy vs. Portfolio Performance\normalfont. These scatterplots compare the annualized Sharpe ratios and the $R_{oos}^2$ obtained from each model across all portfolio themes (left panel) and risk factors (right panel). Each point represents one volatility model applied to one portfolio. The weak correlation demonstrates that superior forecasting accuracy does not guarantee superior portfolio outcomes.}    
	\label{fig:tradeoff accuracy performance}
\end{figure}

Table \ref{tab:oos_r2_comparison} presents the detailed out-of-sample $R_{oos}^2$ results, while Figure \ref{fig:tradeoff accuracy performance} visualizes the relationship between forecasting accuracy and economic performance for each portfolio-method combination.

\begin{table}[hp!]
\centering
\renewcommand{\arraystretch}{1}
\resizebox{.8\textwidth}{!}{
   \begin{tabular}{lrrrrrrrr}
          &       & \multicolumn{7}{c}{Out-of-Sample $R^2$}                                                    \\
                    \cmidrule{3-9}
          &       & \texttt{RV} & \texttt{RV6} & \texttt{RV-AR(1)} & \texttt{HAR} & \texttt{GARCH} & \texttt{SV} & \texttt{SSV} \\
          \midrule 
                    & \multicolumn{1}{l}{MktRF} & 0.082 & 0.177 & 0.276 & 0.026 & 0.158 & 0.100 & 0.091 \\
          & \multicolumn{1}{l}{SMB} & 0.183 & 0.392 & 0.348 & 0.182 & {-0.153} & {-0.015} & 0.052 \\
          & \multicolumn{1}{l}{HML} & 0.441 & 0.524 & 0.480 & 0.368 & {-0.271} & 0.044 & 0.245 \\
          & \multicolumn{1}{l}{RMW} & 0.438 & 0.348 & 0.481 & 0.196 & 0.267 & 0.458 & 0.115 \\
          & \multicolumn{1}{l}{CMA} & 0.289 & 0.377 & 0.390 & 0.358 & 0.013 & 0.210 & 0.131 \\
    Factors & \multicolumn{1}{l}{MOM} & 0.033 & 0.258 & 0.287 & {-0.132} & {-3.332} & {-0.838} & 0.064 \\
          & \multicolumn{1}{l}{St Rev} & {-0.290} & 0.035 & 0.116 & {-0.317} & {-1.148} & {-0.679} & 0.072 \\
          & \multicolumn{1}{l}{Lt Rev} & 0.437 & 0.544 & 0.475 & 0.337 & 0.300 & {-0.456} & 0.361 \\
          & \multicolumn{1}{l}{Bab} & 0.191 & 0.277 & 0.364 & 0.198 & {-0.397} & 0.049 & {-0.160} \\
          & \multicolumn{1}{l}{IA} & 0.253 & 0.329 & 0.373 & 0.328 & 0.321 & 0.334 & 0.128 \\
          & \multicolumn{1}{l}{ROE} & 0.380 & 0.301 & 0.438 & 0.229 & {-0.517} & 0.085 & 0.112 \\
          &       &       &       &       &       &       &       &  \\
                    \midrule 
                           & \multicolumn{1}{l}{Accruals} & {-0.204} & 0.108 & 0.178 & {-0.007} & {-0.358} & {-0.352} & {-1.657} \\
          & \multicolumn{1}{l}{Debt issuance} & 0.279 & 0.303 & 0.412 & 0.190 & 0.409 & 0.320 & 0.136 \\
          & \multicolumn{1}{l}{Investment} & 0.486 & 0.562 & 0.562 & 0.430 & {-1.919} & {-1.814} & 0.090 \\
          & \multicolumn{1}{l}{Low leverage} & 0.488 & 0.482 & 0.531 & 0.342 & {-0.137} & 0.315 & 0.151 \\
          & \multicolumn{1}{l}{Low risk} & 0.263 & 0.276 & 0.415 & 0.242 & 0.356 & 0.354 & 0.082 \\
          & \multicolumn{1}{l}{Momentum} & {-0.025} & 0.156 & 0.279 & {-0.076} & {-2.247} & {-1.141} & 0.112 \\
    Themes & \multicolumn{1}{l}{Profit growth} & 0.143 & 0.288 & 0.304 & 0.259 & {-0.550} & {-0.080} & 0.264 \\
          & \multicolumn{1}{l}{Profitability} & 0.068 & 0.071 & 0.317 & {-0.338} & 0.031 & 0.052 & 0.020 \\
          & \multicolumn{1}{l}{Quality} & 0.369 & 0.418 & 0.485 & 0.285 & {-5.888} & {-4.139} & 0.248 \\
          & \multicolumn{1}{l}{Seasonality} & {-0.011} & 0.253 & 0.283 & {-0.045} & {-0.013} & 0.226 & 0.178 \\
          & \multicolumn{1}{l}{Size} & 0.081 & 0.259 & 0.313 & 0.105 & {-11.810} & {-0.970} & 0.074 \\
          & \multicolumn{1}{l}{St Rev} & 0.040 & 0.156 & 0.208 & 0.114 & 0.055 & 0.375 & 0.151 \\
          & \multicolumn{1}{l}{Value} & 0.357 & 0.333 & 0.458 & 0.203 & {-0.273} & 0.369 & 0.105 \\
\midrule 
Average & 0.203 & 0.312 & 0.380 & 0.162 & {-0.891} & {-0.253} & 0.149 \\
\midrule 
    \end{tabular}}
    \caption{\footnotesize \bf Out-of-Sample $R^2$\normalfont. This table reports the $R_{oos}^2$ across different models for all portfolio themes and risk factors. Higher values indicate better forecasting accuracy relative to a naive average forecast. Negative values indicate performance worse than the naive benchmark.}
\label{tab:oos_r2_comparison}
\end{table}

Beyond the aggregate patterns reported in the main text, several detailed insights emerge. Traditional \texttt{RV} methods show particularly poor forecasting accuracy for momentum-based portfolios (negative $R_{oos}^2$ values), precisely where volatility management delivers the largest economic gains. This suggests that the relationship between statistical accuracy and economic utility depends critically on the underlying return-generating process.

\texttt{GARCH}-based methods exhibit extreme variance in forecasting accuracy across portfolios, ranging from -11.810 (Size portfolio) to 0.409 (Debt issuance). This instability translates to unreliable portfolio performance, explaining why GARCH achieves only moderate economic utility despite reasonable average forecasting accuracy.

Our \texttt{SSV} method demonstrates consistent moderate forecasting accuracy across all portfolios (range: -1.657 to 0.361), avoiding the extreme negative values that plague other methods. This consistency translates to reliable economic performance, supporting the design principle of optimizing implementable utility rather than statistical metrics.

\paragraph{Comparing volatility estimates}

\begin{figure}[!ht]
\centering
	\includegraphics[width=1\textwidth]{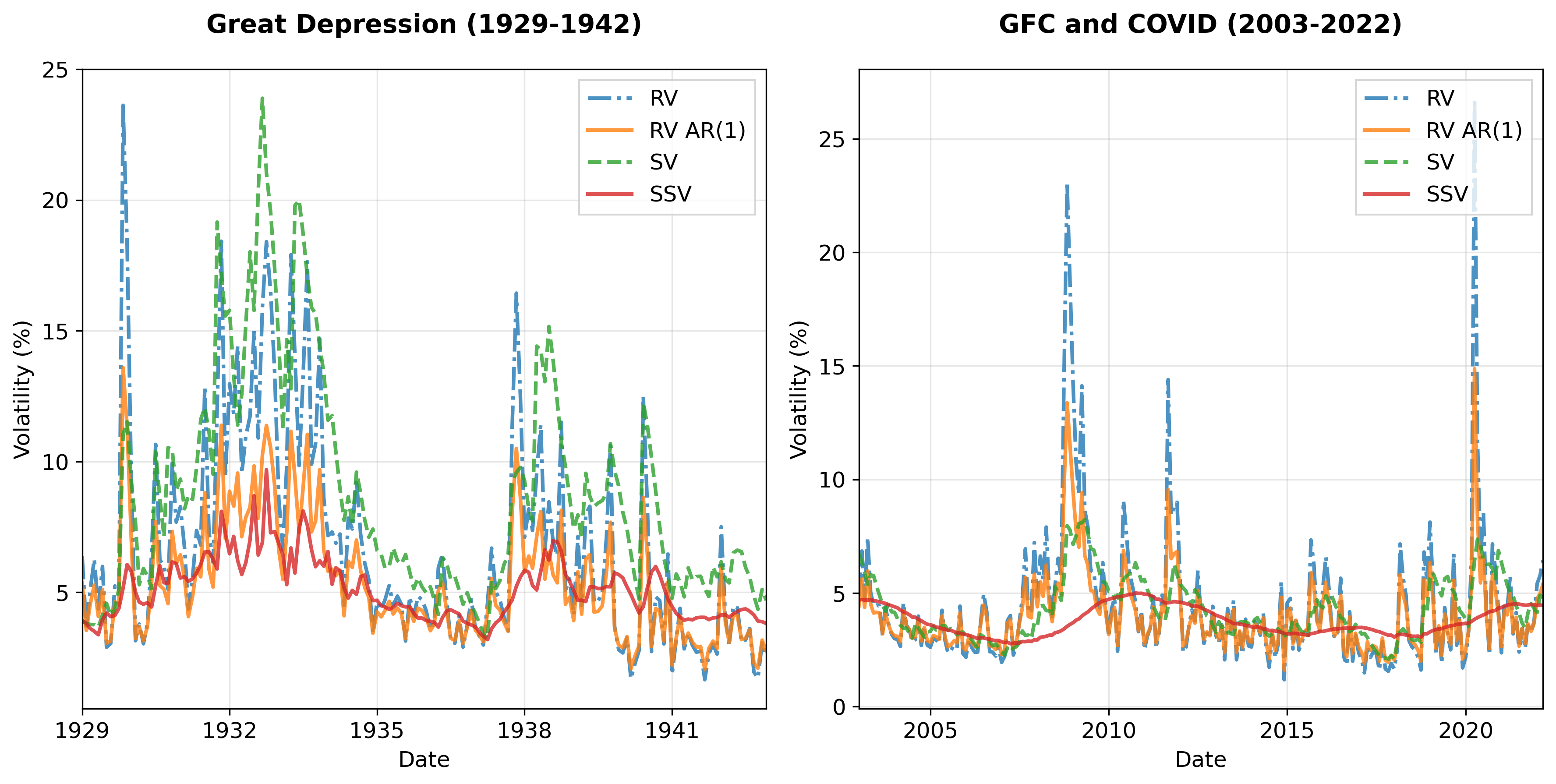}
\caption{\footnotesize \bf Market volatility forecast comparison\normalfont. These figures show the different forecasts of market volatility $\sqrt{\sigma_{t+1|t}^2}$ for the \texttt{RV}, \texttt{RV AR(1)}, \texttt{SV} and \texttt{SSV} models.}    
	\label{fig:volatility comparison}
\end{figure}

Figure \ref{fig:volatility comparison} shows the different forecasts of market volatility $\sqrt{\sigma_{t+1|t}^2}$ for the \texttt{RV}, \texttt{RV AR(1)}, \texttt{SV} and \texttt{SSV} models. The left panel highlights the period around the Great Depression whereas the right panel focuses on the period around the Great Financial Crisis (GFC) and the COVID-19 outbreak.

These plots provide compelling visual evidence for our main empirical findings. The volatility forecasts differ dramatically during crisis periods—precisely when accurate volatility management is most crucial for portfolio performance. During the Great Depression, GFC (2008–2009), and COVID (2020), realized variance (\texttt{RV}) exhibits extreme spikes reaching 20–25\% (annualized standard deviation), while our \texttt{SSV} method maintains much greater stability around 5–10\%.

This difference has direct economic implications. Since volatility-managed portfolio weights are calculated as $\omega_t = \gamma_t/\sigma^2_{t+1|t}$, these standard deviation spikes translate to squared effects on portfolio rebalancing. When \texttt{RV} spikes from 5\% to 20\% (a 4$\times$ increase in volatility), the variance increases 16-fold, triggering massive deleveraging and rebalancing costs. This mechanism directly explains our empirical findings: \texttt{RV}-based strategies generate average portfolio turnover of 1.058 versus \texttt{SSV}'s 0.032 (a 97\% reduction).

During stable periods (mid-1930s, 2010–2019), all methods converge to similar levels around 2–5\%, confirming that the practical advantages of smoothing are concentrated precisely when they matter most—during crisis periods when transaction costs and liquidity constraints are most binding.

The visual evidence demonstrates that \texttt{SSV} effectively filters economically irrelevant volatility fluctuations while preserving responses to meaningful volatility shifts. This supports the design principle underlying our approach: Daubechies wavelets naturally decompose volatility into different frequency components, allowing us to preserve economically relevant low-frequency volatility trends while filtering high-frequency noise that generates excessive turnover.

These implementation characteristics translate to superior net-of-costs performance: \texttt{SSV} reduces average portfolio turnover by 97\% (from 1.058 to 0.032) compared to \texttt{RV}-based approaches, while maintaining competitive risk-adjusted returns. The plots provide compelling visual evidence that smooth volatility forecasts are not merely academically interesting but practically essential for implementable volatility-managed portfolios during periods of market stress.

\subsection{Sub-Sample Performance}
\label{subsec:subsample}

\begin{figure}[!ht]
\centering
	\includegraphics[width=1\textwidth]{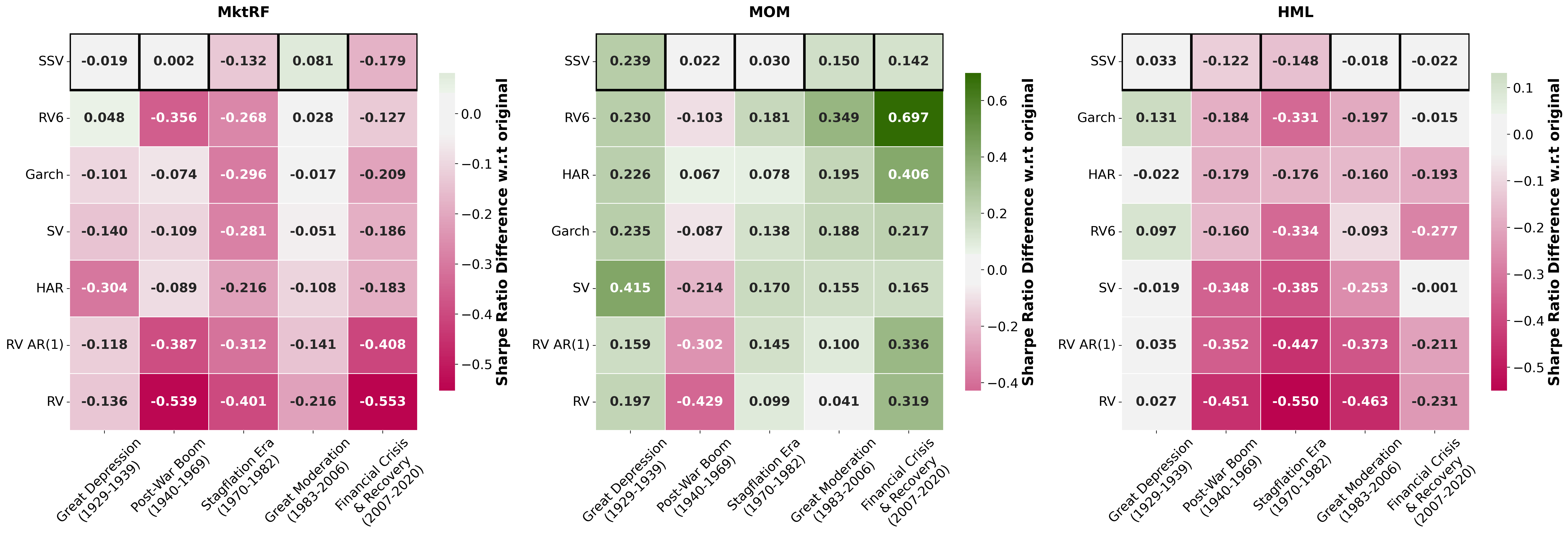}
    \caption{\footnotesize \bf Temporal Evolution of Volatility-Managed Portfolio Performance.\normalfont Sharpe ratio differences between each volatility-targeting strategy and the original portfolio, across major historical market periods. Each subplot corresponds to a different portfolio (MktRF, MOM, HML). Green (positive) values indicate outperformance relative to the original, while red (negative) values indicate underperformance. The {\tt SSV} row is highlighted with black borders for emphasis.}    
	\label{fig:performance per period}
\end{figure}

Figure \ref{fig:performance per period} presents the detailed temporal analysis referenced in the main text, showing Sharpe ratio differences across five historical market periods for representative portfolios (MktRF, MOM, HML).

The complete results confirm the main text findings: \texttt{SSV} maintains consistent positive performance across all periods for momentum strategies, while traditional methods show period-dependent performance with extreme swings. The visual evidence supports our conclusion that \texttt{SSV}'s advantages stem from systematic methodological improvements rather than favorable timing effects.

\end{document}